\pdfoutput=1
\documentclass[12pt]{article}
\newif\ifdraft
\draftfalse 

\usepackage[textwidth=15.5cm,textheight=22.5cm]{geometry}

\usepackage{hyperref}

\usepackage[force]{feynmp-auto}
\usepackage{relsize}

\usepackage{cite}

\usepackage{bm}
\usepackage{amsmath,amssymb,color,graphicx,amscd,amsfonts}
\usepackage{latexsym}
\usepackage{graphicx}
\usepackage{float}
\graphicspath{{fig/}}
\usepackage{subfig}
\usepackage{bbm}
\usepackage{datetime}
\usepackage{pdfpages}

\ifdraft
\usepackage{refcheck}
\usepackage{showlabels}
\fi
\usepackage{amsthm}

\usepackage{empheq}

\usepackage{framed}

\def\lef{\left}
\def\ri{\right}

\newcommand{\linefill}{% a variation on \rightarrowfill
  {-}\mkern-7mu
  \cleaders\hbox{$\mkern-2mu-\mkern-2mu$}\hfill
  \mkern-7mu{-}%
}
\def\<{\langle}
\def\>{\rangle}

\def\bPsi{\overline{\Psi}}

\def\bPhi{\overline{\Phi}}

\def\bmO{\bar{\mathcal{O}}}

\def\mO{\mathcal{O}}
\def\mJ{\mathcal{J}}

\def\a{\alpha}

\def\d{\delta}
\def\e{\epsilon}
\def\l{\lambda}

\def\n{\nu}

\def\t{\tau}

\def\tC{{\tilde{C}}}

\def\der{\partial}

\DeclareMathOperator{\sgn}{sgn}
\DeclareMathOperator{\ph}{ph}

\allowdisplaybreaks[4]

\numberwithin{equation}{section}

\newcommand{\ndt}{\noindent}
\def\sla{\raise.15ex\hbox{$/$}\kern-.57em}

\allowdisplaybreaks[1]

\newcommand{\R}{\mathbbm{R}}

\def\a{\alpha}

\def\G{\Gamma}

\def\d{\delta}
\def\D{\Delta}

\newcommand{\ie}{{\it i.e.}}

\newcommand\vsp[1]{\vspace*{#1 cm}}
\begin{document}

\setcounter{page}{1}

\begin{flushright} 
YITP-21-26\\
%DRAFT
\ifdraft
\yyyymmdddate\today, \currenttime\\[0.2cm]
\fi
%/DRAFT

\end{flushright} 

\vspace{0.1cm}

\begin{center}
{\LARGE
%TITLE
Exact four-point function and OPE \\ for an interacting quantum field theory \\ with
space/time anisotropic scale invariance
}
\end{center}

\vspace*{0.2cm}

\renewcommand{\thefootnote}{\alph{footnote}}

\begin{center}
Hidehiko S{\sc himada}\,$^{\ddag,}$\footnote
{
E-mail address: shimada.hidehiko@gmail.com},
and
Hirohiko S{\sc himada}\,$^{\sharp\,\natural\,}$\footnote
         {
E-mail address: hirohikoshimada@gmail.com} 

\vspace{0.3cm}

${}^{\ddag}$ {\it Yukawa Institute for Theoretical Physics, Kyoto University \\
Kitashirakawa Oiwakecho, Sakyo-ku, Kyoto 606-8502 Japan
}\\[0.2cm]
${}^\sharp$ {\it 
National Institute of Technology, Tsuyama College
\\
624-1, Numa, Tsuyama, Okayama 708-8509 Japan
} \\ [0.2cm]
${}^\natural$ {\it
Mathematical and Theoretical Physics Unit, OIST \\
Graduate University, Onna, Okinawa, 904-0495, Japan
} \\ [0.2cm]
\end{center}

\setcounter{footnote}{0}
\renewcommand{\thefootnote}{\arabic{footnote}}

%%%\vspace{1cm}
\abstract{
We identify a nontrivial yet tractable 
quantum field theory model with
space/time anisotropic scale invariance,
for which one can exactly compute certain four-point correlation functions
and their decompositions via the operator-product expansion(OPE).
The model is the Calogero model, 
non-relativistic particles 
interacting with a pair potential $\frac{g}{|x-y|^2}$
in one dimension, considered as a quantum field theory
in one space and one time dimension via the second quantisation.
This model has the anisotropic scale symmetry
with the anisotropy exponent $z=2$.
The symmetry is also enhanced to the Schr\"odinger symmetry.
The model has one coupling constant $g$ and thus provides
an example of a fixed line in the renormalisation group flow 
of anisotropic theories.

We exactly compute a nontrivial four-point function
of the fundamental fields of the theory.
We decompose the four-point function via OPE in two 
different ways, thereby
explicitly verifying the associativity of OPE for the first time
for an interacting quantum field theory with anisotropic scale invariance.
From the decompositions, one can read off the OPE coefficients 
and the scaling dimensions of the operators appearing in the
intermediate channels.
One of the decompositions is given by a convergent series, 
and only one primary operator and its descendants appear in the OPE.
The scaling dimension of the primary operator we computed depends on the coupling constant.
The dimension correctly reproduces the value expected
from the well-known spectrum of the Calogero model combined with 
the so-called state-operator map which is valid 
for theories with the Schr\"odinger symmetry.
The other decomposition is given by an asymptotic series.
The asymptotic series comes with exponentially small correction terms, 
which also have a natural interpretation in terms of OPE.
}

\tableofcontents

\section{Introduction}

The concept of the renormalisation group underlies
the universality in various critical phenomena~\cite{wilson_renormalization_1974}.
A quantum field theory with (isotropic) scale invariance  
is a fixed point of the renormalisation group flow in the space of 
quantum field theories and represents a universality class.

Quantum field theories invariant 
under space/time {\it anisotropic} scale transformation,
\begin{align}
\begin{split}
\vec{x} &\mapsto \alpha \vec{x},
\\
t  &\mapsto \alpha^{ z} t, 
\end{split}
\end{align}
are also of interest.
The exponent $z\neq 1$ characterises the degree of anisotropy of the system. 
These theories are also 
fixed points of the renormalisation group flow in the generalised 
theory space
of anisotropic quantum field theories.~\footnote{
In general, the renormalisation group flow connects 
two anisotropic theories characterised by different $z$.}

Because of this, these theories are also quite universal.
There are many applications of 
quantum field theory models with anisotropic scale invariance.
To illustrate the richness of the applications,
let us list a few examples:
dynamical critical phenomena in which 
time-dependent fluctuations around a critical point
are considered~\cite{glauber_time-dependent_1963, halperin_renormalization-group_1974},
quantum critical phenomena~\cite{hertz_quantum_1976}, 
more general non-equilibrium critical phenomena
such as the directed percolation
universality class~\cite{migdal_theory_1974,cardy_directed_1980,hinrichsen_nonequilibrium_2000} 
relevant for the onset of turbulence~\cite{pomeau_front_1986,sano_universal_2016},
and the KPZ universality class in the surface growth phenomena~\cite{kardar_dynamic_1986}.
Lucid introductions to these topics can be found in \cite{cardy_scaling_1996,cardy_field_1999}.
Another active area of research, with $z=2$, is the
BEC/BCS crossover (also called the fermions at unitarity), systems of 
non-relativistic spin 1/2 fermions with fine-tuned contact interaction,
which can be experimentally realised in 
cold atom systems~\cite{eagles_possible_1969,leggett_diatomic_1980,nozieres_bose_1985,mehen_conformal_2000,giorgini_theory_2008}.

The operator-product expansion(OPE)
\cite{wilson_non-lagrangian_1969,kadanoff_operator_1969}
\begin{align}
\mO_i(x) \mO_j(0) = \sum_k C^k{}_{ij}(x) \mO_k(x),
\end{align}
where $\mO_i$ are local operators,
summarises the short-distance physics of a quantum field theory,
and is both useful and conceptually important.
Consistency of successive OPEs
imposes constraints on the theory,
called the OPE associativity or the crossing symmetry.
For the isotropic case, in particular,
when the scale symmetry is enhanced into the
conformal symmetry~\cite{polyakov_conformal_1970},
the constraints are often so powerful that 
consideration of them alone almost fixes the theory itself.
This approach,
originally conceived by Polyakov~\cite{polyakov_non-hamiltonian_1974},
is called the conformal bootstrap program.
It had remarkable success for
quantum field theories in two spacetime dimensions 
as pioneered by 
the fundamental work
by Belavin, Polyakov and Zamolodchikov~\cite{belavin_infinite_1984}.
In recent years, 
starting with \cite{rattazzi_bounding_2008},
it has become clear that the program can be successful 
also in higher spacetime dimensions.
See e.g. \cite{rychkov_epfl_2017,poland_conformal_2019} for recent reviews.

It is natural to ask whether a similar bootstrap approach 
can be successful for anisotropic theories.
Theories with $z=2$ would be the first target 
since in this case the scale symmetry can be extended to 
a larger symmetry,
called the Schr\"odinger symmetry~\cite{hagen_scale_1972,niederer_maximal_1972}.
(The basic properties of the Schr\"odinger symmetry are briefly summarised 
in appendix \ref{RSASchrodingerSymmetry}.)
This enhancement is analogous to the enhancement of the
scale invariance to the conformal symmetry which occurs for
many interesting isotropic theories.~\footnote{
See \cite{nakayama_scale_2015} for a review of
the criteria for symmetry enhancement in the isotropic case.
The general criteria for the enhancement of $z=2$ scale invariance 
to the Schr\"odinger symmetry are not understood.
Discussion of this issue for a class of models can be found in 
\cite{nakayama_gravity_2009, nakayama_scale_2015}.}
In particular, if the Schr\"odinger symmetry is present, 
one can classify the local operators
into primary operators and their descendants 
(those operators obtained by acting with spacetime derivatives on the primary operators),
where the primary operators are defined by 
requiring that they commute with certain generators of the Schr\"odinger symmetry.~\footnote{
To be precise, a primary operator $\mO(0, \bm 0)$ can be characterised by the conditions
$[C, \mO(0, \bm 0)]=0$ and $[K_i, \mO(0, \bm 0)]=0$ in the notation explained in appendix \ref{RSASchrodingerSymmetry}.
We note that, in principle,
one can define the concept of primary operators indirectly even if 
both the conformal and Schr\"odinger symmetries
are absent (thus even if $z\neq1$ and $z\neq2$)
by the condition that a primary operator can never
be obtained as a spacetime derivative of other fields.
}
For the isotropic case, 
the representation theory of the conformal symmetry, 
including the classification of operators into primary operators and their descendants,
is a key tool in the conformal bootstrap program.
The analogous representation theory of the Schr\"odinger symmetry 
relevant for the classification of operators can be found in 
\cite{dobrev_lowest_1997, dobrev_non-relativistic_2014} and references therein.
Constraints on the correlation functions
imposed by the Schr\"odinger symmetry,
analogous but less restrictive compared to the isotropic case, are 
derived by Henkel~\cite{henkel_schrodinger_1994,henkel_schrodinger_2003}.

Somewhat surprisingly, the study of OPE for
theories with anisotropic scale invariance started only 
relatively recently~\cite{polyakov_turbulence_1995, braaten_exact_2008}.
We expect the OPE for anisotropic theories
to present new features since the 
short-distance behaviours of isotropic and anisotropic theories are
markedly different. For example, the behaviour
of the two-point function of scalar primary operators in $z=1$ 
conformal field theory (CFT) is 
\begin{align}
\< \mO(x) \mO(0) \> = \frac1{|x|^{2\D}},
\end{align}
whereas in $z=2$ Schr\"odinger invariant theory, it is
\begin{align}
\< \mO(t, \bm x) \bmO(0, \bm 0) \>= 
\begin{cases}
\frac1{t^\D} e^{-\frac{N_{\bmO} \bm x^2}{2 t}} &(t >0)
\\
0 &(t<0)
\end{cases}.
\label{RFTwoPointGeneral}
\end{align}
Here, $\bmO$ is the complex conjugate of the operator $\mO$, 
and $N_{\bmO} >0 $  is the U(1) charge, which is  contained in the 
Schr\"odinger symmetry\footnote{
The U(1) charge is a central charge, {\it i.e.} it commutes 
with all other charges in the Schr\"odinger symmetry.
In some literature, this U(1) charge is called 
the ``mass'' parameter. 
}, of the operator $\mO$.
Thus, the behaviour in the limit $t\to 0, \bm x \to 0$
in the anisotropic theory depends heavily on the precise manner of taking the limit
and is more involved compared to the isotropic case.

Because of this difference, it is important to understand 
general questions regarding the 
OPE in the anisotropic theories such as
``What are the convergence properties of the OPEs?'' and 
``Does the operator associativity hold?''.

For this purpose, 
it would be useful to have exactly solvable yet nontrivial 
examples of quantum field theory models.
For the isotropic case, 
the two-dimensional Ising model and the massless Thirring model
(which is equivalent to the compactified free-boson CFT via bosonisation)
played an instrumental role when the ideas of OPE and the anomalous dimensions
were established~\cite{
wilson_non-lagrangian_1969,wilson_operator-product_1970,
kadanoff_operator_1969, kadanoff_correlations_1969,kadanoff_determination_1971}.
Exactly solvable models also gave substantial support to the development of 
two-dimensional conformal field theory~\cite{belavin_infinite_1984}.
We may hope that study of exactly solvable anisotropic models
may play a similar role in the understanding of 
the $z\neq 1$ fixed points of the renormalisation group.

In this paper, we identify an interacting yet highly tractable model 
with anisotropic $z=2$ scale invariance and its extension to the Schr\"odinger symmetry.~\footnote{
A quantum field theory possessing the Schr\"odinger symmetry is also called 
a ``non-relativistic CFT'' in recent literature.}
The model is the well-known Calogero model~\cite{calogero_solution_1969,
calogero_ground_1969,marchioro_solution_1970,
calogero_solution_1971 ,calogero_erratum_1996} 
considered as a quantum field theory in one space and one time
dimension via the second quantisation.

We exactly compute the nontrivial four-point function
of the fundamental fields of the theory,
$\< \Psi(t_4, x_4) \Psi(t_3, x_3) \bPsi(t_2, x_2) \bPsi(t_1, x_1) \> $.
The result takes a particularly simple form
when $t_1=t_2=0$ and 
$t_3=t_4=t$.
It is 
expressed in terms of the modified Bessel function.
We call these special four-point functions ``pairwise equal-time''.
For the generic case, we give an expression of the four-point function 
in terms of a double convolution integral involving the pairwise equal-time
four-point function and the propagator of non-relativistic
free particles.
The double convolution integral can also be evaluated using 
a generalised hypergeometric function.

We decompose the pairwise equal-time four-point function in two different ways via  OPE,
thereby explicitly verifying the associativity of the OPE for the first time
for an interacting quantum field theory with anisotropic scale invariance.
From the decomposition, one can read off the OPE coefficients 
and the scaling dimensions of the operators appearing in the
intermediate channel.

One of the decompositions is obtained by expanding 
the pairwise equal-time four-point function by the parameter $\frac{x^2}t$,
where $x$ refers collectively to $x_{21}=x_2-x_1, x_{43}=x_4-x_3$.
This decomposition arises from the OPE of $\bPsi(0, x_2)\bPsi(0, x_1) $ and 
of $\Psi(t, x_4) \Psi(t, x_3)$.  
The decomposition can be schematically represented as 
\begin{fmffile}{s-channel}
\begin{equation}
\< \Psi(t, x_4) \Psi(t, x_3) \bPsi(0, x_2) \bPsi(0, x_1) \> 
=
\sum
\begin{gathered}
\begin{fmfgraph*}(16,30)
\fmfleft{i1,i2}
\fmfright{o1,o2}
\fmf{plain}{i1,v1,o1}
\fmf{plain}{i2,v2,o2}
\fmf{plain}{v1,v2}
\fmfv{label=$\mathsmaller\bPsi_1$,label.dist=2}{i1}
\fmfv{label=$\mathsmaller\bPsi_2$,label.dist=2}{o1}
\fmfv{label=$\mathsmaller\Psi_3$,label.dist=2}{i2}
\fmfv{label=$\mathsmaller\Psi_4$,label.dist=2}{o2}
\end{fmfgraph*}
\end{gathered}
\ \ ,
\label{RFSchannelSchematic}
\end{equation}
\end{fmffile}
where the subscripts of $\Psi$ and $\bPsi$ are the labels of the spacetime points.
We will call this expansion the ``s-channel'' decomposition of
the four-point function.
The expansion is convergent.
Only one primary operator (together with its descendants)
appears 
in the intermediate channel.
Thus the four-point function is the analogue of the conformal block 
which plays an important role in the conformal bootstrap program.
The primary operator has U(1) charge $2$.
The scaling dimension of the primary operator depends on the coupling constant of the theory.
The result is consistent with the well-known energy spectrum 
of the Calogero model, combined with the so-called state-operator 
map~\cite{nishida_nonrelativistic_2007,goldberger_ope_2015},
a relation between the scaling dimensions of the operators of a system
with the Schr\"odinger symmetry and the energy spectrum of the theory put in an
external harmonic oscillator potential.

The other decomposition is the expansion of the four-point function 
by $\frac t{x^2}$. This decomposition corresponds to the OPE of  
$\Psi(t, x_3)\bPsi(0, x_1) $
and of
$\Psi(t, x_4)\bPsi(0, x_2) $
(where $x_1<x_2$, $x_3<x_4$ are assumed) and we call it the ``t-channel'' decomposition,
\begin{fmffile}{t-channel}
\begin{equation}
\< \Psi(t, x_4) \Psi(t, x_3) \bPsi(0, x_2) \bPsi(0, x_1) \> 
=
\sum
\ \,\,
\begin{gathered}
\begin{fmfgraph*}(36,14)
\fmfleft{i1,i2}
\fmfright{o1,o2}
\fmf{plain}{i1,v1,i2}
\fmf{plain}{o1,v2,o2}
\fmf{plain}{v1,v2}
\fmfv{label=$\mathsmaller\bPsi_1$,label.dist=2}{i1}
\fmfv{label=$\mathsmaller\bPsi_2$,label.dist=2}{o1}
\fmfv{label=$\mathsmaller\Psi_3$,label.dist=2}{i2}
\fmfv{label=$\mathsmaller\Psi_4$,label.dist=2}{o2}
\end{fmfgraph*}
\end{gathered}
\ \ .
\label{RFTchannelSchematic}
\end{equation}
\end{fmffile}
We found that this decomposition is an asymptotic expansion.

The asymptotic nature may be understood intuitively as follows.
As can be seen, for example, in \eqref{RFTwoPointGeneral},
correlation functions in a Schr\"odinger invariant theory 
generically 
involve exponential factors of the form $e^{-a\frac{x^2}{t}}$,
where $a$ is a numerical constant. These exponential 
factors play the role of the ``instanton effect'' if we 
think about $\frac{t}{x^2}$ as the ``coupling constant''.
As is well-known, the asymptotic nature of a 
perturbation series is inherently related to the
existence of the non-perturbative ``instanton effect''.~\footnote{
For the isotropic case, scale-invariant theories have convergent
OPEs~\cite{luscher_operator_1976,mack_convergence_1977,pappadopulo_operator_2012}
whereas for general quantum field theories without scale invariance
OPEs are asymptotic~\cite{wilson_operator_1972}.
One explanation of this is as follows. (See the discussion below
(2.11) of \cite{ginsparg_applied_1988}.)
If the OPE (which is an expansion in terms of $x$) is asymptotic, it would imply the existence
of the non-perturbative ``instanton'' effect of the form $e^{-\frac{l^2}{x^2}}$ where $l$
is a length scale. This is impossible for scale-invariant theories,
hence OPEs cannot be asymptotic for these theories whereas theories
without the scale invariance have asymptotic OPEs. 
Our intuitive understanding of the asymptotic nature
of the ``t-channel'' decomposition is reminiscent of this explanation.
}
(See, for example,  \cite{le_guillou_large-order_1990}.)
Thus, one could have anticipated 
the asymptotic nature of the expansion in $\frac{t}{x^2}$
from the presence of the factors $e^{-a \frac{x^2}t}$ in Schr\"odinger invariant theories.

The operators appearing in the intermediate channel 
of the ``t-channel'' decomposition have vanishing U(1) charges.
The charge-zero operators are important in particular because they include
currents associated with any internal symmetry
(including the U(1) symmetry in the Schr\"odinger symmetry)
and the energy-momentum tensor. 
But they are elusive since the technique of
the state-operator map is not applicable for them.
The charge-zero sector is studied from the perspective of Schr\"odinger symmetry
(and its infinite extension for specific models, the fermion at unitarity) 
in \cite{golkar_operator_2014} and \cite{bekaert_symmetries_2012}.
We study these charge-zero operators directly via the decomposition of the four-point function.
For example, we will show that some charge-zero operators have non-vanishing
two-point functions only if they are put on the same time slice.

The asymptotic expansion comes with exponentially small correction terms, 
which also can be interpreted naturally in terms of OPE:
we found that the exponentially small terms 
are inherently related to the ``u-channel'' contributions arising from OPEs of
$\Psi(t,x_3) \bPsi(0,x_2)$ 
and of $\Psi(t,x_4) \bPsi(0,x_1)$. These terms can be schematically represented as
\begin{fmffile}{u-channel-intro}
\begin{equation}
\sum
\begin{gathered}
\begin{fmfgraph*}(30,16)
\fmfleft{i1,i2}
\fmfright{o1,o2}
\fmf{plain, tension=3}{i2,v1}
\fmf{phantom, tension=2}{v1,i1}
\fmf{plain, tension=3}{o2,v2}
\fmf{phantom, tension=2}{v2,o1}
\fmf{plain}{v1,v2}
\fmf{plain,tension=0}{v1,o1}
\fmf{plain,tension=0, rubout}{i1,v2}
\fmfv{label=$\mathsmaller\bPsi_1$,label.angle=-90,label.dist=1}{i1}
\fmfv{label=$\mathsmaller\bPsi_2$,label.angle=-90,label.dist=1}{o1}
\fmfv{label=$\mathsmaller\Psi_3$,label.angle=90,label.dist=2}{i2}
\fmfv{label=$\mathsmaller\Psi_4$,label.angle=90,label.dist=2}{o2}
\end{fmfgraph*}
\end{gathered}.
\end{equation}
\end{fmffile}

Some general properties of the OPE in the $z=2$ Schr\"odinger invariant theory have been 
uncovered in recent years~\cite{golkar_operator_2014,goldberger_ope_2015,pal_unitarity_2018}.
Golkar and Son pointed out in \cite{golkar_operator_2014}, 
among other important results, that
the restrictions imposed by the symmetry 
on the correlation functions become much stronger if one of the
operators saturates the unitarity bound.
Goldberger, Khandker and Prabhu proved the convergence of the OPE 
for the case when the operators in the intermediate channel have nonzero U(1) charges~\cite{goldberger_ope_2015}.
Pal studied Schr\"odinger invariant field theories 
focusing on the $SL(2,\R)$ subgroup of the Schr\"odinger symmetry
and uncovered properties of correlation functions of operators
which are aligned on a timelike line~\cite{pal_unitarity_2018}.
In particular, it was shown that the OPE relevant for these correlation functions
converges even when the OPE involves charge-zero operators.

The results in this paper obtained for a particular solvable model confirm and supplement
these general results.
We compute the explicit OPE coefficients and show that the OPE converges 
for the ``s-channel'' OPE decomposition associated with charge-two operators
in the intermediate channel.
This is consistent with the results in \cite{goldberger_ope_2015}.
The spacetime dependence of the three-point function 
we compute by pinching two insertions in 
the four-point function agrees with the result of \cite{golkar_operator_2014}
based on the Schr\"odinger symmetry.

On the other hand, we found novel features which presumably are
shared by general Schr\"odinger invariant theories.
The OPE decomposition associated with the ``t-channel'' OPE 
(involving charge-$0$ operators) is asymptotic, rather than 
convergent.
This does not contradict the results of Pal \cite{pal_unitarity_2018}.
We are studying different correlation functions:
we consider the case where the operators are spatially separated,
whereas in \cite{pal_unitarity_2018} the operators are separated only in the timelike direction. 

The organisation of this paper is as follows.
In section \ref{RSModel}, we discuss the model and establish the notation.
In section \ref{RS4ptfunc}, we describe the computation of
the four-point functions of fundamental fields in the model.
Section \ref{RSOPE} is devoted to what can be read off
from the four-point function. 
We will decompose the four-point function via OPE in two ways (the ``s-channel'' and
``t-channel'' decompositions).
We examine the detailed properties of these decompositions, including 
the identification of the unique primary operator (whose scaling dimension
depends on the coupling constant) and the computation 
of the OPE coefficients in the ``s-channel'' decomposition.
We discuss the asymptotic nature of the ``t-channel'' decomposition
and the exponentially small corrections for the asymptotic series, which can be interpreted as 
the ``u-channel'' contributions.
We also compute a three-point function by starting from 
the four-point function using OPE.
Section \ref{RSConclusion} contains final comments.
Several appendices give auxiliary results.

\section{The model} 
\label{RSModel}
The Hamiltonian of the Calogero 
model (or the Calogero-Marchioro model)~\cite{calogero_solution_1969,
calogero_ground_1969,marchioro_solution_1970,
calogero_solution_1971,calogero_erratum_1996} in the first quantised formulation is~\footnote{
The term ``Calogero model'' often refers to particles 
interacting via a pairwise potential of the form $V(r)=g/r^2+ a r^2$,
or equivalently, particles interacting via a pairwise potential $V(r)=g/r^2$
put in an external harmonic oscillator potential. 
The model we consider can also be considered as
the infinite volume limit (with the total number of particles, not the density, fixed) of 
the Sutherland model\cite{sutherland_quantum_1971,sutherland_quantum_1971-1, sutherland_exact_1971,sutherland_exact_1972}, 
particles on a circle (with radius $R$) interacting with a pairwise potential 
of the form $V(r)=g/(R^2\sin^2{\frac rR})$.
}
\begin{align}
H= - \frac12 \sum_i \der_{i}^2 
+
\sum_{i<j} \frac g{(x_j-x_i)^2}.
\label{RFHamiltonianInverseSquarePotential}
\end{align}
We work in the convention where the mass of the particle is set to unity.

For $x_j -x_i \to 0$, the solution to the Schr\"odinger equation
behaves as  $\Psi \sim \lef|x_j-x_i\ri|^\l$ where 
$g=\l(\l-1)$.
The coupling constant $g$ should satisfy $g \ge -\frac14$
in order that the energy spectrum be bounded below~\cite[section 35]{landau_quantum_1977}.
Solutions with $0\le \l$
are considered as acceptable.
In the regime $-\frac14< g< 0$,
there are two solutions satisfying $\l>0 $ for given $g$.
Corresponding to these two possible boundary conditions, 
we have two different theories.~\footnote{
The possibility of considering the branch with the smaller value of $\l$ was
discussed already in \cite{sutherland_quantum_1971}.
For a review of the Calogero and related models containing
an explanation of this point, see \cite{polychronakos_physics_2006}.  }
Thus $\l\ge0$ provides a good parametrisation of the interacting theory.
For $\l=0$ and $\l=1$, the pair potential vanishes.
These points are equivalent to the free bosons and free fermions (or equivalently, bosons interacting with the infinitely large repulsive $\delta$-function potential), respectively.
It is also convenient (to conform with the convention used for the Bessel functions)
to use another parameter $\n$ defined by
\begin{align}
\n=&\l-\frac12,
\qquad -\frac12 \le \n, \label{RFNuGEQOneHalf}
\\
g=&
\lef(\n-\frac12\ri)
\lef(\n+\frac12\ri).
\end{align}

In the second quantised formulation, 
the action is 
\begin{align}
S=&\int dt d x \lef(
\overline\Psi \der_t \Psi
+
\frac12 \lef|\nabla \Psi\ri|^2
\ri)
+
\frac{g}{2} \int dt dx dy
\lef|\Psi\ri|^2 (x) 
\frac{1}{(x-y)^2} 
\lef|\Psi\ri|^2 (y).
\label{RFActionSecondQuantised}
\end{align}
We consider the Euclidean statistical field theory in this paper.
The canonical (anti-)commutation relations are,
\begin{align}
[\Psi(x), \bPsi(x')]_{\pm} =& \d(x-x'), \label{RFCanonicalCommutationRelation}
\\
[\Psi(x), \Psi(y)]_{\pm}=&0,
\\
[\bPsi(x), \bPsi(y)]_{\pm}=&0.
\end{align}
The signs here are chosen according to whether we consider the bosonic or the fermionic model.

We wish to note however that, as is well known, in the Calogero model, 
the difference between the bosonic and the fermionic theory is not important 
in the following sense.~\footnote{
This was already pointed out in the original papers by Calogero~\cite{calogero_solution_1969,calogero_ground_1969,calogero_solution_1971,calogero_erratum_1996}
and emphasised and explained in detail in \cite{polychronakos_non-relativistic_1989}.
See also the review article \cite{polychronakos_physics_2006}.}
One can solve the Schr\"odinger equation of the model
in the $n$-particle sector with the restriction $x_1<x_2< \cdots < x_n$,
imposing the correct boundary condition $\Psi\sim (x_{i+1}-x_{i})^\l$ when $x_{i+1}-x_i\to +0$.
This is sufficient for the understanding of the properties of the Calogero model.
Note that the boundary condition on $\Psi$ implies that there is no tunnelling amplitude of the particle (for $\l>0$), say, $1$ from the region $x_1<x_2$ to the region $x_1>x_2$;
the wave function vanishes at $x_1=x_2$.
One can define the wave function for the regions where the condition $x_1<x_2< \dots < x_n$
is not satisfied, by complete symmetrisation or anti-symmetrisation for bosons or fermions, 
respectively.
Whether one is dealing with bosons or fermions does not affect physical observables
such as the energy levels (when an external harmonic oscillator potential is present) of the system.
In our analysis, we also found that,
for example, the four-point functions are the same for fermions and bosons, 
provided that the ordering of the particles are properly specified.
We will work both for the bosonic and fermionic models throughout this paper,
except when otherwise explicitly stated.

The Calogero model possesses the Schr\"odinger symmetry
as first shown in \cite{burdet_many-body_1972}.
Thus the model constitutes a fixed line of the renormalisation group 
parametrised by $\n\ge -\frac12$.
The special significance of the potential energy $1/r^2$ 
regarding scale invariance was noted also in 
\cite{gambardella_exact_1975,de_alfaro_conformal_1976}.
The U(1) charge in the Schr\"odinger symmetry is given by
\begin{align}
N=\int \bPsi\Psi dx,
\end{align}
and coincides with the particle number.

Although the Lagrangian \eqref{RFActionSecondQuantised} is non-local, 
we will show that this theory has local OPEs.
This is not too surprising;
there are examples of quantum field theories with non-local interaction, 
which nonetheless exhibit critical 
properties described by a fixed point of the renormalisation group 
and can be studied by OPE and the conformal bootstrap
such as systems with a non-local dipole-dipole 
interaction~\cite{larkin_phase_1969,brezin_critical_1976}
and Ising models with a non-local 
interaction term~\cite{fisher_critical_1972,behan_scaling_2017,behan_long-range_2017,behan_bootstrapping_2019}.

We will study the correlation functions of the model around the true vacuum,
{\it i.e.} the state in which no particles are present.
The simplest of such correlation functions is 
the two-point function of
the fundamental fields,
\begin{align}
\< \Psi(t, x)\bPsi(0, 0)\> 
=
\begin{cases}
\sqrt{\frac{1}{2\pi t}} e^{-\frac{(x-y)^2}{2t}} &(t >0)
\\
0 &(t<0)
\end{cases}.
\label{RFTwoPointFunctionPsibPsi}
\end{align}
The U(1) charges of the fundamental fields are $N_\Psi=-1, N_{\bPsi}=+1$.
The two-point function \eqref{RFTwoPointFunctionPsibPsi}
is not renormalised, {\it i.e.} agrees with the free-theory result.
In particular, the fields $\Psi$, $\bPsi$ have scaling dimension $\frac12$.
See \eqref{RFTwoPointGeneral}.~\footnote{ 
We fix the normalisation of $\Psi, \bPsi$ 
by the canonical (anti-)commutation relation \eqref{RFCanonicalCommutationRelation}.}
This non-renormalisation is a consequence of the fact that 
the two-point functions are associated only with 
one-particle states, and one-particle states by construction
are not affected by the interaction term.
(There are no amplitudes to create virtual particles starting 
from the one-particle states in the model.
Also, there are no vacuum polarisation effects.)
General correlation functions around the true vacuum  
are, of course, nontrivial and contain dynamical information
of the model as we will see in later sections of this paper.

Correlation functions around the true vacuum are different from the 
correlation functions around the ``finite-density vacuum'' 
(the ground state with a constant finite density of particles) of the Calogero model~\footnote{
Correlation functions around the ``finite-density vacuum'' of the Calogero model are 
also equivalent to the correlation functions 
of the Sutherland model in the thermodynamic limit, the large-volume limit with 
the density fixed.
}, which have been extensively studied.
See, for example, \cite{astrakharchik_off-diagonal_2006} and references therein.
The reason we study the correlation functions around the true vacuum in this paper is that
we are interested in the $z=2$ scale-invariant correlation functions;
the presence of the nonzero density breaks the $z=2$ scale invariance spontaneously.

\section{Four-point function} \label{RS4ptfunc}
In this section, we will compute the nontrivial four-point function
of the fundamental fields
\begin{align}
\begin{split}
&\langle
\Psi(t_4, x_4) 
\Psi(t_3, x_3) 
\bPsi(t_2, x_2) 
\bPsi(t_1, x_1) 
\rangle
\\
=&
\langle 0|
T
\Psi(t_4, x_4) 
\Psi(t_3, x_3) 
\bPsi(t_2, x_2) 
\bPsi(t_1, x_1) 
|0\rangle,
\end{split}
\end{align}
of the model described in the previous section.

\subsection{Pairwise equal-time four-point function}
\label{RSS4ptfuncPairwiseEqualTime}
The four-point function can be easily
computed for the special, pairwise equal-time case, {\it i.e.} when
\begin{align}
t_1=&t_2=0,
\\
t_3=&t_4=t >0.
\end{align}
If $t<0$ the four-point function trivially vanishes
since the operator $\Psi(t, x)$ annihilates the vacuum.  
The key observation is that the
pairwise equal-time correlation function
is equivalent to the two-particle Feynman propagator
$K^{(2)}(x_3,x_4; x_1, x_2;t)$
in the first quantised formulation,
\ie\ the transition amplitude of two particles starting at $x_1, x_2$
arriving at $x_3, x_4$ after time $t$ passes,~\footnote{
For the fermionic theory, it is useful to 
consider the four-point function as 
the limit $\lim_{\e_1\to 0, \e_2 \to 0} \< \Psi(t+\e_2, x_4) \Psi(t-\e_2, x_3) \bPsi(\e_1, x_2) \bPsi(-\e_1, x_1) \>$. The limit is well-defined and does not depend on the
sign of $\e_1, \e_2$.
}
\begin{align}
\langle
\Psi(t, x_4) 
\Psi(t, x_3) 
\bPsi(0, x_2) 
\bPsi(0, x_1) 
\rangle
=&
\langle 0|
T
\Psi(t, x_4) 
\Psi(t, x_3) 
\bPsi(0, x_2) 
\bPsi(0, x_1) 
|0\rangle
\nonumber
\\
=&
\langle 0|
\Psi(t, x_4) 
\Psi(t, x_3) 
\bPsi(0, x_2) 
\bPsi(0, x_1) 
|0\rangle
\label{RFPairwiseEqualtimeFourPointIsTwoParticlePropagator}
\\
=&
K^{(2)}(x_3,x_4;x_1, x_2;t).
\nonumber
\end{align}
The propagator $K^{(2)}$ is a solution of the two-body Schr\"odinger equation,
\begin{align}
i\der_t K^{(2)} = \lef(-\frac12 \der_3^2 -\frac12 \der_4^2 + g \frac1{(x_4-x_3)^2} \ri) K^{(2)},
\end{align}
with the initial condition
\begin{align}
\lim_{t\to 0+} K^{(2)} = \d(x_3-x_1) \d(x_4-x_2),
\end{align}
where we assume for simplicity $x_1<x_2$, $x_3<x_4$.

The relation (\ref{RFPairwiseEqualtimeFourPointIsTwoParticlePropagator}) 
follows from the basic feature of the second quantisation.
(See, for example, sections 64 and 65 of \cite{landau_quantum_1977}.)
Let us recall that 
the state,
\begin{align}
\bPsi(x')\bPsi(x'')|0\>,
\end{align}
in the second quantised formulation,
where we use $\bPsi(x)$ to denote the creation operator in the $x$-representation,
is a two-particle state, specified by the wave function,
\begin{align}
\Psi(x_1, x_2) = 
\frac1{\sqrt2}\lef( \d(x_1-x')\d(x_2-x'')
\pm
\d(x_1-x'')\d(x_2-x')
\ri),
\end{align}
in the first quantised formulation. 
(The sign $\pm$ above refers to the bosonic and the fermionic model, respectively.)
The equivalence of the pairwise equal-time four-point function 
and the propagator (\ref{RFPairwiseEqualtimeFourPointIsTwoParticlePropagator}) 
immediately follows.

By separating out the centre of mass motion, 
the computation of the two-particle propagator reduces
to that of the propagator of a particle in an external potential
of the form $1/r^2$.
Defining the relative position $r=x_2-x_1\equiv x_{21}$,
the relevant Hamiltonian is
\begin{align}
H_{\text{rel}}=-\der_r^2+\frac{\l(\l-1)}{r^2}.
\end{align}
We can focus on the region $r>0$.
The propagator for this potential 
was first computed by Peak and Inomata~\cite{peak_summation_1969},
\begin{align}
\langle r'| e^{-H_{\text{rel}}t} |r \rangle
=&
\sqrt{rr'} \frac{1}{2t} e^{-\frac{r^2+r'^2}{4t}} 
I_\n\lef(\frac{rr'}{2t}\ri),
\end{align}
where $\n=\l-\frac12$.
The boundary condition is such that the wave function behaves as $r^\l$ at $r\to0$.
For completeness, we will present a derivation of
this result in appendix \ref{RSAPropagatorInverseSquarePotential}.

The centre of mass contribution to the four-point function is
\begin{align}
\sqrt{\frac1{\pi t}} e^{-\frac{X^2}{t}},
\end{align}
where $X$ is the change of the centre of mass from the initial to the final state,
\begin{align}
X=\frac{x_3+x_4}{2}-\frac{x_1+x_2}{2}
=\frac{x_{31}+x_{42}}{2}
=\frac{x_{32}+x_{41}}{2}.
\end{align}
Hence the full four-point function is~\footnote{
It is easy to check that putting $\n=-\frac12$ in \eqref{RFPairwiseEqualtime4ptFunction} 
reproduces the four-point function of
the free bosonic theory. See appendix \ref{RSAFree}.}
\begin{align}
\begin{split}
&
\langle
\Psi(t, x_4) 
\Psi(t, x_3) 
\bPsi(0, x_2) 
\bPsi(0, x_1) 
\rangle
\\
=&
e^{-\frac{
        x_{21}^2+x_{43}^2+\lef(x_{3}+x_{4}-x_{1}-x_{2}\ri)^2}
        {4t}}
\times
\sqrt{
    \frac{x_{21}x_{43}}
        {4\pi t^3}}
I_\n\lef(
    \frac{x_{21} x_{43}}{2t}\ri).
\end{split}
\label{RFPairwiseEqualtime4ptFunction}
\end{align}
Here $t>0$ is assumed; 
if $t<0$ the correlation function trivially vanishes.
Also, the conditions
\begin{align}
&x_{21}>0,
\\
&x_{43}>0,
\end{align}
are assumed, which come from the assumption that the relative position $r$ is
positive. 
The expression (\ref{RFPairwiseEqualtime4ptFunction}) is valid for
both the bosonic and fermionic cases under these conditions.

It is easy to obtain the four-point function for the generic case. 
The results for the bosonic and the fermionic theory differ by a sign factor.
For the bosonic theory, we have
\begin{align}
&
\langle
\Psi(t, x_4) 
\Psi(t, x_3) 
\bPsi(0, x_2) 
\bPsi(0, x_1) 
\rangle
\nonumber
\\
=&
e^{-\frac{
        x_{21}^2+x_{43}^2+\lef(x_{3}+x_{4}-x_{1}-x_{2}\ri)^2}
        {4t}}
\times
\sqrt{
    \frac{\lef|x_{21}x_{43}\ri|}
        {4\pi t^3}}
I_\n\lef(
    \frac{\lef|x_{21} x_{43}\ri|}{2t}\ri),
\label{RFPairwiseEqualtime4ptFunctionBosonic}
\end{align}
and, for the fermionic theory,
we have
\begin{align}
&
\langle
\Psi(t, x_4) 
\Psi(t, x_3) 
\bPsi(0, x_2) 
\bPsi(0, x_1) 
\rangle
\nonumber
\\
=&
\sgn(x_{43}) \sgn(x_{21})
e^{-\frac{
        x_{21}^2+x_{43}^2+\lef(x_{3}+x_{4}-x_{1}-x_{2}\ri)^2}
        {4t}}
\times
\sqrt{
    \frac{\lef|x_{21}x_{43}\ri|}
        {4\pi t^3}}
I_\n\lef(
    \frac{\lef|x_{21} x_{43}\ri|}{2t}\ri),
\label{RFPairwiseEqualtime4ptFunctionFermionic}
\end{align}
where $\sgn(x) = \frac{x}{|x|}$.

\subsection{Double integral formula for the general four-point function}
\label{RSSForPointDoubleIntegral}

The four-point function in a generic position 
can be computed by a convolution integral of the
free particle propagator and the pairwise
equal-time correlation function computed in the previous subsection.
We consider the four-point function,
\begin{align}
&\langle \Psi(t_4, x_4) \Psi(t_3, x_3) 
 \bPsi(t_2, x_2) \bPsi(t_1, x_1) \rangle,
\end{align}
assuming $t_1<t_2<t_3<t_4$ without loss of generality. 
If $t_1<t_3<t_2<t_4$, for example, then the four-point function 
trivially factorises into a product of two-point functions.

\begin{figure}
    \centering
    \def\svgwidth{0.5\columnwidth}
    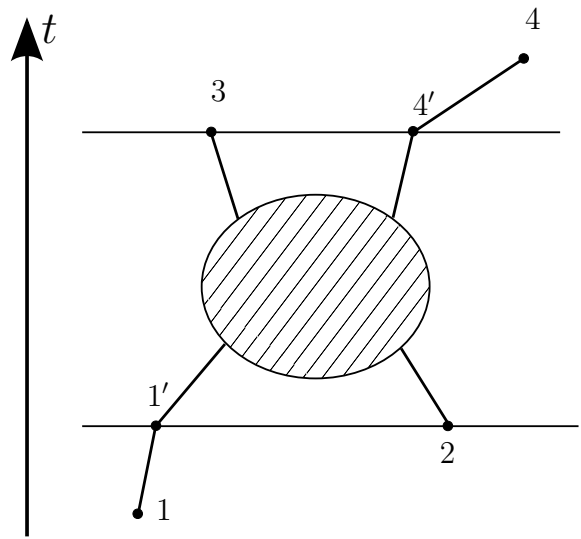
\caption{
Horizontal lines are $t=t_3$ and $t=t_2$.
The coordinates of the points $1'$ and $4'$ 
are  
$(t_1'=t_2, x_1')$ and 
$(t_4'=t_3, x_4')$ respectively. 
Both $x_1'$ and $x_4'$ are to be integrated from $-\infty$ to $+\infty$.
}
\label{RPDoubleIntegral}
\end{figure}

The double integral formula is
\begin{align}
&\langle \Psi(t_4, x_4) \Psi(t_3, x_3) 
 \bPsi(t_2, x_2) \bPsi(t_1, x_1) \rangle
\nonumber
\\
=&
\int_{-\infty}^{+\infty} \int_{-\infty}^{+\infty} K(x_4; x_{4}'; t_4-t_3) 
K^{(2)}(x_3, x_4'; x_1', x_2; t_3-t_2) K(x_1';x_1;t_2-t_1) dx_1' dx_4',
\label{RFDoubleIntegral}
\end{align}
where $K(x;y;t)$ is the free one-particle propagator,
\begin{align}
K(x;y;t)=\frac{1}{\sqrt{2\pi t}} e^{-\frac{(x-y)^2}{2t}},
\label{RFFreePropagator}
\end{align}
and the pairwise four-point function or equivalently the two-particle propagator, 
\begin{align}
K^{(2)}(x_3, x_4; x_1, x_2; t)=\<\Psi(t, x_4) \Psi(t, x_3) \bPsi(0, x_2) \bPsi(0, x_1) \>,
\end{align}
is given by \eqref{RFPairwiseEqualtime4ptFunctionBosonic} 
or \eqref{RFPairwiseEqualtime4ptFunctionFermionic} according 
to whether the theory is bosonic or fermionic.
The formula \eqref{RFDoubleIntegral} holds because in the intervals $t_1<t<t_2$ and
$t_3< t < t_4$ there is only a single particle as shown in Fig. \ref{RPDoubleIntegral}.

\begin{figure}
    \centering
    \def\svgwidth{0.5\columnwidth}
    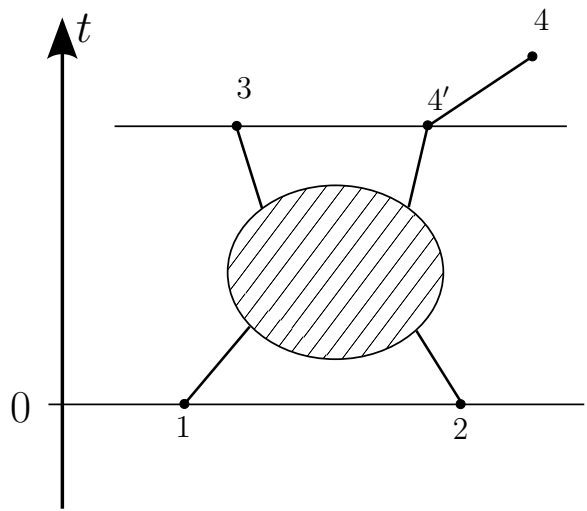
\caption{
Horizontal lines are $t=t_3$ and $t=t_1=t_2=0$.
The coordinates of the point $4'$ 
are $(t_4'=t_3, x_4')$. 
$x_4'$ is integrated from $-\infty$ to $+\infty$.
}
\label{RPSingleIntegral}
\end{figure}

There is, of course, also a simpler integral formula to 
compute a four-point function where only $\bPsi$'s are 
inserted on the same time slice as shown in Fig. \ref{RPSingleIntegral},
\begin{align}
\begin{split}
&\langle \Psi(t_4, x_4) \Psi(t_3, x_3) 
 \bPsi(0, x_2) \bPsi(0, x_1) \rangle
\\
=&
\int_{-\infty}^{+\infty} K(x_4; x_{4}'; t_4-t_3) K^{(2)}(x_3, x_4'; x_1, x_2; t_3)
dx_4',
\end{split}
\label{RFSingleIntegral}
\end{align}
where $0=t_1=t_2<t_3<t_4$.
This formula will be used later in section \ref{RSSThreePointGeneralPosition}
when we compute a three-point function.

\subsection{Four-point function in general position via generalised
hypergeometric function}
\label{RSSFourPointHyperGeometric}
\begin{sloppypar}
The double integral
(\ref{RFDoubleIntegral}) 
representing 
the four-point function $\< \Psi(t_4, x_4) \Psi(t_3, x_3) \bPsi(t_2, x_2) \bPsi(t_1, x_1) \>$
in general position 
can be expressed in terms of the generalised hypergeometric function.
The detailed derivation of the formula and its consequences 
will be discussed in a separate publication.
In this paper, we give the expression and discuss a few of its basic properties.
We focus on the bosonic theory. We consider the nontrivial case, $t_1< t_2< t_3<t_4$.
\end{sloppypar}

The result is
\begin{align}
&\< \Psi(t_4, x_4) \Psi(t_3, x_3) \bPsi(t_2, x_2) \bPsi(t_1, x_1) \>\nonumber \\ =& 
\frac{e^{-\frac{x_{43}^2}{2t_{43}} - \frac{x_{32}^2}{t_{32}} - \frac{x_{21}^2}{2t_{21}}}}{(2\pi)^{\frac{3}{2}}\sqrt{t_{43} t_{21}}} %&
\frac{ \Gamma^2\lef(\frac{\nu}{2}+\frac{3}{4}\ri)}{2^{\nu-1}\Gamma(\nu+1)}
\tau^{\frac{\nu}{2}+\frac{3}{4}}   %%%\nonumber \\
\mathcal{F}(v_{123},v_{234},\tau),
\label{full4pt}
\end{align}
where the function $\mathcal{F}$ is a generalised hypergeometric 
function with three variables defined by the following triple series expansion,
\begin{align}
\mathcal{F}(v_{123},v_{234},\tau)= %%%&
\sum _{\substack{p=0 \\
m, n=0
}}^{\infty }
\frac{\lef( \tau \ri)^p}{p!}
\frac{\lef(v_{123}\ri)^m}{m!}
\frac{\lef(v_{234}\ri)^n}{n!}
%\lef[\Lambda_{0} \lef(pmn\ri)+\lef(4\nu + 6\ri) \sqrt{ \tau v_{234} v_{123}}  \Lambda_{1}\lef(pmn\ri)\ri], \label{nontrivialF}\\
\lef[
\Lambda^{(0)}_{pmn} +
\sqrt{\tau v_{123} v_{234}} 
  \Lambda^{(1)}_{pmn} % \sqrt{ \tau v_{234} v_{123}}%
\ri], \label{nontrivialF}
\end{align}
with the coefficients $\Lambda^{(j)}_{pmn}$ ($j=0,1$) given by
\begin{align}
\Lambda^{(j)}_{pmn} %\lef(p,m,n\ri)
= %%%&
\lef( \nu+\tfrac{3}{2}\ri)^{2j}
\frac{\lef(\tfrac{\nu}{2}+\tfrac{3}{4}+j\ri)_{p+m}
\lef(\tfrac{\nu}{2}+\tfrac{3}{4}+j\ri)_{p+n}
\lef(\nu+\frac{1}{2}+j\ri)_{2p}
}{
\lef(\tfrac{1}{2}+j\ri)_{m}\lef(\tfrac{1}{2}+j\ri)_{n}
\lef(\tfrac{1}{2}+j\ri)_{p}
\lef(2\nu+1+j\ri)_{2p}
}, 
\end{align}
where we used the Pochhammer symbol $(x)_n=x(x+1)\cdots(x+n-1)=\frac{\G(x+n)}{\G(x)}$.

The quantities $\t$ and $v$ are 
Schr\"odinger invariant quantities defined by
\begin{align}
\tau=\frac{t_{21}t_{43}}{t_{31}t_{42}},\quad 
v_{123}=\frac{\left( t_{21}x_{32} - t_{32}x_{21} \right)^2}{2t_{21}t_{32}t_{31}},\quad 
v_{234}=\frac{\left( t_{32}x_{43} - t_{43}x_{32} \right)^2}{2t_{32}t_{43}t_{42}}.
\label{crossratios}
\end{align}
They may be considered as the analogue of the cross-ratios, quantities invariant 
under the conformal symmetry, in usual CFT.

We note that an ansatz for Schr\"odinger invariant four-point functions is given in
\cite{volovich_correlation_2009}
which contains an arbitrary function of 
four Schr\"odinger invariant ``cross-ratios'' ($\t$ and three $v$'s).
For one space and one time dimension, the number of independent Schr\"odinger invariant cross-ratios is 
decreased by one~\footnote{
Concretely, $\sqrt{v_{134}}$ can be written as a linear combination
of $\sqrt{v_{123}}$ and $\sqrt{v_{234}}$ for one space and one time dimension.
}, so that there are three  independent cross-ratios 
($\t$ and two $v$'s appearing in \eqref{full4pt}).

The function \eqref{nontrivialF} is symmetric in $v_{123}$ and $v_{234}$.
The invariant quantity $v$, say, $v_{123}$ is proportional to the squared ``area'' of the 
triangle spanned by the space-time points 1, 2, 3. Thus $v$ may be considered 
as measuring the degree of ``non-collinearity" of the three space-time points.
For instance, if the points $1$-$2$-$3$ are collinear, one has $v_{123}=0$.
In this case, \eqref{nontrivialF} reduces to a double hypergeometric function,
\begin{align}
\mathcal{F}(0,\,v_{234},\,\tau)=&F{}^{1:0:3}_{0:1:3}
  \left[
   \setlength{\arraycolsep}{0pt}% local assignment
  \begin{array}{c@{{}:{}}c@{;{}}c}
\frac{\nu}{2}+\frac{3}{4}& \linefill & \frac{\nu}{2}+\frac{3}{4},\,\frac{\nu}{2}+\frac{3}{4},\,\frac{\nu}{2}+\frac{1}{4}\\[1ex]
   \linefill& \frac{1}{2} &\frac{1}{2},\quad \nu+1,\quad \nu + \frac{1}{2} 
   \end{array}
   \;\middle|\;
   v_{234},\, \tau  \right],\label{Kampe}
\end{align}
where the Kamp\'{e} de F\'{e}riet series \cite[p.27]{srivastava_multiple_1985} is used.
If all the four points lie along a line, it further reduces to a hypergeometric function of the single cross-ratio $\tau$,
\begin{align}
\mathcal{F}(0,\,0,\,\tau)=&
{}_4F_{3}\lef(\begin{array}{c} 
\frac{\nu}{2}+\frac{3}{4},\,\frac{\nu}{2}+\frac{3}{4},\,\frac{\nu}{2}+\frac{3}{4},\,\frac{\nu}{2}+\frac{1}{4}
 \\ \frac{1}{2},\, \nu+1,\, \nu + \frac{1}{2} \end{array}\;\middle|\; 
 \tau \ri). \label{4F3}
\end{align}

\section{OPE decomposition of the four-point function} \label{RSOPE}

By taking various limits of the four-point function
we have computed in the previous section, 
one can extract the information of OPE coefficients and 
a three-point function.

In section \ref{RSSOPEPairwiseFourPoint} we consider 
two different decompositions of the pairwise equal-time four-point function
(\ref{RFPairwiseEqualtime4ptFunction}) using OPE.
One of the decompositions, the ``s-channel'' decomposition,
arises from $\Psi\Psi$ and $\bPsi\,\bPsi$ OPEs and
is represented by a convergent series.
The operators appearing in the intermediate channel 
have anomalous dimensions, \ie\ their scaling dimensions depend on the coupling constant.
The other decomposition, the ``t-channel'' decomposition,
arises from two $\Psi \bPsi$ OPEs and is represented by an asymptotic series.
By representing the same four-point function in two ways by OPE, we prove
the operator associativity of the model, for this particular four-point function.
The asymptotic series for the ``t-channel'' decomposition has also exponentially small correction terms, which also have an interpretation via OPE.

In section \ref{RSSDetailedSchannel} we study in detail the ``s-channel'' decomposition
and show that  only one primary operator $\Phi$ appears in the $\Psi\Psi$ OPE.
We fix the forms of the descendant operators of $\Phi$ appearing in the OPE and
compute all the relevant OPE coefficients.
In section \ref{RSSThreePoint} we compute the three-point function $\Psi\Psi\bPhi$
from the four-point function.
In section \ref{RSSPeculiarPropertyJ} we discuss a peculiar property of the two-point functions
between operators arising in the $\Psi \bPsi$-OPE.

\subsection{Decomposition of pairwise equal-time four-point function} 
\label{RSSOPEPairwiseFourPoint}

\subsubsection{``s-channel'' decomposition} 
\label{RSSSDecompositionSChannel}

We consider the expansion of the pairwise equal-time four-point function 
\eqref{RFPairwiseEqualtime4ptFunction} in the parameter $x^2/t$,
where $x$ refers to both  $x_{21}$ and  $x_{43}$.
As it turns out, this expansion has an infinite convergence radius, so that 
the expansion is valid for an arbitrarily large value of $\frac{x^2}t$.
The expansion becomes more useful when $\frac{x^2}t\ll 1$ since 
the first few terms will then dominate the series.
The smaller the value of $\frac{x^2}t$, the closer are the operators
$ \Psi(t, x_4), \Psi(t, x_3) $ and 
$\bPsi(0, x_2), \bPsi(0, x_1)$ respectively.
Therefore, considering this expansion should amount to considering the OPE between 
$ \Psi(t, x_4) \Psi(t, x_3) $ and 
$\bPsi(0, x_2) \bPsi(0, x_1)$.
(See \eqref{RFSchannelSchematic}.)
Since the argument of the modified Bessel function is 
$z=\frac{x_{21} x_{43}}{2t}\sim \frac{x^2}t$,
one can use the definition of the modified Bessel function by 
the series expansion,
\begin{align}
I_{\nu}\lef(z\ri)=
(\tfrac{1}{2}z)^{\nu}\sum_{k=0}^{\infty}\frac{(\tfrac{1}{4}z^{2})^{k}}{k!\Gamma\lef(\nu+k+1\ri)}.
\label{RFBesselExpansionAroundZero}
\end{align}
This series is convergent for any value of the argument $z$.
Substituting \eqref{RFBesselExpansionAroundZero} into \eqref{RFPairwiseEqualtime4ptFunction}, 
we obtain,
\begin{align}
\begin{split}
&
\langle
\Psi(t, x_4) 
\Psi(t, x_3) 
\bPsi(0, x_2) 
\bPsi(0, x_1) 
\rangle
\\
=&
e^{-\frac{X^2}{t}}
\sqrt{
    \frac{x_{21}x_{43}}
        {4\pi t^3}}
\lef(
\frac{x_{21}
x_{43}}
{4 t} 
\ri)^\n
e^{-\frac{x_{21}^2+x_{43}^2}{4t}}
\sum_{k=0}^{\infty}\frac{\lef(
\frac{x_{21} x_{43}}{4t}
\ri)^{2 k}}{k!\Gamma\lef(\nu+k+1\ri)},
\end{split}
\label{RFSchannelDecomposition}
\end{align}
where $X\equiv\frac{x_3+x_4-x_1-x_2}{2}$.

This decomposition of the four-point function emerges 
from the equal-time OPEs
\begin{align}
\bPsi(0, x_2)
\bPsi(0, x_1) 
=&\sum_{k=0}^{\infty} C_{k} x_{21}^{\D_k-1} \bPhi_k \lef(0, \frac{x_1+x_2}{2}\ri),
\label{RFbPsibPsiOPESimple}
\\
\Psi(t, x_4)
\Psi(t, x_3)
=&\sum_{k=0}^{\infty} C_{k} x_{43}^{\D_k-1} \Phi_k \lef(t, \frac{x_3+x_4}{2}\ri).
\label{RFPsiPsiOPESimple}
\end{align}
Here $C_k$'s are the OPE coefficients,~\footnote{
We take $C_k$ to be real by choosing the phases of $\Phi_k$'s appropriately.} 
and $\D_k$'s are dimensions of the operator $\Phi_k$ (and $\bPhi_k$).
The operators $\bPhi_k$ are charge-$2$ operators, $N_{\bPhi_k}=2$.
The powers of $x_{21}$ and $x_{43}$ in \eqref{RFbPsibPsiOPESimple} and \eqref{RFPsiPsiOPESimple}
are fixed by scale invariance.
For simplicity, we will set $(x_1+x_2)/2=0$, $(x_3+x_4)/2=X$,
using translational invariance.
Equation \eqref{RFSchannelDecomposition} is valid provided $x_{21}>0, x_{43}>0$ for both the bosonic and fermionic models.
We assume in this subsubsection,
without loss of generality, that these conditions are met.

In \eqref{RFbPsibPsiOPESimple} and 
\eqref{RFPsiPsiOPESimple},
we are not distinguishing primary and descendant operators.
We will see that $\Phi_0\equiv\Phi$ is the only primary operator appearing in the OPE
and all other operators $\Phi_k (k=1,2,\cdots)$ are its descendants in section \ref{RSSDetailedSchannel}, where we also compute all OPE coefficients $C_k$'s.

It is easy to read off
the dimensions $\D_k$ by comparing 
the powers of $x_{21}$ and $x_{43}$ in the formulae
\eqref{RFbPsibPsiOPESimple} and 
\eqref{RFPsiPsiOPESimple} with \eqref{RFSchannelDecomposition}.
We obtain,
\begin{align}
\D_k= \frac32+ \n +2k.
\label{RFScalingDimensionsSchannel}
\end{align}
We see that the scaling dimensions depend on the coupling constant $\n$;
the operators $\Phi_k$'s have anomalous dimensions.

Let us consider the leading order contribution from
the lowest-dimension operator $\Phi$
with dimension $\D_0=\frac32+\n$.
From \eqref{RFbPsibPsiOPESimple} and \eqref{RFPsiPsiOPESimple} we obtain
\begin{align}
\begin{split}
&\<
\Psi(t, x_4)
\Psi(t, x_3)
\bPsi(0, x_2)
\bPsi(0, x_1) 
\>
\\
\approx &
C_{0}^2
x_{43}^{\n+\frac12} 
x_{21}^{\n+\frac12} 
\lef\<  
\Phi \lef(t, X\ri)
\bPhi \lef(0, 0\ri)
\ri\>.
\end{split}
\end{align}
Comparing this to the leading order term (both in 
the expansion by $x_{43}$ and $x_{21}$) of \eqref{RFSchannelDecomposition},
\begin{align}
\<
\Psi(t, x_4) 
\Psi(t, x_3) 
\bPsi(0, x_2) 
\bPsi(0, x_1) 
\>
\approx
\frac1{t^{\frac32+\n}}
e^{ - \frac{X^2}{t}}
\times
\frac1{4^\n \sqrt{4\pi}\G(\n+1)}
x_{21}^{\n+\frac12}
x_{43}^{\n+\frac12},
\end{align}
we can read off the leading OPE coefficient to be
\begin{align}
C_0=&
\frac1{2^\n (4\pi)^{\frac14}\sqrt{\G(\n+1)}},
\label{RFCZero}
\end{align}
together with the two-point function 
\begin{align}
\langle 
\Phi(t,X) \bPhi(0,0)
\rangle
=&
\frac1{t^{\frac32+\n}}e^{-\frac{X^2}{t}}.
\label{RFTwoPointPhibPhi}
\end{align}
The spacetime dependence 
agrees with the general form of the two-point function of the primary operator
\eqref{RFTwoPointGeneral} with $N_{\bPhi}=2$ and $\D=\D_0=\frac32+\n$.
We fixed the normalisation of $\Phi$ by \eqref{RFTwoPointPhibPhi}.
We will see presently that $\Phi$ is indeed a primary operator.
The differences in scaling dimensions of $\Phi_k$ $(k>0)$ 
and $\Phi$ are integers. This suggests that $\Phi_k$ $(k>0)$ are descendants of 
$\Phi$. We will see later in section \ref{RSSDetailedSchannel} that this
is the case.

In section \ref{RSS4ptfuncPairwiseEqualTime}, we computed the four-point function
by identifying it with the two-particle propagator.
In this identification, the expansion parameters $x_{43}$ and $x_{21}$ are the relative coordinates, 
and $X$ appearing above is the difference in the final and the initial 
centre of mass position.
Thus, the ``s-channel'' OPE decomposition described here
may be interpreted as representing the separation of the centre of mass and 
relative motions.

\paragraph{The state-operator map}
The spectrum of operators \eqref{RFScalingDimensionsSchannel} is consistent with the
state-operator map introduced by Nishida and Son 
for Schr\"odinger invariant theories~\cite{nishida_nonrelativistic_2007}.
(See also \cite{goldberger_ope_2015}.)
The state-operator map is a one-to-one correspondence 
between an operator $\bmO$ (with positive
U(1) charge $N_{\bmO}>0$) 
of a Schr\"odinger invariant theory 
and a state $|\bmO\>$ of the model which is obtained by adding an
external harmonic oscillator potential to the theory.~\footnote{
The state-operator map first appeared in \cite{niederer_maximal_1973} 
for the free-field theory.
For a specific interacting model (the fermion at unitarity),
Werner and Castin applied a similar map between
the theory with and without the external harmonic oscillator potential~\cite{werner_unitary_2006}.
We note also that the state-operator maps given in \cite{nishida_nonrelativistic_2007} 
and \cite{goldberger_ope_2015} are slightly different.
The map given in \cite{goldberger_ope_2015} has
the advantage that the operator $[K_i, \bmO]$
is directly mapped to the state $K_i |\mO\>$.
}
For our model, the extra harmonic oscillator term can be represented as
an additional contribution
\begin{align}
S_{\text{ext}}=-\int dt dx \frac12 x^2 \lef|\Psi\ri|^2,
\end{align}
to the action \eqref{RFActionSecondQuantised}.
The state-operator map has the property that 
the scaling dimension of $\bmO$ equals the energy of the state $|\bmO\>$ in the deformed model.
The map also preserves the U(1) charge: the state $|\bmO\>$ is an $N_{\bmO}$-particle state
in the deformed model.

The energy spectrum of the deformed model can be exactly computed.
This is the celebrated result by Calogero~\cite{calogero_solution_1971}. 
For $N$-particle state, it is,
\begin{align}
E=\frac{N}{2}+\lef(\n+\frac12\ri)\frac{N(N-1)}{2}+\sum_{i=1}^N n_i,
\label{RFEnergySpectrumCalogero}
\end{align}
where $n_i$ $(i=1,\cdots, N)$ are integers satisfying 
\begin{align}
0 \leq n_1 \leq n_2 \leq \cdots \leq n_N,
\end{align}
for both the bosonic and fermionic models.
(See, for example, (16) of \cite{polychronakos_physics_2006}.)
Thus, via the state-operator map, we have the complete tabulation of 
operators (with positive U(1) charge~\footnote{
Of course, negatively charged operators are also classified since they are the
complex conjugates of positively charged operators.
The state-operator map fails to capture, importantly, charge-zero operators
(operators with $N_\mO=0$).
})
including their correct multiplicities.
The scaling dimensions are given simply by \eqref{RFEnergySpectrumCalogero},
with the identification $\D=E$.

Let us consider the simplest case, $N=1$. 
The operators are labelled by a single non-negative integer $n$
and their dimensions are 
\begin{align}
\D= \frac12 + n.
\end{align}
The lowest dimension operator necessarily is a primary operator,
which is nothing but the fundamental field $\bPsi$.
(Recall that the scaling dimension of the fundamental field is protected
and equals $\frac12$. See the explanation below \eqref{RFTwoPointFunctionPsibPsi}.)
The operators with $n>0$ are descendants of the fundamental field, 
$\der_x^n \bPsi$. It is not necessary to consider descendants produced 
by acting with $\der_t$'s on $\bPsi$. This is because 
it is redundant to consider the null operator $\lef(\der_t -\frac12 \der_x^2\ri)\Psi$ 
in the OPE. See \cite{golkar_operator_2014,dobrev_lowest_1997,dobrev_non-relativistic_2014}. 

For the $N=2$ case, which is relevant for the ``s-channel'' OPE we are considering,
\eqref{RFEnergySpectrumCalogero} becomes
\begin{align}
\D=\frac32+ \n+ n_1 +n_2,
\label{RFGeneralChargeTwoScalingDimensions}
\end{align}
with $0\leq n_1 \leq n_2$.~\footnote{
For the free-field theory case, $\n=-\frac12$, one can understand this spectrum as that of 
the operators $\der_x^{n_1} \bPsi \der_x^{n_2} \bPsi$.}
This includes the spectrum found from the OPE, \eqref{RFScalingDimensionsSchannel},
consistently with the correct coupling constant dependence of the scaling dimensions.
We now see that the operator $\bPhi$ with dimension $\D_0=\frac32+\n$
has the lowest scaling dimension in the charge-$2$ sector,
and must therefore be a primary operator.
(In \eqref{RFScalingDimensionsSchannel} the scaling dimensions are separated by
even integers whereas in \eqref{RFGeneralChargeTwoScalingDimensions} the separations
are general integers. This difference arises because we are defining the OPE at the symmetric points
$\frac{x_1+x_2}2, \frac{x_3+x_4}2$
in \eqref{RFbPsibPsiOPESimple} and 
\eqref{RFPsiPsiOPESimple}.)

One can also characterise primary operators
using the state-operator map:
they correspond to the states annihilated by the charges $C, K$.~\footnote{For our notation about
the Schr\"odinger algebra, see appendix \ref{RSASchrodingerSymmetry}. 
}
It should be possible to directly study this condition in the
Calogero model (with the harmonic oscillator external potential).
This will lead to a complete classification of the primary operators
(with nonzero charge, since the zero-charge sector defies the
use of the state-operator map).
The operator technique developed in \cite{brink_explicit_1992} seems to be well-suited
for this purpose. This problem will be addressed in a separate publication.

\subsubsection{``t-channel'' decomposition} 
\label{RSSSDecompositionTChannel}

\begin{sloppypar}
Next, we consider the pairwise equal-time four-point function
$\<\Psi(t, x_4) \Psi(t, x_3) \bPsi(0, x_2) \bPsi(0, x_1)\>$
in the regime in which $x^2/t$ is large, 
where $x$ refers collectively to $x_{21}>0$ and $x_{43}>0$. 
In this regime, the spacetime points $2, 4$ and $1, 3$ can be made 
close to each other, respectively. 
Hence we expect that this regime should be understood from 
the OPEs $\Psi(t, x_4) \bPsi(0, x_2)$ and $\Psi(t, x_3) \bPsi(0, x_1)$.
(See \eqref{RFTchannelSchematic}.)
\end{sloppypar}

One can use,
for the four-point function \eqref{RFPairwiseEqualtime4ptFunction},
the asymptotic expansion of 
the modified Bessel function~\cite[(10.40.1), (10.17.1)]{olver_nist_nodate}
\begin{align}
I_{\nu}\lef(z\ri)\sim\frac{e^{z}}{(2\pi z)^{\frac{1}{2}}}
\sum_{p=0}^{\infty}(-1)^{p}\frac{a_{p}(\nu)}{z^{p}},
\label{RFBesselAsymptoticExpansion}
\end{align}
where $a_0(\n)=1$ and
\begin{align}
a_{p}(\nu)=\frac{(4\nu^{2}-1^{2})(4\nu^{2}-3^{2})\cdots(4\nu^{2}-(2p-1)^{2})}{p!8^{p}}.
\label{RFDefaNu}
\end{align}
The asymptotic expansion is valid in the limit $z\to\infty$ for $|\ph z|\le \frac\pi 2 - \e$,
which includes $z>0$ relevant for us. ($\ph z$ is the argument of $z$
and $\e$ is a positive infinitesimal quantity.)
Substituting \eqref{RFBesselAsymptoticExpansion} to 
\eqref{RFPairwiseEqualtime4ptFunction},
we obtain
\begin{align}
&
\<\Psi(t, x_4) \Psi(t, x_3) \bPsi(0, x_2) \bPsi(0, x_1)\>
\\
=&
e^{-\frac{
        x_{21}^2+x_{43}^2+\lef(x_{1}+x_{2}-x_{3}-x_{4}\ri)^2}
        {4t}}
\times
\sqrt{
    \frac{x_{21}x_{43}}
        {4\pi t^3}}
e^{\frac{x_{21} x_{43}}{2t}}
\frac1{\sqrt{2\pi \frac{x_{21} x_{43}}{2t}}}
\sum_{p=0}^{\infty}(-1)^{p}\frac{a_{p}(\nu)}{
\lef({\frac{x_{21} x_{43}}{2t}}\ri)^{p}}.
\end{align}
It is essential that the exponential factors in this formula combine to yield
\begin{align}
e^{-\frac{
        x_{21}^2+x_{43}^2+\lef(x_{3}+x_{4}-x_{1}-x_{2}\ri)^2}
        {4t}}
e^{\frac{x_{21} x_{43}}{2t}}
=&
e^{-\frac{x_{31}^2+x_{42}^2}{2t}}.
\label{RFTChannelDecompositionExponentialFactorCombines}
\end{align}
We obtain
\begin{align}
\begin{split}
&
\langle
\Psi(t, x_4) 
\Psi(t, x_3) 
\bPsi(0, x_2) 
\bPsi(0, x_1) 
\rangle
\\
=&
\frac1{2\pi}e^{-\frac{x_{31}^2+x_{42}^2}{2t}}
\times
\frac1t
\sum_{p=0}^{\infty}(-1)^{p}
2^p a_p(\n)
\frac{t^{p}}{x_{21}^p x_{43}^p}.
\label{RFTChannelDecomposition}
\end{split}
\end{align}
To clarify the connection to the OPE, we define
\begin{align}
X'=\frac{x_4+x_2}2-\frac{x_3+x_1}2.
\end{align}
Then we have, using 
$x_{21}=X'-\frac{x_{42}-x_{31}}2$ and 
$x_{43}=X'+\frac{x_{42}-x_{31}}2$,
\begin{align}
\begin{split}
&
\< \Psi(t, x_4) \Psi(t, x_3) \bPsi(0, x_2) \bPsi(0, x_1) \>
\\
=&
\frac1{2\pi}e^{-\frac{x_{31}^2+x_{42}^2}{2t}}
\times
\frac1t
\sum_{p=0}^{\infty}(-1)^{p}
2^p a_p(\n)
\frac{t^p}{\lef(X'^2- 
\frac{(x_{42}-x_{31})^2}4\ri)^p
}.
\end{split}
\label{RFTChannelDecompositionUsingXPrime}
\end{align}
This ``t-channel'' decomposition of the four-point function is 
valid for $x_{21}>0, x_{43}>0$ for both the bosonic and fermionic models.
Hereafter in this subsubsection, 
to be specific, we consider the bosonic model.

The decomposition 
\eqref{RFTChannelDecompositionUsingXPrime}
indeed has the form which arises from 
the OPEs $\Psi(t, x_4) \bPsi(0, x_2)$ and $\Psi(t, x_3) \bPsi(0, x_1)$.
In particular, it shows that the OPE $\Psi(t, x) \bPsi(0, x')$ is well-defined
in the limit $t\to 0$ with fixed $\frac{(x-x')^2}t$.
More precisely, the OPE has the form
\begin{align}
\begin{split}
\Psi(t, x) \bPsi(0, x') =& \sum_{k=0}^{\infty} \tC_k(t, x-x') \mathcal{J}_k\lef(\frac t2, \frac{x+x'}2\ri)
\\
=& \sum_{k=0}^{\infty} (x-x')^{\tilde{\D}_k-1}f_k\lef(\frac{(x-x')^2}t\ri) \mJ_k\lef(\frac t2, \frac{x+x'}2\ri),
\end{split}
\label{RFPsibPsiOPE}
\end{align}
where $\mJ_k$ (with scaling dimension $\tilde{\D}_k$) are the operators in the intermediate 
channel. 
In the second line, we have used the scale invariance to constrain the OPE coefficients
$\tC_k$.
The two-point functions of $\mJ_k$'s are
\begin{align}
\<\mJ_k(0, X') \mJ_{k'}(0, 0)\> = \frac{D_{kk'}}{X'^{\tilde{\D}_k+\tilde{\D}_{k'}}}.
\label{RFTwoPointJJ}
\end{align}
Some of the coefficients $D_{kk'}$ can be absorbed into the
normalisation of the operators $\mJ_k$.
Here, we are not specifying whether $\mJ_k$ is a primary or a descendant operator.
(They can be a linear combination of primary and descendant operators in general.)
The operators $\mJ_k$ have vanishing U(1) charges.
Note that the primary and descendant operators for the charge-zero sector 
behave differently from those in other sectors~\cite{golkar_operator_2014}. 
This is because the generators $K$ and $P$, 
which act as the ``ladder operators'' for the sectors 
with the nonzero U(1) charge, commute for the charge-zero sectors.
(The commutation relations of the Schr\"odinger algebra are given in appendix 
\ref{RSASchrodingerSymmetry}.)

From the OPE \eqref{RFPsibPsiOPE} and the two-point functions \eqref{RFTwoPointJJ},
we obtain
\begin{align}
\begin{split}
&
\<
\Psi(t, x_4) 
\Psi(t, x_3) 
\bPsi(0, x_2) 
\bPsi(0, x_1) 
\>
\\
=&
\sum_{m=0}^{+\infty}\sum_{n=0}^{+\infty}
x_{42}^{\tilde{\D}_m-1}
x_{31}^{\tilde{\D}_n-1}
f_m\lef(\frac{x_{42}^2}{t}\ri)
f_n\lef(\frac{x_{31}^2}{t}\ri)
D_{mn} \frac1{X'^{\tilde{\D}_m+\tilde{\D}_n}}.
\end{split}
\label{RFPWETFourPointFromTChannelOPE}
\end{align}
The powers of $X'$, $x_{21}$, and $x_{43}$
are all integers in \eqref{RFTChannelDecomposition}. 
Comparing \eqref{RFTChannelDecomposition} and \eqref{RFPWETFourPointFromTChannelOPE}, 
we see that this strongly suggests that 
$\tilde{\D}_k$'s are also integers.
There are ambiguities, however, which stem from the fact that one
can insert 
\begin{align}
1=\lef(\frac{x_{42}^2}{t}\ri)^\d \left(\frac{t}{x_{31}^2}\ri)^\d
\lef(\frac{x_{31}^2}{x_{42}^2}\ri)^{\d}
,
\label{RFAmbiguityTChannel}
\end{align}
where $\d$ is an arbitrary number, 
into \eqref{RFPWETFourPointFromTChannelOPE}.
This leads to a redefinition of $f_m$ and $f_n$ and 
a shift of the dimensions $\tilde{\D}_m$ and $\tilde{\D}_n$ by $\pm 2\d$.
We will return to a possible resolution of this ambiguity towards the end of this subsubsection.
Although it seems unlikely that a set of consistent OPE coefficients exist with non-integer valued $\tilde{\D}_m>0$, we have not succeeded in ruling this possibility out.
We hereafter assume that the scaling dimensions are integers and write 
\begin{align}
\tilde{\D}_k=k.
\end{align}

We can fix the first two OPE coefficients, $f_0$ and $f_1$, by 
comparing \eqref{RFTChannelDecomposition} and \eqref{RFPWETFourPointFromTChannelOPE}
under this assumption. 
Firstly, we see that the lowest dimension operator $\mJ_0$ 
has scaling dimension $0$,~\footnote{
This conclusion is not affected by the ambiguity associated with \eqref{RFAmbiguityTChannel}.}
and hence should be identified with the identity operator $\mJ_0 = 1$.
This implies $D_{00}=1$ and $D_{0n}=0$ $(n>0)$
since the one-point function of any operator with nonzero dimension vanishes
because of scale invariance.
The OPE coefficient is
\begin{align}
f_0\lef(\frac{x^2}{t}\ri)  = e^{-\frac{x^2}{2t}} \frac x{\sqrt{2\pi t}},
\end{align}
and hence the leading order term in the $\Psi\bPsi$ OPE is
\begin{align}
\Psi(0,t) \bPsi(0,0) = e^{-\frac{x^2}{2t}} \frac1{\sqrt{2\pi t}} \times 1+\dots
\end{align}
This is an expected result in view of the two-point function \eqref{RFTwoPointFunctionPsibPsi}.

The subleading OPE coefficient can also be read off.
We obtain
\begin{align}
f_1^2 D_{11} = -\frac1\pi a_1(\n)
e^{-\frac{x^2}{t}},
\end{align}
where $a_1(\n)$ is given by \eqref{RFDefaNu}.
The operator $\mJ_1$ has dimension $1$ and it is natural to identify 
it with the density of the U(1) charge.
We have not fixed the normalisation of $\mJ_1$; this is the reason why
both $f_1$ and $D_{11}$ appear in the above formula.
By redefining the operator $\mJ_1$ appropriately by multiplying it by a phase 
factor, we can choose $f_1$ to be real and $\mJ_1$ to be hermitian.

Note that $a_1(\n)$ flips its sign at $\n=\frac12$.
The coefficient of the two-point function $\<\mJ_1(0,x) \mJ_1(0,0)\>$, $D_{11}$,
is positive for $\n<\frac12$ and negative for $\n>\frac12$.
This does not contradict the unitarity of the theory.
(The unitarity of the theory is guaranteed as the Hamiltonian of the model is hermitian.)
We recall that, for the isotropic case,
the positivity of the two-point function in a unitary CFT is proven via 
the state-operator map. The analogous state-operator map in $z=2$ Schr\"odinger invariant theory
is not applicable to the charge-zero sector.
Furthermore, one cannot invoke the positivity of the norm of the Hilbert space via 
$\<\mJ_m(t, x) \mJ_n(0, 0)\> = \<0|\mJ_m(t, x) \mJ_n(0,0)|0\>$,
since one can show that this equation holds for $t>0$ but not for $t=0$, 
which is the case of interest here. 
This point will be explained in section \ref{RSSPeculiarPropertyJ}.

The special point $\n=\frac12$, where $f_1^2 D_{11}=0$, corresponds to the point at which the Calogero model coincides with the system of free fermions;
The asymptotic series \eqref{RFBesselAsymptoticExpansion} truncates at that point.
Similar truncations of the asymptotic series occur also at $\n=-\frac12$ and at $\n=\frac32, \frac52, \cdots$.
The $\Psi\bPsi$-OPE appears to be degenerate for these special points.  

It is not possible to fix the higher OPE coefficients
$f_n$ $(n=2, 3, \cdots)$ unambiguously from the
pairwise equal-time four-point function.
This is because of the ambiguity associated with \eqref{RFAmbiguityTChannel}
(where $\d$ is chosen to be an integer).
Starting from the
pairwise equal-time point-function, 
we are forced to take the coincident limit of both pairs of the spacetime points 
$(1,3)$ and $(2,4)$.
(Note that we have to take the limit $(t_{31}, x_{31}) \to 0, (t_{42}, x_{42})\to 0$ 
with fixed $\frac{x_{31}^2}{t_{31}}$, $\frac{x_{42}^2}{t_{42}}$ in order to have a well-defined 
$\Psi\bPsi$ OPE.)
We should obtain more information 
on OPE coefficients from the general four-point function
discussed in sections \ref{RSSForPointDoubleIntegral} and \ref{RSSFourPointHyperGeometric},
since then we can, say, pinch the spacetime points $(1,3)$ while 
keeping $(2,4)$ un-pinched.
By studying this type of limit, we expect to get the complete OPE coefficients
and understanding of the primary/descendant structure of operators $\mJ_k$.
This will be left as a future problem.

It is possible to read off some properties of the operators $\mJ_k$ without going into the details of the expression of the general four-point function. 
In particular, we find that the operators $\mJ_k$ $(k\geq1)$ have a rather unusual property, namely, that their two-point functions 
vanish unless the two operators are inserted on the same time slice.
This will be shown in section \ref{RSSPeculiarPropertyJ}.

To summarise this subsubsection, we have shown that the
pairwise equal-time four-point function \eqref{RFPairwiseEqualtime4ptFunction}
can be decomposed by using two $\Psi \bPsi$ OPEs. 
The decomposition \eqref{RFTChannelDecompositionUsingXPrime} is represented by an
asymptotic series rather than a convergent series.
The charge-zero operators $\mJ_m$ appearing in the intermediate channels
appear to have integer-valued scalar dimensions. $\mJ_0$ is the identity operator.
We have computed the leading and the next-to-leading OPE coefficients 
associated with $\mJ_0$ and $\mJ_1$.

Since we have represented the same four-point function now in two ways as ``the s-channel'' decomposition (in section \ref{RSSSDecompositionSChannel}) and ``the t-channel'' decomposition here,
we have thereby shown the operator associativity for this model, for the particular four-point function. 
Schematically, we have shown

\begin{fmffile}{s=t}
\begin{equation}
\sum
\begin{gathered}
\begin{fmfgraph*}(16,30)
\fmfleft{i1,i2}
\fmfright{o1,o2}
\fmf{plain}{i1,v1,o1}
\fmf{plain}{i2,v2,o2}
\fmf{plain}{v1,v2}
\fmfv{label=$\mathsmaller\bPsi_1$,label.dist=2}{i1}
\fmfv{label=$\mathsmaller\bPsi_2$,label.dist=2}{o1}
\fmfv{label=$\mathsmaller\Psi_3$,label.dist=2}{i2}
\fmfv{label=$\mathsmaller\Psi_4$,label.dist=2}{o2}
\end{fmfgraph*}
\end{gathered}
\ \,
=
\< \Psi(t, x_4) \Psi(t, x_3) \bPsi(0, x_2) \bPsi(0, x_1) \> 
=
\,
\sum
\ \,\,
\begin{gathered}
\begin{fmfgraph*}(36,14)
\fmfleft{i1,i2}
\fmfright{o1,o2}
\fmf{plain}{i1,v1,i2}
\fmf{plain}{o1,v2,o2}
\fmf{plain}{v1,v2}
\fmfv{label=$\mathsmaller\bPsi_1$,label.dist=1.5}{i1}
\fmfv{label=$\mathsmaller\bPsi_2$,label.dist=2}{o1}
\fmfv{label=$\mathsmaller\Psi_3$,label.dist=2}{i2}
\fmfv{label=$\mathsmaller\Psi_4$,label.dist=2}{o2}
\end{fmfgraph*}
\end{gathered}
\ \ \  
.
\label{RFSchematicOperatorAssociativity}
\end{equation}
\end{fmffile}
The second equality has to be understood as the representation of a function
by an asymptotic series.

\subsubsection{The exponentially small corrections and ``u-channel'' contributions}
\label{RSSSUChannel}

The asymptotic expansion of the modified Bessel function \eqref{RFBesselAsymptoticExpansion},
and hence the ``t-channel'' decomposition of the four-point function,
comes with exponentially small contributions.
Here we will show that these correction terms can also be 
interpreted using the OPE.

As we shall see below, it is necessary to analytically continue the time variable $t$.
We write
\begin{align}
t= e^{i\a} t', \label{RFAnalyticContinuationTime}
\end{align}
where $\a \in \R$ and $t' > 0$.
We will only consider the regime $0 \le \a\le \frac\pi2$.
We consider the analytically continued pairwise equal-time four-point function 
defined by
\begin{align}
\<0|
\Psi(t, x_4) 
\Psi(t, x_3) 
\bPsi(0, x_2) 
\bPsi(0, x_1) 
|0\>
\end{align}
where the operator in the Heisenberg picture, 
$\Psi(t,x)$, is given by
\begin{align}
\Psi(t,x) = e^{Ht} \Psi(0,x) e^{-Ht}.
\end{align}
For $\a=0$,
this definition coincides with the four-point function with the Euclidean time we have been considering in this paper.
The computation in section \ref{RSS4ptfuncPairwiseEqualTime} 
(and appendix \ref{RSAPropagatorInverseSquarePotential}) goes through for the analytically continued case, and hence the result \eqref{RFPairwiseEqualtime4ptFunction} is unaffected in form,
\begin{align}
&\<0|
\Psi(t, x_4) 
\Psi(t, x_3) 
\bPsi(0, x_2) 
\bPsi(0, x_1) 
|0\>
\nonumber
\\
=&
e^{-\frac{
        x_{21}^2+x_{43}^2+\lef(x_{3}+x_{4}-x_{1}-x_{2}\ri)^2}
        {4t}}
\times
\sqrt{
    \frac{x_{21}x_{43}}
        {4\pi t^3}}
I_\n\lef(
    \frac{x_{21} x_{43}}{2t}\ri).
\end{align}
where we assume $x_{21}>0, x_{43}>0$.

By gradually increasing $\a$ from $0$ to $\frac{\pi}2$, we have $t=i t'$ and 
we obtain the four-point function of the theory with the ``Minkowski'' time:
the analytically continued four-point function, considered as a function of $t'$, 
coincides with the four-point function of the theory with the ``Minkowski'' time $t'$.
Let us write explicitly the four-point function for this case,
\begin{align}
\begin{split}
& \<\Psi(t', x_4) 
\Psi(t', x_3) 
\bPsi(0, x_2) 
\bPsi(0, x_1) \>_{\text M}
\\
=&
\<0|
\Psi_{\text M}(t', x_4) 
\Psi_{\text M}(t', x_3) 
\bPsi(0, x_2) 
\bPsi(0, x_1) 
|0\>
\\
=&
e^{i\frac{
        x_{21}^2+x_{43}^2+\lef(x_{3}+x_{4}-x_{1}-x_{2}\ri)^2}
        {4t'}}
\times
e^{-i \frac34 \pi}
\sqrt{ 
    \frac{x_{21}x_{43}}
        {4\pi t'^3}}
I_\n\lef( -i 
    \frac{x_{21} x_{43}}{2t'}\ri)
\\
=&
e^{i\frac{
        x_{21}^2+x_{43}^2+\lef(x_{3}+x_{4}-x_{1}-x_{2}\ri)^2}
        {4t'}}
\times
e^{-i \frac{\pi}{2} \lef(\n + \frac32\ri) }
\sqrt{ 
    \frac{x_{21}x_{43}}
        {4\pi t'^3}}
J_\n\lef( \frac{x_{21} x_{43}}{2t'}\ri).
\end{split}
\label{RFPWETFourPointMinkowski}
\end{align}
Here, the subscript M refers to the theory in the ``Minkowski'' signature and
the Heisenberg operator assumes the usual quantum mechanical form, 
$\Psi_{\text M}(t', x_4) =  e^{iHt'} \Psi(0,x) e^{-iHt'}$.
We used 
$I_\n(z) =e^{-\frac{i\pi \n }2} J_\n\lef( 
e^{i\frac{\pi}2}
z
\ri)$
\cite[(10.27.6)]{olver_nist_nodate}.
The expression \eqref{RFPWETFourPointMinkowski}, of course, can also be obtained directly for the Minkowski theory without relying on the analytic continuation.

In the regime $0<\a\le \frac{\pi}2$, 
the asymptotic expansion accompanied with exponentially small 
correction terms \cite[(10.40.5)]{olver_nist_nodate},
\begin{align}
I_{\nu}\lef(z\ri)\sim
\frac{e^{z}}{(2\pi z)^{\frac{1}{2}}}\sum_{p=0}^{\infty}(-1)^{p}\frac{a_{p}(\nu)}{z^{p}}
- ie^{-\nu\pi i}\frac{e^{-z}}{(2\pi z)^{\frac{1}{2}}}\sum_{p=0}^{\infty}\frac{a_{p}(\nu)}{z^{p}},
\label{RFModifiedBesselAsymptoticExpansionWithExponentialCorrections}
\end{align}
captures the modified Bessel function $I_\n(z)$ accurately.
The coefficients $a_p(\n)$ are defined in \eqref{RFDefaNu}.
The first term in \eqref{RFModifiedBesselAsymptoticExpansionWithExponentialCorrections} 
coincides with 
the asymptotic expansion 
\eqref{RFBesselAsymptoticExpansion} 
used for the ``t-channel'' decomposition in 
section \ref{RSSSDecompositionTChannel}.

The expression \eqref{RFModifiedBesselAsymptoticExpansionWithExponentialCorrections} is 
not valid for the Euclidean theory ($\a=0$).
The reason is that $\a=0$ corresponds to a Stokes 
line of the modified Bessel function $I_\n(z)$,
where the first term in \eqref{RFModifiedBesselAsymptoticExpansionWithExponentialCorrections}
is maximally dominating over the second term.~\footnote{
The Minkowski case $\a=\frac\pi2$ corresponds to an anti-Stokes line where the
second and first terms are of comparable size. For $\a>\frac\pi2$,
the second term becomes exponentially large compared to the first term.
}
As is well known, across the Stokes line, the coefficients of the smaller terms 
change almost discontinuously albeit in a controlled manner \cite{berry_uniform_1989}.
One needs to use a specially tailored expansion formula 
to study the behaviour of a function exactly on the Stokes line.
Such a formula for $I_\n(z)$ was derived in \cite{paris_note_2017}.
We found a natural interpretation in terms of OPE for the formula
\eqref{RFModifiedBesselAsymptoticExpansionWithExponentialCorrections}
rather than the expansion valid exactly on the Stokes line given in \cite{paris_note_2017}.
This may suggest that it is useful to define the Euclidean theory not exactly 
at $\a=0$ but rather using the limit $\a \to 0$.
Note that, for large $|z|$, one needs only  small $\a>0$ to make
the expansion \eqref{RFModifiedBesselAsymptoticExpansionWithExponentialCorrections} 
accurate.

\begin{sloppypar}
Substituting \eqref{RFModifiedBesselAsymptoticExpansionWithExponentialCorrections} into 
\eqref{RFPairwiseEqualtime4ptFunction}, we find that the exponentially small corrections
to the four-point function $\< \Psi(t, x_4) \Psi(t, x_3) \bPsi(0, x_2) \bPsi(0, x_1) \>$
are
\begin{align}
-i e^{ -i \pi\n } 
\times
\frac{1}{2\pi} 
e^{-\frac{
        x_{41}^2+x_{32}^2}
        {2 t}}
\frac1{t}
\lef(\sum_{p=0}^\infty 
2^p
a_{p}(\n)
\frac{t^p}{
x_{21}^p x_{43}^p}
\ri).
\label{RFExponentiallySmallTermsPWETFourPoint}
\end{align}
This formula is valid for $x_{21}>0, x_{43}>0$ for both the bosonic and fermionic models.
Hereafter in this subsubsection, we focus on the bosonic model.
\end{sloppypar}

The similarity of \eqref{RFExponentiallySmallTermsPWETFourPoint} with the 
``t-channel'' decomposition \eqref{RFTChannelDecomposition} is clear.
In particular, the exponential factor in \eqref{RFModifiedBesselAsymptoticExpansionWithExponentialCorrections} and \eqref{RFPairwiseEqualtime4ptFunction} combines in a similar manner to 
\eqref{RFTChannelDecompositionExponentialFactorCombines} and yields the exponential factor 
$e^{-\frac{ x_{41}^2+x_{32}^2} {2 t}}$ in
\eqref{RFExponentiallySmallTermsPWETFourPoint}. Comparing this exponential factor 
with the corresponding factor,
$e^{-\frac{ x_{31}^2+x_{42}^2} {2 t}}$,
in \eqref{RFTChannelDecomposition}, 
we find that the roles of spacetime points $1$ and $2$ (or equivalently $3$ and $4$) 
are interchanged.
This leads us to identify the exponentially small contributions \eqref{RFExponentiallySmallTermsPWETFourPoint} as arising from the OPEs $\Psi(t,x_3) \bPsi(0,x_2)$ and $\Psi(t,x_4) \bPsi(0,x_1)$.
Schematically these contributions can be represented as,

\begin{fmffile}{u-channel-maintext}
\begin{equation}
\sum
\begin{gathered}
\begin{fmfgraph*}(30,16)
\fmfleft{i1,i2}
\fmfright{o1,o2}
\fmf{plain, tension=3}{i2,v1}
\fmf{phantom, tension=2}{v1,i1}
\fmf{plain, tension=3}{o2,v2}
\fmf{phantom, tension=2}{v2,o1}
\fmf{plain}{v1,v2}
\fmf{plain,tension=0}{v1,o1}
\fmf{plain,tension=0, rubout}{i1,v2}
\fmfv{label=$\mathsmaller\bPsi_1$,label.angle=-90,label.dist=1}{i1}
\fmfv{label=$\mathsmaller\bPsi_2$,label.angle=-90,label.dist=1}{o1}
\fmfv{label=$\mathsmaller\Psi_3$,label.angle=90,label.dist=2}{i2}
\fmfv{label=$\mathsmaller\Psi_4$,label.angle=90,label.dist=2}{o2}
\end{fmfgraph*}
\end{gathered}.
\end{equation}
\end{fmffile}

This interpretation can be made more precise.
To clarify the connection to the OPE, we define
\begin{align}
X''=\frac{x_4+x_1}2-\frac{x_3+x_2}2.
\end{align}
Then \eqref{RFExponentiallySmallTermsPWETFourPoint} becomes, using 
$x_{21}=-X''+\frac{x_{41}-x_{32}}2$ and 
$x_{43}=X''+\frac{x_{41}-x_{32}}2$,
\begin{align}
\begin{split}
e^{-i\pi\lef(\n+\frac12\ri)}
\times
\frac1{2\pi}e^{-\frac{x_{41}^2+x_{32}^2}{2t}}
\times
\frac1t
\sum_{p=0}^{\infty}
(-1)^p 2^p a_p(\n)
\frac{t^p}{\lef(X''^2- 
\frac{(x_{41}-x_{32})^2}4\ri)^p
}.
\end{split}
\label{RFUChannelDecompositionUsingXPP}
\end{align}
Note that an extra factor $(-1)^p$ appeared in the summand,
compared to \eqref{RFExponentiallySmallTermsPWETFourPoint},
due to the rewriting in terms of the variable $X''$.
Now we see that \eqref{RFUChannelDecompositionUsingXPP} have precisely the same form,
except for the overall phase factor $e^{-i\pi\lef(\n+\frac12\ri)}$,
to the ``t-channel'' decomposition \eqref{RFTChannelDecompositionUsingXPrime}.
This is natural since both terms originate from the $\Psi \bPsi$-OPE.

The overall phase factor has a natural interpretation within the framework 
of the generalised statistics~\cite{polychronakos_non-relativistic_1989,polychronakos_physics_2006} for the Calogero model.
The generalised statistics is an interesting way of understanding various properties of 
the Calogero model as a consequence of the phase factor $e^{-i \pi\lef(\n+\frac12\ri)}$ associated with each exchange of two particles.
We indeed see that the ``u-channel'' terms which 
are obtained by the exchange of, say, the two particles at the spacetime points 1 and 2,
acquire precisely that phase factor relative to the  ``t-channel'' terms.

The successful interpretation of the exponentially small terms as the ``u-channel'' contributions relies on the fact that the coefficients of the first and the second terms of \eqref{RFModifiedBesselAsymptoticExpansionWithExponentialCorrections} are closely related. 
(Both are given in terms of $a_p(\n)$ defined by \eqref{RFDefaNu}.)
This connection is an example of the so-called resurgence phenomenon.
(See, for example, \cite{dunne_introduction_2018}.)
Thus the resurgence property of the modified Bessel function represents the fact that both ``t-channel'' and ``u-channel'' contributions arise from the $\Psi \bPsi$ OPE.

There is another way of understanding the necessity of the resurgence property 
and the role of the ``u-channel'' terms from the point of view of the OPE.
When $\nu$ is a half-odd integer (\ie\ $\nu=-\frac12, \frac12, \frac32, \frac52, \cdots$),
the asymptotic series \eqref{RFModifiedBesselAsymptoticExpansionWithExponentialCorrections} truncates and becomes exact. (The Stokes phenomenon does not occur for these values of $\n$.)
The four-point function (in Euclidean time) becomes, writing $\n=n+\frac12$ with $n=0, 1, \cdots$,~\footnote{
Some formulae for the free-boson case, $\n=-\frac12$, are presented in appendix \ref{RSAFree}.
}
\begin{align}
\begin{split}
&
\langle
\Psi(t, x_4) 
\Psi(t, x_3) 
\bPsi(0, x_2) 
\bPsi(0, x_1) 
\rangle
\\
=&
e^{-\frac{
        x_{21}^2+x_{43}^2+\lef(x_{3}+x_{4}-x_{1}-x_{2}\ri)^2}
        {4t}}
\times
\frac{x_{21}x_{43}}{2\pi t^2}
\times
\mathsf{i}^{(1)}_{n}
\lef(
    \frac{x_{21} x_{43}}{2t}\ri)
\end{split}
\label{RFFourPointSphericalBessel}
\end{align}
where ${\mathsf{i}^{(1)}_{n}}\lef(z\ri)$ is a modified spherical Bessel function 
defined by \cite[(10.49.8)]{olver_nist_nodate}
\begin{align}
{\mathsf{i}^{(1)}_{n}}\lef(z\ri)=\tfrac{1}{2}e^{z}
\sum_{k=0}^{n}(-1)^{k}\frac{a_{k}(n+\frac{1}{2})}{z^{k+1}}
+(-1)^{n+1}\*\tfrac{1}{2}e^{-z}\sum_{k=0}^{n}\frac{a_{k}(n+\frac{1}{2})}{z^{k+1}},
\label{RFDefModifiedSphericalBessel}
\end{align}
which is related to $I_{\n}(z)$ by \cite[(10.47.7)]{olver_nist_nodate},
\begin{align}
{\mathsf{i}^{(1)}_{n}}\left(z\right)=\sqrt{\tfrac{1}{2}\pi/z}I_{n+\frac{1}{2}}\left(z\right).
\end{align}
The first and the second finite sum in \eqref{RFDefModifiedSphericalBessel} 
correspond to the exponentially large and small contributions
in \eqref{RFModifiedBesselAsymptoticExpansionWithExponentialCorrections}, respectively.
Since these formulae are valid for all $z=\frac{x_{43}x_{21}}{2t}$,
one can in particular consider the limit $z\to 0$.
This limit corresponds to the limit where spacetime points $(1,2)$ or $(3,4)$ become
coincident (related to the ``s-channel'' decomposition studied in section \ref{RSSSDecompositionSChannel}).
Although each term in the first and second sum in \eqref{RFDefModifiedSphericalBessel}
diverges, there are cancellations between these terms such that
${\mathsf{i}^{(1)}_{n}}\left(z\right)\sim z^n$ for $z\to 0$.
This must be the case.
Consider, say, the limit $x_{43} \to 0$, 
in which $z$ also goes to zero, $z\sim x$.
In this limit, the four-point function is controlled by the OPE \eqref{RFPsiPsiOPESimple}, $\Psi(0,x) \Psi(0,0)\sim x^{n+1} \Phi$.
(Note that the scaling dimensions of the operators $\Psi$ and $\Phi$ are $\frac12$ and $n+2$, respectively.)
Hence the four-point function behaves as $x_{43}^{n+1}$.
This agrees with \eqref{RFFourPointSphericalBessel} and \eqref{RFDefModifiedSphericalBessel} 
together with ${\mathsf{i}^{(1)}_{n}}\left(z\right)\sim z^n$.
The consistency of the four-point function with the $\Psi\bPsi$ OPE
relies on the cancellations, which in turn occur
because of  the resurgence relations, \ie\ 
the relations between the coefficients of the
exponentially small and large terms of \eqref{RFModifiedBesselAsymptoticExpansionWithExponentialCorrections}.

It is intriguing that the interpretation of the exponentially small terms as the ``u-channel'' contributions means  that the operator associativity relation  \eqref{RFSchematicOperatorAssociativity}  can be made more accurate by including ``u-channel'' contributions. Schematically, we have,

\begin{fmffile}{s=t+u}
\begin{equation}
\sum
\begin{gathered}
\begin{fmfgraph*}(16,30)
\fmfleft{i1,i2}
\fmfright{o1,o2}
\fmf{plain}{i1,v1,o1}
\fmf{plain}{i2,v2,o2}
\fmf{plain}{v1,v2}
\fmfv{label=$\mathsmaller\bPsi_1$,label.dist=2}{i1}
\fmfv{label=$\mathsmaller\bPsi_2$,label.dist=2}{o1}
\fmfv{label=$\mathsmaller\Psi_3$,label.dist=2}{i2}
\fmfv{label=$\mathsmaller\Psi_4$,label.dist=2}{o2}
\end{fmfgraph*}
\end{gathered}
\ 
=
\,
\sum
\ \,\,
\begin{gathered}
\begin{fmfgraph*}(36,14)
\fmfleft{i1,i2}
\fmfright{o1,o2}
\fmf{plain}{i1,v1,i2}
\fmf{plain}{o1,v2,o2}
\fmf{plain}{v1,v2}
\fmfv{label=$\mathsmaller\bPsi_1$,label.dist=1.5}{i1}
\fmfv{label=$\mathsmaller\bPsi_2$,label.dist=2}{o1}
\fmfv{label=$\mathsmaller\Psi_3$,label.dist=2}{i2}
\fmfv{label=$\mathsmaller\Psi_4$,label.dist=2}{o2}
\end{fmfgraph*}
\end{gathered}
\ \ \,\,
+
\,\,
\sum
\begin{gathered}
\begin{fmfgraph*}(30,16)
\fmfleft{i1,i2}
\fmfright{o1,o2}
\fmf{plain, tension=3}{i2,v1}
\fmf{phantom, tension=2}{v1,i1}
\fmf{plain, tension=3}{o2,v2}
\fmf{phantom, tension=2}{v2,o1}
\fmf{plain}{v1,v2}
\fmf{plain,tension=0}{v1,o1}
\fmf{plain,tension=0, rubout}{i1,v2}
\fmfv{label=$\mathsmaller\bPsi_1$,label.angle=-90,label.dist=1}{i1}
\fmfv{label=\ $\mathsmaller\bPsi_2$,label.angle=-90,label.dist=1}{o1}
\fmfv{label=$\mathsmaller\Psi_3$,label.angle=90,label.dist=2}{i2}
\fmfv{label=$\mathsmaller\Psi_4$,label.angle=90,label.dist=2}{o2}
\end{fmfgraph*}
\end{gathered}\ .
\end{equation}
\end{fmffile}

\subsection{Detailed analysis of ``s-channel'' decomposition}
\label{RSSDetailedSchannel}

In this subsection, we take a closer look into 
the ``s-channel'' OPE decomposition  
of the pairwise equal-time four-point function (\ref{RFPairwiseEqualtime4ptFunction})
which arises when we consider the OPE of 
$\bPsi(0, x_2)\bPsi(0, x_1)$ and of $\Psi(t, x_4) \Psi(t, x_3)$.
In section \ref{RSSSDecompositionSChannel}, we have seen that the operators
$\Phi_k$ appearing in the $\Psi\Psi$ OPE, \eqref{RFPsiPsiOPESimple},
have dimensions,
\begin{align}
\D_k = \frac32 + \n + 2k, \quad (k=0, 1, 2, \cdots),
\end{align}
and the lowest dimension operator $\Phi=\Phi_0$ is 
a primary operator. We also obtained the leading OPE coefficient
\eqref{RFCZero} involving $\Phi$.

We will now study the subleading operators $\Phi_k$  ($k=1, 2, \cdots$)
in the $\Psi\Psi$ OPE and show that they coincide with 
the following special descendants of 
the primary operator $\Phi$,
\begin{align}
\Phi^{(k)}=&(\der_x^2 -4\der_t)^k \Phi.
\end{align}
The corresponding special descendants of $\bPhi$ are
\begin{align}
\bPhi^{(k)}=&(\der_x^2 +4\der_t)^k \bPhi.
\end{align}
Thus the $\Psi\Psi$ OPE 
involves only one primary operator $\Phi$.
This will be shown in the following steps.
Firstly, in section \ref{RSSSOPESchannelDescendants},
we fix the form of the special descendants 
$\Phi^{(k)}$ appearing in the OPE by studying a part of the decomposition of the four-point function 
\eqref{RFSchannelDecomposition}.
Next, we compute the coefficients 
of the $\Psi\Psi$ OPE involving the $\Phi^{(k)}$'s.
Finally, we show that there are no subleading 
operators other than $\Phi^{(k)}$ appearing in the $\Psi\Psi$ OPE.
(For example, a primary operator $\Phi'$ 
with dimension $\frac32+\n+ 2n$, where $n$ is
a positive integer, could appear on the RHS of 
\eqref{RFPsiPsiOPESimple}. We have to exclude
this type of possibilities.)
This is done in section \ref{RSSSOPESchannelReproduce} 
by completely reproducing the
full pairwise equal-time four-point function \eqref{RFPairwiseEqualtime4ptFunction}
just by summing up
contributions from the primary operator $\Phi$ 
together with $\Phi^{(k)}$.
This shows in particular that the $\Psi\Psi$ OPE is
exhausted by the primary operator $\Phi$ and its special descendants $\Phi^{(k)}$.
(In other words, one can put $\Phi_k= \Phi^{(k)}$ 
in \eqref{RFPsiPsiOPESimple}.)

Throughout section \ref{RSSDetailedSchannel} we will assume $x_{21}>0, x_{43}>0$
without loss of generality.
Under this assumption, all formulae are valid for both the bosonic and fermionic theories.

\subsubsection{The contribution from the descendants of $\bPhi, \Phi$} 
\label{RSSSOPESchannelDescendants}
We will fix the descendants of $\bPhi, \Phi$ appearing 
in the OPE \eqref{RFbPsibPsiOPESimple} and 
\eqref{RFPsiPsiOPESimple}.
We will see that the following observation is essential: each term in 
the ``s-channel'' decomposition \eqref{RFSchannelDecomposition} of the four-point function 
contains $X$ only in the exponent and not in the prefactor of the exponential factor
$e^{-\frac{X^2}t}$.

We consider a part of the ``s-channel'' decomposition 
\eqref{RFSchannelDecomposition}, namely, the leading order terms in the expansion in terms of
$x_{21}$ (keeping all subleading terms in the expansion by $x_{43}$),
\begin{align}
\begin{split}
&\<
\Psi(t, x_4) 
\Psi(t, x_3) 
\bPsi(0, x_2) 
\bPsi(0, x_1) 
\>
\\
\approx
&e^{-\frac{x_{43}^2}{4t}}
e^{-\frac{X^2}{t}}
\times
\sqrt{
    \frac{x_{21}x_{43}}
        {4\pi t^3}}
\lef(\frac{x_{21}x_{43}}{4t}\ri)^\n
\frac1{\G(\n+1)}
\\
=
&
\sum_{k=0}^{+\infty}
\frac{1}{k!} \lef(-\frac{x_{43}^2}{4t}\ri)^k
e^{-\frac{X^2}{t}}
\times
\sqrt{
    \frac{x_{21}x_{43}}
        {4\pi t^3}}
\lef(\frac{x_{21}x_{43}}{4t}\ri)^\n
\frac1{\G(\n+1)}.
\end{split}
\label{RFPWET4ptLeadingX12}
\end{align}
These terms should arise from the lowest dimension operator $\bPhi$ in the
$\bPsi(0, x_2)\bPsi(0, x_1)$ OPE.
Each term in this series corresponds to each operator  
contained in the $\Psi(t, x_4)\Psi(t, x_3)$ OPE. 
Now, in a theory with $z=2$ Schr\"odinger symmetry, primary operators
with different scaling dimensions have
vanishing two-point functions~\cite{henkel_schrodinger_1994}.
This means that $k\geq1$ terms in \eqref{RFPWET4ptLeadingX12} 
must all come from the descendants of $\Phi$ appearing in the $\Psi\Psi$ OPE.

In order to obtain the expression for these descendant operators, we need to 
know the two-point functions between a primary operator and its descendants.
We will set $(x_1+x_2)/2=0$, $(x_3+x_4)/2=X$, for simplicity.
Let us first consider the $k=1$ case.
The relevant descendant operators should have dimension $\D_0+2$;
they are $\der_x^2 \Phi$ and $\der_t \Phi$. 
(We recall that $\D_0=\frac32+\n$.)
Taking spacetime derivatives of the two-point function (\ref{RFTwoPointPhibPhi}), we obtain 
\begin{align}
\langle \der_x^2 \Phi (t,X) \bPhi(0,0)\rangle 
=& \frac1{t^{\D_0}} \lef(-\frac{2}{t}+ \frac{4X^2}{t^2} \ri) e^{-\frac{X^2}{t}},
\label{RFTwoPointDerXSquaredPhibPhi}
\\
\langle \der_t \Phi (t,X) \bPhi(0,0)\rangle 
=& \frac1{t^{\D_0}} \lef(-\frac{\D_0}{t}+ \frac{X^2}{t^2} \ri) e^{-\frac{X^2}{t}}.
\label{RFTwoPointDerTPhibPhi}
\end{align}
Notice that each of the expressions contains $X$
in the prefactor of $e^{-\frac{X^2}{t}}$.
However, we see that the $k=1$ term (in fact, all terms) 
in \eqref{RFPWET4ptLeadingX12} does not 
contain $X$ in the prefactor of $e^{-\frac{X^2}{t}}$.
Therefore, 
the special linear combination of the descendant operators $\der_x^2\Phi$ and $\der_t \Phi$,
\begin{align}
\Phi^{(1)}= \lef(\der_x^2-4\der_t\ri)\Phi,
\end{align}
must be responsible for the $k=1$ term in \eqref{RFPWET4ptLeadingX12}.
The linear combination $\Phi^{(1)}$ is constructed so that the two-point function
\begin{align}
\langle 
\Phi^{(1)} (t,X) \bPhi(0,0)\rangle 
=
4\lef(\D_0-\frac12\ri)\times \frac1{t^{\D_0+1}}e^{-\frac{X^2}{t^2}}.
\end{align}
does not contain $X$ in the prefactor of $e^{-\frac{X^2}{t}}$.
Thus, the first subleading term in the $\Psi\Psi$ OPE should contain descendants of $\Phi$ 
only in the form of $\Phi^{(1)}$.

One can repeat this process of forming linear combinations of descendant operators further 
to construct  special descendant operators $\Phi^{(k)}$; we observe that 
the necessary computations are the same, except that $\D_0$ should be replaced by
$\D_0 + 1$ and then by $\D_0+2$, and so forth.~\footnote{
To construct the special descendant operators by linear combinations,
the operators arising at each step by applying $\der_x^2$ and $\der_t$
should be linearly independent. This is assured for 
$\D_0> \frac12$.
}
We obtain the special descendant operators 
\begin{align}
\Phi^{(k)} = 
\lef(\der_x^2-4\der_t\ri)^k
\Phi,
\end{align}
with the two-point functions
\begin{align}
\begin{split}
\langle
\lef(\der_x^2-4\der_t\ri)^k 
\Phi\lef(t, X\ri)
\bPhi\lef(0, 0\ri)
\rangle
=&
4^{k}\lef(\D_0-\frac12+k-1\ri)\cdots\lef(\D_0-\frac12\ri)
\frac1{t^{\D_0+k}} e^{-\frac{X^2}{t}}
\\
=&
4^{k}
\frac{\G\lef(\n+k+1\ri)}{\G\lef(\n+1\ri)}
\frac1{t^{\frac32+\n+k}} e^{-\frac{X^2}{t}},
\end{split}
\label{RFTwoPointSpecialDescendantSpecial}
\end{align}
which do not contain $X$ in the prefactor of $e^{-\frac{X^2}{t}}$.
(We used $\D_0=\frac32+\n$ above.)

The coefficients before $\Phi^{(k)}$ in the $\Psi\Psi$ OPE can be read off
from \eqref{RFPWET4ptLeadingX12} using
\eqref{RFTwoPointSpecialDescendantSpecial} and \eqref{RFCZero}.
We obtain,
\begin{align}
\begin{split}
&
\Psi(t, x_4)
\Psi(t, x_3)
\\
=&
\sum_{k=0}^{+\infty}
\frac1{2^\n (4\pi)^{\frac14}\sqrt{\G(\n+1)}}
\times
(-1)^k
\frac{1}{k!}
\frac1{4^{2k}}
\frac{\G\lef(\n+1\ri)}{\G\lef(\n+k+1\ri)}
\times
x_{43}^{\n+\frac12 + 2k}
\times
\Phi^{(k)}\lef(t, \frac{x_3+x_4}2\ri).
\end{split}
\label{RFEqualTimeOPEPsiPsi}
\end{align}
The term with $k=0$ of course is the leading order term in the OPE 
we have already seen in \eqref{RFbPsibPsiOPESimple} and 
\eqref{RFCZero}, $\Phi=\Phi^{(0)}$.

We have shown that the descendants of $\Phi$ should appear in the $\Psi\Psi$ OPE 
in the way given in \eqref{RFEqualTimeOPEPsiPsi}.
However, there could be another primary operator,
say, $\Phi'$ with dimension $\D_0+ 2n$ where $n$ is a non-negative integer,
which enters the $\Psi\Psi$ OPE together with its descendants.
(In other words, $\Phi_k$ in \eqref{RFPsiPsiOPESimple} may be a
linear combination of $\Phi^{(k)}$ and $\Phi'$ itself or its descendants.)
We will exclude this possibility in section \ref{RSSSOPESchannelReproduce}.
Once this is done, we can conclude that \eqref{RFEqualTimeOPEPsiPsi} is complete and coincides with \eqref{RFPsiPsiOPESimple}
with $\Phi_k=\Phi^{(k)}$ and
\begin{align}
C_k= \frac1{2^\n (4\pi)^{\frac14}\sqrt{\G(\n+1)}}
\times
(-1)^k
\frac{1}{k!}
\frac1{4^{2k}}
\frac{\G\lef(\n+1\ri)}{\G\lef(\n+k+1\ri)}.
\label{RFCk}
\end{align}

By repeating the same argument 
starting from the leading order terms in $x_{43}$ of \eqref{RFSchannelDecomposition},
we obtain similar results for the $\bPsi\,\bPsi$ OPE.
Thus, descendants of $\bPhi$ must enter the  $\bPsi\,\bPsi$ OPE in the 
following special linear combinations,~\footnote{
We note that the sign flip before $\der_t$ of $\bPhi^{(k)}$ compared to $\Phi^{(k)}$ 
is due to our use of Euclidean time, 
$\Phi(t, x)= e^{Ht}\Phi(0, x)e^{-Ht}$,
$\bPhi(t, x)= e^{Ht}\bPhi(0, x)e^{-Ht}$.
Thus we have $\overline{\Phi^{(k)}(0, x)}= \bPhi^{(k)}(0, x)$ because
$\overline{\der_t \Phi(0, x)} = \overline{[H, \Phi(0, x)]} = -[H, \bPhi(0, x)]
=-\der_t \bPhi(0, x)$.
}
\begin{align}
\bPhi^{(k)}=&(\der_x^2 +4\der_t)^k \bPhi,
\end{align}
which are constructed so that  the two-point  functions
\begin{align}
\begin{split}
\langle
\lef(\der_x^2+4\der_t\ri)^k
\bPhi\lef(t, X\ri)
\Phi\lef(0, 0\ri)
\rangle
=
4^{k}
\frac{\G\lef(\n+k+1\ri)}{\G\lef(\n+1\ri)}
\frac1{t^{\frac32+\n+k}} e^{-\frac{X^2}{t}},
\end{split}
\end{align}
do not contain $X$ in the prefactor of $e^{-\frac{X^2}{t}}$.
The $\bPsi\,\bPsi$ OPE becomes
\begin{align}
\begin{split}
&
\bPsi(0, x_2)
\bPsi(0, x_1)
\\
=&
\sum_{n=0}^{+\infty}
\frac1{2^\n (4\pi)^{\frac14}\sqrt{\G(\n+1)}}
\times
(-1)^n
\frac{1}{n!}
\frac1{4^{2n}}
\frac{\G\lef(\n+1\ri)}{\G\lef(\n+n+1\ri)}
\times
x_{21}^{\n+\frac12 + 2n}
\times
\bPhi^{(n)} \lef(0, \frac{x_1+x_2}2\ri).
\end{split}
\label{RFEqualTimeOPEBPsiBPsi}
\end{align}
Again we will see in section \ref{RSSSOPESchannelReproduce} that
\eqref{RFEqualTimeOPEBPsiBPsi} is complete and coincides with \eqref{RFbPsibPsiOPESimple}
with $\bPhi_k=\bPhi^{(k)}$ and \eqref{RFCk}.

The important property of the special descendants $\bPhi^{(m)}$, $\Phi^{(n)}$
is that their mutual two-point functions
\begin{align}
\begin{split}
\<
\bPhi^{(m)}\lef(t, X\ri)
\Phi^{(n)}\lef(0, 0\ri)
\>
=
4^{m+n}
\frac{\G\lef(\n+m+n+1\ri)}{\G\lef(\n+1\ri)}
\frac1{t^{\frac32+\n+m+n}} e^{-\frac{X^2}{t}},
\end{split}
\label{RFTwoPointSpecialDescendantGeneral}
\end{align}
do not contain $X$ in the prefactor of $e^{-\frac{X^2}{t}}$.
This reflects the absence of $X$ 
in the prefactor of $e^{-\frac{X^2}{t}}$ for all terms contained in
\eqref{RFSchannelDecomposition}.

\subsubsection{Reproducing full four-point function from OPE}
\label{RSSSOPESchannelReproduce}

Here we shall prove that the OPEs
\eqref{RFEqualTimeOPEPsiPsi} and 
\eqref{RFEqualTimeOPEBPsiBPsi}
are complete by
showing that they fully reproduce 
the pairwise equal-time four-point function (\ref{RFPairwiseEqualtime4ptFunction}).

From the OPEs \eqref{RFEqualTimeOPEPsiPsi} and 
\eqref{RFEqualTimeOPEBPsiBPsi}
and the two-point functions \eqref{RFTwoPointSpecialDescendantGeneral},
we obtain 
\begin{align}
\begin{split}
&
\langle
\Psi(t, x_4) 
\Psi(t, x_3) 
\bPsi(0, x_2) 
\bPsi(0, x_1) 
\rangle
\\
=&
\frac1{4^\n (4\pi)^{\frac12}\G(\n+1)}
\\
\times&
\sum_{m=0}^\infty
\sum_{n=0}^\infty
\frac{(-1)^n}{n!}
\frac1{4^{2n}}
\frac{\G\lef(\n+1\ri)}{\G\lef(\n+n+1\ri)}
x_{43}^{\n+\frac12 + 2n}
\times
\frac{(-1)^m}{m!}
\frac1{4^{2m}}
\frac{\G\lef(\n+1\ri)}{\G\lef(\n+m+1\ri)}
x_{21}^{\n+\frac12 + 2m}
\\
\times&
4^{m+n}
\frac{\G\lef(\n+m+n+1\ri)}{\G\lef(\n+1\ri)}
\frac1{t^{\frac32+\n+m+n}} e^{-\frac{X^2}{t}}
\\
%%%%%%%%%%%%%%%%%%%%%%%%%%%%%%%%%%%%%%%%
=&
e^{-\frac{X^2}{t}}
\sqrt{
\frac{x_{21} x_{43}}{4\pi t^3}
}
\lef(
\frac{x_{21}
x_{43}}
{4 t} 
\ri)^\n
\sum_{m=0}^\infty
\sum_{n=0}^\infty
\frac{(-1)^{m+n}}{m! n!}
\frac{\G\lef(\n+m+n+1\ri)}{\G\lef(\n+m+1\ri)\G\lef(\n+n+1\ri)}
\frac{
x_{21}^{2m}
x_{43}^{2n}
}
{(4t)^{m+n}}.
\end{split}
\end{align}

We wish to show this formula agrees with the ``s-channel'' decomposition 
\eqref{RFSchannelDecomposition}.
Factoring out the common factor, the identity we have to show is
\begin{align}
\begin{split}
\sum_{m=0}^\infty
\sum_{n=0}^\infty
\frac{(-1)^{m+n}}{m! n!}
%%%\frac{\lef(\n+m+1\ri)_n}{\G\lef(\n+n+1\ri)} %%%Pochhammer
\frac{\G\lef(\n+m+n+1\ri)}{\G\lef(\n+m+1\ri)\G\lef(\n+n+1\ri)}
\frac{
x_{21}^{2m}
x_{43}^{2n}
}
{(4t)^{m+n}} 
%\stackrel{!}{=}
=
e^{-\frac{x_{21}^2+x_{43}^2}{4t}}
\sum_{k=0}^{\infty}\frac{\lef(
\frac{x_{21} x_{43}}{4t}
\ri)^{2 k}}{k!\Gamma\lef(\nu+k+1\ri)},
\end{split}
\end{align}
which is equivalent to
\begin{align}
\frac{1}{m! n!}
\frac{\G\lef(\n+m+n+1\ri)}{\G\lef(\n+m+1\ri)\G\lef(\n+n+1\ri)}
=
%\stackrel{!}{=}
\sum_{k=0}^{\min(m,n)}
\frac1{k!\Gamma\lef(\nu+k+1\ri)}
\frac1{(m-k)!} 
\frac1{(n-k)!}.
\label{RFGammaIdentity}
\end{align}
Now, 
using the Pochhammer symbol,
we have
\begin{align}
\begin{split}
\text{(RHS)}=&
\sum_{k=0}^{\min(m,n)}
\frac1{k!}
\frac1{(\n+1)_k \Gamma\lef(\nu+1\ri)}
(-1)^k\frac{(-m)_k}{m!}
(-1)^k\frac{(-n)_k}{n!}
\\
=&
\sum_{k=0}^{+\infty}
\frac1{k!}
\frac1{(\n+1)_k \Gamma\lef(\nu+1\ri)}
\frac{(-m)_k}{m!}
\frac{(-n)_k}{n!}
\\
=&
\frac1{\G\lef(\n+1\ri) m! n!}
\ {}_2 F_1(-m,-n; \n+1;1)
=\text{(LHS)},
\end{split}
\end{align}
where we used a well-known identity for hypergeometric functions,
\begin{align}
{}_2 F_1(a,b; c;1)
=\frac{\G(c)\G(c-a-b)}{\G(c-a)\G(c-b)}.
\end{align}
Thus the OPEs \eqref{RFEqualTimeOPEPsiPsi} and 
\eqref{RFEqualTimeOPEBPsiBPsi}
reproduce the pairwise equal-time four-point function (\ref{RFPairwiseEqualtime4ptFunction})
fully and hence are complete.

\subsection{Three-point function $\<\Psi\Psi\bPhi\>$}
\label{RSSThreePoint}
We have seen in section \ref{RSSDetailedSchannel}
that 
there is only one primary operator, $\bPhi$,
involved in the OPE $\bPsi(x) \bPsi(0)$.
By pinching the two insertion points of $\bPsi$ 
of the four-point functions obtained in section \ref{RS4ptfunc},
we can compute the three-point function $\<\Psi\Psi\bPhi\>$.

In \cite{golkar_operator_2014}, Golkar and Son 
showed that the constraint from Schr\"odinger symmetry alone 
fixes the spacetime dependence of the three-point function
(except, of course, the overall coefficient which contains the dynamical information
of the theory considered) when one of the operators involved saturates the unitarity bound,
which, in one space dimension, is $\D\geq \frac12$.
The field $\Psi$ saturates the unitarity bound.
The form of the three-point function we obtained
is consistent with Golkar and Son's analysis.
Since their analysis is done in Minkowski signature,
and the continuation to Euclidean signature is not entirely trivial, 
we give the analysis done for theories with Euclidean time in appendix 
\ref{RSAGolkarSonEuclid}. The appendix also contains a discussion
of the boundary conditions necessary to fix the spacetime dependence.
We point out that the boundary conditions give different constraints for one
space dimension compared to other cases.

\subsubsection{Three-point functions with two operators at equal-time}
\label{RSSSThreePointEqualTime}

First, we consider the case in which two $\Psi$'s are inserted at the same time,
\begin{align}
\<
\Psi(t, x_3)
\Psi(t, x_2)
\bPhi(0, 0)
\>.
\end{align}
We will consider the case $t>0$; If $t<0$ the three-point function vanishes trivially since the operator $\Psi(x)$ annihilates the vacuum.

We keep the leading order term in the expansion in $x_{21}=x_2-x_1$ in 
the pairwise equal-time four-point function (\ref{RFPairwiseEqualtime4ptFunction}) using
\eqref{RFBesselExpansionAroundZero}.
Putting $\frac{x_1+x_2}{2}=0$, we obtain
\begin{align}
\begin{split}
&
\langle
\Psi(t, x_4) 
\Psi(t, x_3) 
\bPsi(0, x_2) 
\bPsi(0, x_1) 
\rangle
\\
\approx&
e^{-\frac{x_{43}^2+\lef(x_{3}+x_{4}\ri)^2}
        {4t}}
\times
\sqrt{
    \frac{x_{21}x_{43}}
        {4\pi t^3}}
\frac1{\Gamma\lef(\n+1\ri)}
\lef(\frac{x_{21} x_{43}}{4t}\ri)^\n
+\cdots.
\end{split}
\end{align}
This is valid when $x_{43}>0, x_{21}>0$ for both the bosonic and fermionic models.
Comparing this with the leading term of the OPE (\ref{RFEqualTimeOPEBPsiBPsi})
\begin{align}
\bPsi(0, x_2)
\bPsi(0, x_1)
\approx &
\frac1{2^\n (4\pi)^{\frac14}\sqrt{\G(\n+1)}}
x_{21}^{\n+\frac12 }
\times
\bPhi\lef(0, 0\ri),
\end{align}
we obtain, after relabelling,
\begin{align}
\begin{split}
&
\langle
\Psi(t, x_3) 
\Psi(t, x_2) 
\bPhi\lef(0, 0\ri)
\rangle
\\
=&
\frac1{2^\n (4\pi)^{\frac14}\sqrt{\G(\n+1)}}
\times
x_{32}^{\n+\frac12}
t^{-\lef(\n+\frac32\ri)}
\times 
e^{-\frac{x_{3}^2+x_{2}^2}{2t}}.
\end{split}
\label{RFThreepointSemiEqualTime}
\end{align}
This expression is valid
when $x_{32}>0$ for both the bosonic and fermionic models.
The formula valid for $x_{32}<0$ can be obtained easily as is done for the equal-time four-point function, \eqref{RFPairwiseEqualtime4ptFunctionBosonic} and
\eqref{RFPairwiseEqualtime4ptFunctionFermionic}.
The normalisation conditions for the operators $\Psi$ and $\Phi$
are fixed by the two-point functions, (\ref{RFTwoPointFunctionPsibPsi}) and
(\ref{RFTwoPointPhibPhi}).

\begin{sloppypar}
The result \eqref{RFThreepointSemiEqualTime} is,
apart from an overall factor,
the product of two free propagators
$\< \Psi(t, x_3) 
\bPsi\lef(0, 0\ri)\>\< 
\Psi(t, x_2) 
\bPsi\lef(0, 0\ri)\> \sim
\frac1t e^{-\frac{x_{3}^2+x_{2}^2}{2t}}$ 
dressed with a factor (depending on $x_{32}$ and $t$)
$ x_{32}^{\n+\frac12} t^{-\lef(\n+\frac12\ri)}$.
\end{sloppypar}

\subsubsection{Three-point functions in general position}
\label{RSSThreePointGeneralPosition}

\paragraph{Integral representation}
We consider the three-point function in general position,
\begin{align}
\<
\Psi(t_3, x_3)
\Psi(t_2, x_2)
\bPhi(t_1, x_1)\>.
\end{align}
Here we will only consider the bosonic model.
We set $x_1=0, t_1=0$ using translational invariance.
We assume $t_3>t_2$ without loss of generality.
We consider the case $t_2>t_1$ since otherwise the three-point function
vanishes trivially, the operator $\Psi(t_2, x_2)$ annihilating the vacuum.

We begin with the integral representation (\ref{RFSingleIntegral})
for the four-point function
$\<\Psi(t_4, x_4) \Psi(t_3, x_3) \bPsi(0, x_2) \bPsi(0, x_1) \>$, with
$t_4>t_3>0$. (The labels $3,4$ will be replaced respectively by $2,3$ later.)
We consider the limit $x_2 \to x_1$ and
keep the leading order term in the expansion in terms of $x_{21}$.
Writing $t_3=t, t_{43}=t'$ 
and setting $\frac{x_1+x_2}{2}=0$, 
we obtain, using the leading order term of (\ref{RFBesselExpansionAroundZero})
and \eqref{RFPairwiseEqualtime4ptFunctionBosonic},
\begin{align}
\begin{split}
&\langle
\Psi(t+t', x_4)
\Psi(t, x_3)
\bPsi(0, x_2)
\bPsi(0, x_1)
\rangle
\\
\approx&
\int
e^{-\frac{
        x_{4'3}^2+\lef(x_{3}+x_{4'}\ri)^2}
        {4t}}
\times
\sqrt{
    \frac{|x_{21}x_{4'3}|}
        {4\pi t^3}}
\times
\lef|\frac{x_{21} x_{4'3}}{4t}\ri|^\n
\frac1{\G(\n+1)}
\times
\sqrt{\frac{1}{2\pi t'}} e^{-\frac{1}{2}\frac{x_{44'}^2}{t'}}
dx_4'.
\end{split}
\label{RFSingleIntegralFormula12Pinched}
\end{align}
Note that we used \eqref{RFPairwiseEqualtime4ptFunctionBosonic} valid for the bosonic model and
applicable for both positive and negative $x_{4'3}$.

Comparing (\ref{RFSingleIntegralFormula12Pinched})
with the leading order term in the OPE (\ref{RFEqualTimeOPEBPsiBPsi}),
\begin{align}
\bPsi(0, x_2)
\bPsi(0, x_1)
\approx &
\frac1{2^\n (4\pi)^{\frac14}\sqrt{\G(\n+1)}}
\lef|x_{21}\ri|^{\n+\frac12 }
\times
\bPhi\lef(0, 0\ri),
\end{align}
we obtain an integral representation of the three-point function in general position,
\begin{align}
\begin{split}
&\langle
\Psi(t+t', x_4)
\Psi(t, x_3)
\bPhi\lef(0, 0\ri)
\rangle
\\
=&
\frac1{2^{\n+1}\pi^{\frac34}\sqrt{\G(\n+1)}}
\frac1{\sqrt{t'}}
\frac1{t^{\frac32+\n}}
\int
e^{-\frac{
        x_{4'3}^2+\lef(x_{4'}+x_{3}\ri)^2}
        {4t}}
\times
|x_{4'3}|^{\n+\frac12}
\times
e^{-\frac{1}{2}\frac{x_{44'}^2}{t'}}
dx_{4'}.
\end{split}
\label{RFThreePointIntegralRep}
\end{align}

\paragraph{Three-point function in general position}
This integral can be worked out, separating contributions from $x_{4'3}>0$ and $x_{4'3}<0$.
The result can be expressed in terms of the parabolic cylinder functions or
the confluent hypergeometric functions.
The details of the computation, including the comparison to the
generic form of the three-point function found by Henkel\cite{henkel_schrodinger_1994, henkel_schrodinger_2003},
are given in appendix \ref{RSAThreePoint}.
The final result expressed via the confluent hypergeometric function $M(a,b,z)$ (in the notation of \cite{olver_nist_nodate})
is,
\begin{align}
\begin{split}
&\langle
\Psi(t_3, x_3)
\Psi(t_2, x_2)
\bPhi\lef(t_1, x_1\ri)
\rangle
\\
=&
\frac{\G\lef(\n+\frac32\ri)}{2^{\frac32 \n+\frac34} \pi^{\frac14}\sqrt{\G(\n+1)}\Gamma\lef(\frac{\nu}2+ \frac{5}{4}\ri)}
\sqrt{\frac{{t_{32}}^{\n+\frac12}}{t_{31}^{\n+\frac32}t_{21}^{\n+\frac32}}}
e^{-\frac{x_{21}^2}{2t_{21}}}
e^{-\frac{x_{31}^2}{2t_{31}}}
\\&
\times
e^{-\frac{w^2}{2}}
M\lef(\frac\nu 2 +\frac34,\frac{1}{2},\frac{1}{2}w^{2}\ri).
\end{split}
\label{RFThreePointConfluentHypergeometricWithExp}
\end{align}
where $w$ is a quantity which is invariant under the Schr\"odinger symmetry,
\begin{align}
w
=
\lef(\frac{x_{21}}{t_{21}}
-\frac{x_{32}}{t_{32}}\ri)
\sqrt{\frac{t_{21}t_{32}}{t_{31}}}.
\end{align}
We have chosen $t_3>t_2>t_1$ and hence $w\in \R$.
(In the notation of  section \ref{RSSFourPointHyperGeometric}, $\frac12 w^2=v$).

The spacetime dependence is consistent with 
the analysis based on the Schr\"odinger symmetry by Golkar and Son~\cite{golkar_operator_2014}. 
See appendix \ref{RSAGolkarSonEuclid} for a detailed comparison.

By taking the limit $t_{32}\to 0$ of \eqref{RFThreePointConfluentHypergeometricWithExp},
we recover the result of section \ref{RSSSThreePointEqualTime}.
See appendix \ref{RSSACheckEqualTime}.
The special case $\nu=-\frac12$ agrees with the result for the
free-boson. See appendix \ref{RSASFreeThreePoint}.

\subsection{Two-point function of the charge-zero operators appearing in the ``t-channel'' decomposition}
\label{RSSPeculiarPropertyJ}
In this subsection, we will deduce a peculiar property of 
the charge-zero operators $\mJ_m$ $(m=1,2,\cdots)$ 
appearing in the $\bPsi \Psi$ OPE.
Namely, we will show that the two-point functions
\begin{align}
\< \mJ_m (t_1, x_1) \mJ_n(t_2, x_2)\> \quad (m>0, n>0)
\end{align}
vanish if $t_1\neq t_2$. Note that the two-point functions 
are non-vanishing and finite in general for $t_1=t_2$, \eqref{RFTwoPointJJ}.
Our argument is fairly general and is not restricted to the Calogero model.
The assumptions are the existence of the OPE, the scale invariance, the U(1) symmetry, 
and the uniqueness of the vacuum. Hence the argument will apply in particular
to any theory with Schr\"odinger symmetry and with a unique vacuum.

The basis of the argument is the following property
of the four-point function 
$
\< \Psi(t_4, x_4) \Psi(t_3, x_3) \bPsi(t_2, x_2) \bPsi(t_1, x_1) \>
$.
Depending on the time-order of the operators,
the four-point function (i) has a nontrivial form,
(ii) factorises into a product of two-point functions,
or (iii) vanishes.
The first possibility occurs when the time-ordered product of the operators has the form
$\Psi \Psi \bPsi\, \bPsi$, {\it i.e.} when 
$t_4>t_1, t_4>t_2, t_3 > t_2$, and $t_3> t_1$ hold.
The pairwise equal-time four-point function derived in section \ref{RSS4ptfuncPairwiseEqualTime} 
is a particular case of this possibility.
The second possibility occurs when the time-ordered product has the 
form $\Psi \bPsi \Psi \bPsi$, {\it i.e.} when 
$t_4>t_2>t_3>t_1$, 
$t_3>t_2>t_4>t_1$,
$t_4>t_1>t_3>t_2$, or 
$t_3>t_1>t_4>t_2$ hold.
The third possibility occurs when the operator with the smallest time is $\Psi$
or the largest time is $\bPsi$
(because of $\Psi|0\>=0$ and $\<0|\bPsi=0$).

Let us consider the second possibility;
to be specific, we focus on the case $t_4>t_2>t_3>t_1$.
Then we have
\begin{align}
\begin{split}
&\< \Psi(t_4, x_4) \Psi(t_3, x_3) \bPsi(t_2, x_2) \bPsi(t_1, x_1) \>
\\
=&
\<0|T \Psi(t_4, x_4) \Psi(t_3, x_3) \bPsi(t_2, x_2) \bPsi(t_1, x_1)|0 \>
\\
=&
\<0|\Psi(t_4, x_4) \bPsi(t_2, x_2) \Psi(t_3, x_3) \bPsi(t_1, x_1)|0 \>
\\
=&
\<0|\Psi(t_4, x_4) \bPsi(t_2, x_2)|0\>\<0| \Psi(t_3, x_3) \bPsi(t_1, x_1)|0 \>
\\
=
&\< \Psi(t_4, x_4) \bPsi(t_2, x_2)\> \<\Psi(t_3, x_3)\bPsi(t_1, x_1) \>.
\end{split}
\label{RF4ptFactorisation}
\end{align}
In going from the third to the fourth line, 
we inserted a complete set of eigenstates between $\bPsi(t_2, x_2)$ and $\Psi(t_3, x_3)$.
Then  we 
used the fact
that the state $\Psi(t_3, x_3) \bPsi(t_1, x_1)|0 \>$ has
vanishing U(1) charge and hence should coincide with the vacuum $|0\>$
up to a constant factor.

Let us consider the limit where both pairs of spacetime points $(4,2)$
and $(3,1)$ become coincident.
(More precisely, the limit $x_{31}\to 0$, $t_{31}\to 0$ with fixed
$\frac{x_{31}^2}{t_{31}}$, and the similar 
coincident limit for the points $(4,2)$ should be taken.)
In this limit, we can use the OPE 
\eqref{RFPsibPsiOPE} of $\Psi(t_4,x_4)\bPsi(t_2,x_2)$ and of $\Psi(t_3,x_3)\bPsi(t_1,x_1)$
which we have studied in section \ref{RSSSDecompositionTChannel},
\begin{align}
\Psi(t, x) \bPsi(0, 0) 
=&\sum_{k=0}^\infty x^{k-1}f_k\lef(\frac{x^2}t\ri) \mJ_k\lef(\frac t2, \frac{x}2\ri)
.
\end{align}
We recall that $\mJ_0$ is the identity operator and
\begin{align}
x^{-1} f_0\lef(\frac{x^2}{t}\ri)  = \frac1{\sqrt{2\pi t}} e^{-\frac{x^2}{2t}} =
\<\Psi (t,x) \bPsi(0, 0) \>
.
\end{align}
Because of the factorisation property \eqref{RF4ptFactorisation}, 
the four-point function depends on $(t_{31}, x_{31})$ and $(t_{42}, x_{42})$
but not on the relative position between 
$(t_2, x_2) \approx (t_4, x_4)$ and $(t_1, x_1) \approx (t_3, x_3)$.
This implies the vanishing of the two-point functions,
\begin{align}
\< \mJ_m(t_2,x_2)  \mJ_n(t_1, x_1)\>=0,
\end{align}
for $t_2 >t_1$,
except for the case when both the operators are the identity operator
($m=0, n=0$).
Indeed, the identity operator appearing in the $\Psi\bPsi$ OPE 
completely reproduces the factorised four-point function.

The situation is quite different from that considered in 
section \ref{RSSSDecompositionTChannel}, 
where we need infinitely many nonzero equal-time two-point functions
$\< \mJ_m(t, x) \mJ_n(t, x')\> = \frac{D_{mn}}{(x-x')^{m+n}}$,
in order to reproduce the pairwise equal-time four-point function
(except for the special cases where $\n$ is a half odd integer).

We can make this argument more precise by considering three-point functions
$\<\Psi \mJ_k \bPsi\>$.
We start from the non-factorised four-point function,
$t_4>t_3>t_2>t_1$. We then take the coincident limit of the spacetime points $(3,2)$ 
and use the OPE $\bPsi(t_3,x_3) \Psi(t_2,x_2)$. 
Focusing on each term in the OPE expansion, one obtains 
the 
three-point functions where 
$\bPsi$ is inserted at the spacetime point $4$, $\mJ_k$ is inserted at $3=2$, and $\Psi$ at $1$.

Now, if we start with a different time-ordering, say, $t_3>t_2>t_4>t_1$,\footnote{
Note that the points $3$ and $2$ should be adjacent in the time-ordering 
in order that one can take the coincident limit.}
the factorisation \eqref{RF4ptFactorisation} implies that 
the three-point function vanishes except for the special case where 
the operator appearing from the $\Psi \bPsi$ OPE is the identity operator.
Thus, the three-point function 
\begin{align}
\< \Psi(t_3, x_3) \mJ_m(t_2, x_2) \bPsi(t_1, x_1)\> \quad (m=1,2,\cdots)
\end{align}
is non-vanishing only when $t_3>t_2>t_1$.

We then take the coincident limit of the points $3$ and $1$.
Unless we maintain the time-ordering $t_3>t_2>t_1$ during the coincident limit,
the result vanishes (when $\mJ_m$ is not an identity operator). 
We again reach the conclusion that the two-point function
\begin{align}
\<\mJ_m(t_1, x_1) \mJ_n(t_2, x_2)\> \quad (m>0, n>0)
\end{align}
can have a nonzero value only if $t_1=t_2$.

The vanishing of the two-point functions can be deduced from the following more formal 
argument.
One may rewrite the two-point function as the vacuum expectation value
of the time-ordered product,
\begin{align}
\<\mJ_m(t_1, x_1) \mJ_n(t_2, x_2)\>
=
\<0|T \mJ_m(t_1, x_1) \mJ_n(t_2, x_2)|0\>.
\end{align}
Assume that $t_1 >t_2$. Then we have
\begin{align}
\<\mJ_m(t_1, x_1) \mJ_n(t_2, x_2)\>
=
\<0| \mJ_m(t_1, x_1) \mJ_n(t_2, x_2)|0\>.
\label{RFJJDifferentTRewritingUsingOperator}
\end{align}
We can insert a complete set of states between $\mJ_m(t_1, x_1)$ and
$\mJ_m(t_2, x_2)$. Since $\mJ_n(t_2, x_2)|0\>$ has vanishing U(1) charge and
the only state with zero charge is the vacuum, we obtain
\begin{align}
\<\mJ_m(t_1, x_1) \mJ_n(t_2, x_2)\>
=
\<0| \mJ_m(t_1, x_1)|0\> \<0| \mJ_n(t_2, x_2)|0\>.
\end{align}
By the scale invariance, the one-point function of any operator should vanish, 
unless the operator is an identity operator.
Therefore we see that the two-point function 
$\<\mJ_m(t_1, x_1) \mJ_n(t_2, x_2)\>$ should vanish unless $m=0, n=0$.
This argument illustrates the subtlety involved in the two-point function for the case 
$t_1=t_2$. In order to have the finite equal-time two-point functions \eqref{RFTwoPointJJ},
which are required by the ``t-channel'' decomposition of the nontrivial four-point function
and the $\Psi\bPsi$ OPE discussed in section \ref{RSSSDecompositionTChannel}, 
one must conclude that \eqref{RFJJDifferentTRewritingUsingOperator} does not hold when $t_1=t_2$,
\begin{align}
\< \mJ_m(t, x_1) \mJ_n(t, x_2)\>
\neq
\<0| \mJ_m(t, x_1) \mJ_n(t, x_2)|0\>.
\end{align}
as otherwise the equal-time two-point functions would also vanish by the same argument.

In this subsection, we 
deduced features of charge-0 operators 
arising from the $\bPsi\Psi$ OPE.
It is clearly important to pursue this direction further.
For example, by using the four-point function for general positions
derived in section \ref{RSSFourPointHyperGeometric}, one should be able to compute
the three-point function $\< \Psi \mJ_m \bPsi\>$. This will in turn 
give us information about the $\mJ_m \bPsi$ OPE. This is important in understanding
the nature of the operators $\mJ_m$. We expect them to include the energy-momentum tensor
and the symmetry currents. The $\mJ_m \bPsi$ OPE should tell us 
what kind of symmetries, if any, are associated with the operators $\mJ_m$.

\section{Conclusion and Discussion} \label{RSConclusion}
In this paper, we have pointed out that the Calogero model considered as
a quantum field theory in one space and one time dimension via the second quantisation
is a tractable yet nontrivial example of $z=2$ anisotropic scale invariant 
theory. 
We obtained the expression of the four-point function of the elementary fields 
for the special pairwise equal-time case \eqref{RFPairwiseEqualtime4ptFunction}.
The general four-point function can also be expressed either in terms of a double
convolution integral \eqref{RFDoubleIntegral} 
or of a generalised hypergeometric function \eqref{full4pt}.

We have obtained new insights into the $z=2$ theories, 
exploiting the exact expression of the four-point function.
We decomposed it in two different ways (the ``s-channel'' and ``t-channel'' decompositions 
studied in sections \ref{RSSSDecompositionSChannel} and
\ref{RSSSDecompositionTChannel}), corresponding to two different ways of applying the OPE.
In this way, we have verified the OPE associativity 
for the model in the case of the particular four-point function.
The ``t-channel'' decomposition is asymptotic rather than convergent.
The exponentially small corrections to the asymptotic series also 
can be interpreted using the OPE (section \ref{RSSSUChannel}).
The asymptotic nature is inherently connected to the presence of 
the terms behaving as $e^{-a x^2/t}$ in Schr\"odinger invariant theories.
This makes us suspect that the asymptotic nature of the ``t-channel'' decomposition 
and the interpretation of the exponentially small correction terms by the OPE are universal features
of Schr\"odinger invariant theories rather than being specific to our model.

Our analysis suggests the importance of the equal-time observables
(e.g., the pair-wise equal-time four-point function).
They have particularly simple forms but yet contain interesting
dynamical information of the model such as the scaling dimensions
and the OPE coefficients depending on the coupling constant.

The ``s-channel'' decomposition turns out to involve only one primary operator 
(section \ref{RSSDetailedSchannel}). Thus we have obtained an analogue of the conformal block
in isotropic theories. We may call it the ``Schr\"odinger block''. 
We have obtained a special case of the Schr\"odinger block (in two spacetime dimensions).
It is special in that it is restricted to the four-point functions of operators with $\D=\frac12$.
The scaling dimension of the operator running in the intermediate channel 
can be controlled by tuning the coupling constant $\nu$.
We hope this result may serve as a building block in the bootstrap program of $z=2$ Schr\"odinger invariant theories.

By taking a certain limit of the four-point function we have computed a three-point function 
(section \ref{RSSThreePoint}) and have found peculiar properties of correlation functions
involving certain charge-zero operators (section \ref{RSSPeculiarPropertyJ}).

The reason we are able to uncover these new features is because 
our model allows us to explicitly compute the four-point function.
Previously obtained exact results for genuine
interacting Schr\"odinger invariant field theories 
are restricted, to our knowledge, to computations of three-point functions and the associated OPE.~\footnote{
For holographic computations, see for example~\cite{volovich_correlation_2009}.
}
These include the exact computation of a three-point function~\cite{fuertes_correlation_2009}
and OPEs~\cite{goldberger_ope_2015}
for the fermion at unitarity (and for a related bosonic theory) in general space dimensions.
For the computation of the OPE in systems with contact interactions in one space and one time dimension(which are generally not scale invariant), see, for example, \cite{sekino_field-theoretical_2021} and references therein.
For a review of computations of observables related to the three-point correlation functions with Schr\"odinger symmetry in statistical models, see \cite{henkel_dynamical_2017}.

\paragraph{Four-point function in general position}
We have presented the exact expression
\eqref{nontrivialF}
of the four-point function in general position 
using a generalised hypergeometric function with
three variables. 
The three variables are quantities  invariant under the 
Schr\"odinger symmetry and thus are the analogues of the cross-ratios in the
standard CFT.

The exact expression is worth further  investigation.
Firstly, by studying a certain limit of the expression, we should get
a better understanding of the ``t-channel'' OPE 
and hence of the important charge-zero operators.

Secondly, the generalised hypergeometric function 
should obey certain connection formulae, 
analogous to those satisfied by the ordinary Gaussian hypergeometric functions. 
The connection formulae relate different expansions of the function
valid for different limits one can take in their arguments.
These different limits should correspond to the various ways
of decomposing the four-point function by the OPE.
Hence, the connection formulae should be a rather direct manifestation of the
OPE associativity.
A good example which shows the relevance of the hypergeometric functions and their connection formulae in the conformal bootstrap program
is the Liouville CFT.
A four-point function in the Liouville CFT is written directly in terms of the Gaussian hypergeometric function of the cross-ratio, and a connection formula between the hypergeometric functions indeed represents the OPE associativity~\cite{teschner_liouville_1995}.

Finally,
we have seen that Schr\"odinger invariant theories
have an intricate structure:
if looked at from a certain perspective they are described by functions analogous to 
the confluent hypergeometric function (which can be represented by an asymptotic series
when its argument goes to infinity),
and from another perspective, they are described by functions analogous to 
the hypergeometric function (which can be expanded everywhere, even including the point at infinity, and represented as a convergent series).
On the one hand, the pairwise equal-time four-point function is given in terms of a modified Bessel function, which is a special case of the confluent hypergeometric function. 
Also, the three-point function obtained as a limit of the four-point function is written by a confluent hypergeometric function.
On the other hand, if considered as a function of one of the Schr\"odinger invariant ``cross-ratios'' \eqref{crossratios}, $\tau=\frac{t_{21}t_{43}}{t_{31}t_{42}}$,
the four-point function should have features analogous to the hypergeometric function,
consistent with the SL(2,R) subgroup of the Schr\"odinger symmetry discussed in \cite{pal_unitarity_2018}.
The expression of the four-point function via a generalised hypergeometric function should 
embody this mixed feature.
Expressed as a multiple series of a certain set of combinations of the variables
valid for certain limits, the series should be of hypergeometric type.
When another set of combinations of its variables is used, the series should 
have degenerate parameters, and have properties closer to the confluent hypergeometric functions rather than the hypergeometric functions.

\paragraph{Analogy to 2D CFT and the sine-Gordon model}
The model we have considered in this paper, 
the Calogero model in the second-quantised formulation, 
has features analogous to the compactified free-boson CFT in two spacetime dimensions.
Both the Calogero model and the compactified free-boson CFT
are theories  parametrised by a single parameter (the coupling constant and
the compactification radius $R$ respectively). 
The scaling dimensions of the charged operators are dependent on that single parameter, 
e. g. the operator $\Phi$ (arising from the $\Psi\Psi$ OPE)
in the Calogero model and $e^{i \frac1R X}$
of the compactified free-boson CFT, where $X$ is the fundamental scalar field.
That the $\Psi \Psi$ OPE involves only one primary operator $\Phi$ is
reminiscent of the fact that the OPE of $e^{i \frac1R  X} e^{i \frac1R  X}$
involves only one primary operator, $e^{i \frac2R X}$.

This analogy may be more than superficial:
both the Calogero model and the compactified free-boson CFT
can be embedded into the sine-Gordon model.
As is well-known, the IR limit of the sine-Gordon model
(for a range of the coupling constant)
is described by the compactified free-boson CFT.
(See, for example, \cite{amit_renormalisation_1980} and references therein.)
On the other hand, one can first take the non-relativistic limit
of the sine-Gordon model~\cite{korepin_above-barrier_1978,zamolodchikov_factorized_1979} 
\footnote{More precisely, in order to retain the nontrivial S-matrix, 
the non-relativistic limit should be defined as a scaling limit
in which the energies of the particles and 
a parameter of the sine-Gordon theory are going to zero,
while the ratios between them are fixed.
The precise form of the scaling limit can be found in 
\cite{kapustin_non-relativistic_1994}.} 
to obtain a model of two kinds of interacting non-relativistic particles 
(the solitons and the anti-solitons of the original sine-Gordon model).
The pair potential between solitons (or anti-solitons) in this limit has the form
$\sim 1/\sinh^2 (r/r_0)$, and that between a soliton and an anti-soliton has the form
$\sim -1/\cosh^2 (r/r_0)$. By taking a further limit where the length scale of the
non-relativistic model vanishes, one finds that the solitons and anti-solitons decouple
from each other,
and the interactions among each of them are described by the Calogero model.
Thus, the Calogero model and the compactified free-boson CFT
can be realised as different limits of the sine-Gordon model.

As is well known, it is possible to compute correlation functions
of minimal model CFTs,
applying a certain projection 
to the compactified free-boson CFT. (See for example chapter 9 of \cite{di_francesco_conformal_1997}.)
In particular, correlation functions of
the critical two-dimensional Ising model can be calculated by taking the ``square root''
of the compactified free-boson CFT with a special coupling~\cite{luther_calculation_1975, bander_quantum-field-theory_1977, zuber_quantum_1977, schroer_direct_1978,schroer_orderdisorder_1978,schroer_relativistic_1978, kadanoff_correlation_1979, di_francesco_critical_1987,boyanovsky_field_1989}.
It may be possible to obtain correlation functions of various $z=2$ Schr\"odinger
invariant theories starting from the correlation functions of the Calogero model.
In particular, correlation functions of the Glauber model~\cite{glauber_time-dependent_1963},
a model describing the dynamical critical behaviour of the Ising model,
in one space and one time dimension at criticality may be computed
starting from the Calogero model.
The Glauber model in $d=1+1$ at criticality 
has the $z=2$ scale invariant behaviour. (See for example section 10.2 of \cite{cardy_scaling_1996}.)
The model is exactly solvable in the sense that its partition function
can be computed~\cite{felderhof_note_1971} via the mapping to free fermions.
This is analogous to the mapping of the two-dimensional Ising model 
to Majorana fermions~\cite{schultz_two-dimensional_1964}.
For the Ising model, one can calculate the 
correlation functions by further rewriting the Majorana fermions
as the ``square root'' of massless Dirac fermions, which in turn is equivalent 
to the compactified free-boson CFT (with a specific coupling constant) via bosonisation.
One may be able to compute general correlation functions of the Glauber model
in a similar manner using the Calogero model.
We note that the two-point functions of 
the fundamental spin operator of the Glauber model in $1+1$-dimension has been
computed~\cite{glauber_time-dependent_1963} and 
verified to have the form dictated by the Schr\"odinger symmetry
at criticality~\cite{henkel_schrodinger_1994}.
Some correlation functions related to the three-point functions were computed
and it was found that there exists an operator with dimension $\D=3$ 
(in addition to the fundamental spin operator with dimension $\D=\frac12$) \cite{godreche_response_2000,henkel_dynamical_2017}.
It is tempting to conjecture that the Calogero model with $\n=\frac32$,
in which case the dimension of $\Phi$ becomes $\D=\frac32+\n=3$, is 
relevant for the Glauber model, just like the compactified free-boson CFT
with a specific compactification radius is relevant for the two-dimensional Ising model.

Note that $\n=\frac32$ is one of the special ``degenerate'' cases of the Calogero model ($\n=-\frac12, \frac12, \frac32, \dots$)
in which the asymptotic series associated with the ``t-channel'' decomposition
truncates.
The relation of the Calogero model to the sine-Gordon model (and the system of particles
interacting with a $1/\cosh^2 r$ pair potential) may shed light on these special points and
the spectrum of zero-charge operators. 
As is well-known, in the sine-Gordon model, a soliton and an anti-soliton can form a bound state.
The number of bound states takes the value $n=0,1,2,\dots$,
depending on the parameter of the sine-Gordon theory. 
At the special values of the parameter where the number of bound states changes 
discontinuously, the reflection coefficients between a soliton and an anti-soliton vanish.
These special values of the parameters are reminiscent of 
the special cases, $\n=-\frac12, \frac12, \frac32,\dots$, of the Calogero model where 
the asymptotic series truncates. We speculate that 
at these special points the ``multiplicity'' of the zero-charge operators also change 
discontinuously.

\paragraph{Generalisations}
We computed the four-point function by reducing it to the two-particle problem.
The integrability of the Calogero model means that 
one has a certain analytic control over the three- (or more) particle sector.
Exploiting the integrability, therefore, it should be possible to 
calculate six-point functions of the fundamental fields 
(more precisely the correlation functions with three $\Psi$'s and three $\bPsi$'s),
and to study the $\Psi \Phi$ OPE. 

One can introduce a three-body interaction to 
the Calogero model without destroying the integrability in the three-particle problem~\cite{wolfes_one-dimensional_1974,calogero_exact_1974}.
Studying such a deformation would be interesting.
The deformation will not affect the physics of the two-particle sector, and hence the results of our paper. However, the six-point functions of the fundamental fields and hence the $\Psi \Phi$ OPE will be deformed. 

Another interesting variant of the Calogero model is the so-called B$_N$-type Calogero model.
See, for reviews, \cite{olshanetsky_classical_1981,olshanetsky_quantum_1983}.
The model can be considered as the Calogero model put on a semi-infinite line with an appropriate boundary condition, which preserves the integrability of the model.
We expect that the B$_N$-type Calogero model (around the true vacuum) 
will exhibit a $z=2$ anisotropic surface critical behaviour 
and provide a nontrivial yet tractable example of the $z=2$ analogue of a CFT with a boundary.

The integrability of the Calogero model allows one to compute the correlation functions
around the finite-density vacuum. (See, for example, \cite{astrakharchik_off-diagonal_2006} and references 
therein.) 
The finite-density vacuum breaks the $z=2$ scale invariance spontaneously.
It would be interesting to study the finite-density correlation functions
from the point of view of the broken $z=2$ scale invariance and Schr\"odinger invariance.
(For a review of spontaneous breaking of
the Schr\"odinger symmetry, see \cite{semenoff_dilaton_2018}.) 
The IR limit of the Calogero model at finite-density 
is described by a $c=1$ CFT~\cite{kawakami_finite-size_1991,kawakami_application_1991,iso_long_1994, iso_collective_1995,iso_anyon_1995,caracciolo_w_1_1995}. 
Thus the finite-density correlation functions of the Calogero model 
should interpolate between the $z=2$ scale invariant correlation functions studied in this paper in the UV limit and the $z=1$, $c=1$ CFT in the IR limit.

The Calogero model is inherently related to a system of anyons, which is a $z=2$ Schr\"odinger invariant model in one time and two space dimensions. (See, for example, \cite{nishida_nonrelativistic_2007} and references therein).
In particular, the Calogero model is equivalent to a
system of anyons restricted to the lowest Landau levels~\cite{leinaas_intermediate_1988, polychronakos_exact_1991,dunne_exact_1992, hansson_dimensional_1992, brink_calogero_1993,azuma_explicit_1994,iso_long_1994,polychronakos_quantum_2001, ouvry_mapping_2018}.
It would be interesting to study the implications of our exact four-point function 
for the system of anyons. 

In this paper we focused on the case with one space dimension.
However, the Schr\"odinger symmetry exists for any number of space dimensions 
when non-relativistic particles are interacting with a pair potential of the form $1/r^2$~\cite{burdet_many-body_1972}.
We do not expect these models in general to be integrable in the conventional sense.
However, since our analysis of the four-point function is associated
only with the two-particle sector of the model, 
the computation of the four-point function in higher space dimensions appears feasible.
It would be interesting to consider the properties of the OPE, including the
OPE associativity, for this higher dimensional system with the Schr\"odinger symmetry.

Finally, finding an anisotropic scale invariant quantum field theory model
with $z\neq 2$ but with exactly computable OPEs is an interesting open problem.

\vspace{1cm}

We hope that our analysis provides a starting point 
of better understanding of fixed points of the renormalisation group for anisotropic theories,
and of uncovering a rich structure of solvable models with $z=2$ scale invariance.

\vsp{0.7}
\ndt
{\bf Acknowledgements}

\vsp{0.3}
\ndt
We would like to thank 
Sinya~Aoki,
Yasuyuki~Hatsuda,
Shinobu~Hikami,
Yasuaki~Hikida,
Masazumi~Honda,
Masaru~Hongo,
Stefano~Kovacs, 
Yoichi~Kazama,
Shota~Komatsu,
Wenliang~Li,
Jonathan~Miller,
Takeshi~Morita,
Keita~Nii,
Yoshitaka~Okuyama,
Norisuke~Sakai,
Yuta~Sekino,
Shigeki~Sugimoto,
Kenta~Suzuki,
Kotaro~Tamaoka,
Tadashi~Takayanagi,
Yuya~Tanizaki,
Seiji~Terashima,
Tomonori~Ugajin
for encouragement, discussions and useful comments.

This work was supported by JSPS KAKENHI (Grant Nos. KAKEN-19H01896, -20K03955, and -20K03796).

%%%%%%%%%%%%%%%%%%%%%%%%%%%%%%%%%%%%%%%
%%%%%%%%%%%%%%%%%%%%%%%%%%%%%%%%%%%%%%%
%%%%%%%%%%%%%%%%%%%%%%%%%%%%%%%%%%%%%%%

\appendix
\section{Schr\"odinger symmetry}
\label{RSASchrodingerSymmetry}
We list here all nonzero commutators in 
the algebra associated with the Schr\"odinger symmetry.
The members of the algebra are the time translation, 
the space translations, the angular momenta, a U(1) charge,
the dilation, and the spacelike and timelike ``special conformal transformations'':
$H, P_i, D, M_{ij}=-M_{ji}, N, K_i, C$.
Here we used the label $i=1, \cdots, d$ where $d$ is the number of space dimensions.
For $d=1$, generators $M_{ij}$ do not exist.
The scaling dimensions of these generators are reflected in,
\begin{align}
[D, H] =& 2i H,
\\
[D, P_i]=& i P_i,
\\
[D, K_i]=& -i K_i,
\\
[D, C] =& -2i C.
\end{align}
The nonzero commutators involving $M_{ij}$ 
show the transformation properties of the generators under
the spatial rotation,
\begin{align}
[M_{ij}, P_k]=& i\lef(\d_{ik}P_j -\d_{jk}P_i\ri),
\\
[M_{ij}, K_k]=& i\lef(\d_{ik}K_j -\d_{jk}K_i\ri),
\\
[M_{ij}, M_{kl}]=&
i\lef(
 \d_{ik}M_{jl} -\d_{jk}M_{il}
-\d_{il}M_{jk} +\d_{jl}M_{ik}
\ri).
\end{align}
The remaining non-vanishing commutation relations are
\begin{align}
[H, C]=& -i D,
\\
[C, P_i]=& i K_i,
\\
[H, K_i]=& -i P_i,
\\
[K_i, P_j]=& i N \d_{ij}.
\end{align}

\section{Propagator in $1/r^2$ potential}
\label{RSAPropagatorInverseSquarePotential}
In this appendix, we compute the propagator for the Hamiltonian,
\eqref{RFHamiltonianInverseSquarePotential}
\begin{align}
H_{\text{rel}}=-\der_r^2+\frac{\l(\l-1)}{r^2},
\end{align}
corresponding to a particle in an external potential $1/r^2$ where $r>0$.
The boundary condition for $r\to 0$ is $\Psi\sim r^\l$.

The Schr\"odinger equation for an energy eigenstate
$\Psi(r)$  with energy $E=k^2$ ($k>0$) is
\begin{align}
-\Psi'' +\frac{\l(\l-1)}{r^2}\Psi=k^2\Psi.
\end{align}
For $r\to +\infty$,
$\Psi(r)$ asymptotes to a linear combination of $e^{\pm ikr}$.
 
A simple redefinition 
\begin{align}
\Psi=&\sqrt{r}w,
\\
z=&kr,
\end{align}
leads to 
\begin{align}
z^2\frac{d^2w}{dz^2}+z\frac{dw}{dz}+\lef(z^2-\lef(\l-\frac12\ri)^2\ri) w=0,
\end{align}
which is Bessel's equation~\cite[(10.2.1)]{olver_nist_nodate}
with $\n=\l-\frac12$.
Hence the solution to the Schr\"odinger equation
with the desired behaviour at $r\to0$, $\Psi\sim r^\l$, is
\begin{align}
\Psi(r)=N_k \sqrt{r} J_{\n}(kr).
\end{align}
We shall use the bra-ket notation,
\begin{align}
\langle r | k \rangle = N_k \sqrt{r} J_{\n}(kr),
\end{align}
where $k>0$.

The normalisation constant $N_k$ is fixed by 
the requirement that 
$|k\rangle$'s should give a complete orthonormal basis
(with the correct boundary condition)
\begin{align}
\<k'|k\> =\d(k'-k).
\end{align}
Using an integral formula~\cite[(10.22.67)]{olver_nist_nodate}
\begin{align}
\int_{0}^{\infty}
t \exp(-p^{2}t^{2}) J_{\nu}\lef(at\ri) J_{\nu}\lef(bt\ri)\mathrm{d}t
=
\frac{1}{2p^{2}}
\exp\lef(-\frac{a^{2}+b^{2}}{4p^{2}}\ri)
I_{\nu}\lef(\frac{ab}{2p^{2}}\ri),
\label{RFJJIntegral}
\end{align}
we see that
\begin{align}
\langle k' | k \rangle = 
\overline{N_{k'}}
N_k
\int_0^\infty
\sqrt{r} J_{\n}(k'r)
\sqrt{r} J_{\n}(kr)
dr
\end{align}
vanishes for $k\neq k'$ and is IR divergent for $k=k'$.
A natural IR cut-off can be introduced:
\begin{align}
&
\overline{N_{k'}}
N_k
\int_0^\infty
\sqrt{r} J_{\n}(k'r)
\sqrt{r} J_{\n}(kr)
e^{-\a r^2}
dr,
\end{align}
where we are interested in the limit $\a\rightarrow 0$ in the end.
This equals 
\begin{align}
\begin{split}
&
\overline{N_{k'}} N_k \times
\frac{1}{2\a}
e^{-\frac{k^2+k'^2}{4\a}}
\times
I_\n\lef(\frac{kk'}{2\a}\ri)
\\
\approx
&
\overline{N_{k'}} N_k \times
\frac{1}{2\a}
e^{-\frac{k^2+k'^2}{4\a}}
\times
e^{\frac{kk'}{2\a}}
\frac1{\sqrt{2\pi\frac{kk'}{2\a}}}
\\
=
&
\overline{N_{k'}} N_k \times
\frac{1}{\sqrt{2\pi kk'}} \frac1{\sqrt{2\a}}
e^{-\frac{(k-k')^2}{4\a}}
\\
\approx 
&
|N_k|^2 \times
\frac{1}{k} 
\d(k-k'),
\end{split}
\end{align}
using \eqref{RFJJIntegral} and \eqref{RFBesselAsymptoticExpansion}.
Thus we obtain
\begin{align}
N_k=\sqrt{k}.
\end{align}

Finally,  the propagator in $1/r^2$ potential is,
\begin{align}
\begin{split}
\langle r'| e^{-H_{\text{rel}}t} |r \rangle
=&
\int_0^{+\infty} 
\langle r'| k\rangle e^{-k^2t} \langle k |r \rangle
dk
\\
=&
\int_0^{+\infty} 
 k e^{-k^2 t}
\sqrt{r'}J_\n(kr') \sqrt{r}J_\n(kr) dk
\\
=&
\sqrt{rr'} \frac{1}{2t} e^{-\frac{r^2+r'^2}{4t}} 
I_\n\lef(\frac{rr'}{2t}\ri),
\end{split}
\end{align}
using \eqref{RFJJIntegral}.

If $t$ is very small, the propagator reduces  to
\begin{align}
\sqrt{\frac{1}{4\pi t}} 
e^{-\frac{\lef(r'-r\ri)^2}{4t}},
\end{align}
which coincides with the free particle propagator
with the reduced mass $m=\frac12$, as it should be.

\section{Details of the computation of the three-point function $\<\Psi\Psi\bPhi\>$}
\label{RSAThreePoint}
In this appendix, we supply the details of the computation of the
three-point function $\<\Psi\Psi\bPhi\>$ in general position.
We also give the comparison to the generic form of the three-point 
function derived by Henkel~\cite{henkel_schrodinger_1994, henkel_schrodinger_2003}
and elaborate on the properties of the Schr\"odinger invariant quantity
which we denote $w$.

\subsection{Evaluation of the integral representation}
We begin with the integral representation of the three-point function \eqref{RFThreePointIntegralRep},
\begin{align}
\begin{split}
&\langle
\Psi(t+t', x_4)
\Psi(t, x_3)
\bPhi\lef(0, 0\ri)
\rangle
\\
=&
\frac1{2^{\n+1}\pi^{\frac34}\sqrt{\G(\n+1)}}
\frac1{\sqrt{t'}}
\frac1{t^{\frac32+\n}}
\int
e^{-\frac{
        x_{4'3}^2+\lef(x_{4'}+x_{3}\ri)^2}
        {4t}}
\times
|x_{4'3}|^{\n+\frac12}
\times
e^{-\frac{1}{2}\frac{x_{44'}^2}{t'}}
dx_{4'}.
\end{split}
\nonumber
\end{align}
We evaluate this integral as follows.
First, we extract the dependence of the
integrand on $x=x_{34'}$, 
\begin{align}
\begin{split}
&\langle
\Psi(t+t', x_4)
\Psi(t, x_3)
\bPhi\lef(0, 0\ri)
\rangle
\\
=&
\frac1{2^{\n+1}\pi^{\frac34}\sqrt{\G(\n+1)}}
\frac1{\sqrt{t'}}
\frac1{t^{\frac32+\n}}
e^{-\frac{x_{3}^2}{t}}
e^{-\frac{x_{43}^2}{2t'}}
\int_{-\infty}^\infty
|x|^{\n+\frac12}
e^{-\frac12\lef(\frac1t+\frac1{t'}\ri)x^2}
e^{\lef(\frac{x_{3}}{t}
-\frac{x_{43}}{t'}\ri)x}
dx.
\end{split}
\end{align}
Separating the contribution from $x>0$ and $x<0$, we get,
\begin{align}
\begin{split}
&\langle
\Psi(t+t', x_4)
\Psi(t, x_3)
\bPhi\lef(0, 0\ri)
\rangle
\\
=&
\frac1{2^{\n+1}\pi^{\frac34}\sqrt{\G(\n+1)}}
\sqrt{\frac{{t'}^{\n+\frac12}}{(t+t')^{\n+\frac32}t^{\n+\frac32}}}
e^{-\frac{x_{3}^2}{t}
-\frac{x_{43}^2}{2t'}}
\\
&\times
\Bigg(
\quad \int_{0}^\infty
e^{-\frac 12 y^2+ w y}
y^{\n+\frac12}
dy
+
\int_{0}^\infty
e^{-\frac 12 y^2- w y}
y^{\n+\frac12}
dy
\Bigg),
\label{RFIntegralFormThreePointBeforeRelabelling}
\end{split}
\end{align}
where 
\begin{align}
y=&\sqrt{\frac{t+t'}{tt'}}x,
\\
w=&
\sqrt{\frac{tt'}{t+t'}}
\lef(\frac{x_{3}}{t}
-\frac{x_{43}}{t'}\ri).
\end{align}

\subsection{Relabelling and properties of $w$}
We will write the integrals in the last line of 
(\ref{RFIntegralFormThreePointBeforeRelabelling}) using 
the parabolic cylinder functions \cite[section 12]{olver_nist_nodate},
which in turn can be expressed using 
the confluent hypergeometric functions.
Before doing so, we will relabel 
the $x, t$ coordinates and check 
whether the result obtained is consistent with
the general form~\cite{henkel_schrodinger_1994,henkel_schrodinger_2003}
of three-point functions dictated by the Schr\"odinger symmetry.
To do this we slightly modify our notation to bring 
the three-point function 
$\langle
\Psi(t+t', x_4)
\Psi(t, x_3)
\bPhi\lef(0, 0\ri)
\rangle
$
into the form
$\langle
\Psi(t_3, x_3)
\Psi(t_2, x_2)
\bPhi\lef(t_1, x_1\ri)
\rangle
$. Thus, we relabel as, $x_{4} \rightarrow x_{31},
x_{3} \rightarrow x_{21},
t' \to t_{32},
t \to t_{21}$. 
The result is, 
\begin{align}
\begin{split}
&\langle
\Psi(t_3, x_3)
\Psi(t_2, x_2)
\bPhi\lef(t_1, x_1\ri)
\rangle
\\
=&
\frac1{2^{\n+1}\pi^{\frac34}\sqrt{\G(\n+1)}}
\sqrt{\frac{{t_{32}}^{\n+\frac12}}{t_{31}^{\n+\frac32}t_{21}^{\n+\frac32}}}
e^{-\frac{x_{21}^2}{t_{21}}
-\frac{x_{32}^2}{2t_{32}}}
\\
&\times
\Bigg(
\quad \int_{0}^\infty
e^{-\frac 12 y^2+ w y}
y^{\n+\frac12}
dy
+
\int_{0}^\infty
e^{-\frac 12 y^2- w y}
y^{\n+\frac12}
dy
\Bigg),
\end{split}
\label{RFThreePointRelabelledIntegralAsIs}
\end{align}
with
\begin{align}
w
=
\lef(\frac{x_{21}}{t_{21}}
-\frac{x_{32}}{t_{32}}\ri)
\sqrt{\frac{t_{21}t_{32}}{t_{31}}}.
\end{align}
The integral converges since $\n\ge -\frac12$, (\ref{RFNuGEQOneHalf}).
We recall that we chose  $t_3>t_2>t_1$ and hence $w\in \R$.
It is worthwhile to discuss some properties of the 
``cross-ratio'' $w$ which is invariant under the Schr\"odinger symmetry.
For $t_3>t_2>t_1$ we have $w^2\geq 0$, and $w^2=0$ holds if and only if
the spacetime points $1,2,3$ are aligned on a straight line. 
It is easy to show the identity
\begin{align}
x_{12} t_{23} - x_{23} t_{12}=x_{23} t_{31} - x_{31} t_{23}
=
x_{31} t_{12} - x_{12} t_{31},
\label{RFIdentityUnderlyingW}
\end{align}
by direct computation. It is amusing to note that 
the quantity in \eqref{RFIdentityUnderlyingW} 
is twice the ``area'' of a triangle spanned by the spacetime points $1,2,3$
up to sign.
It is completely anti-symmetric  in
the labels $1,2,3$.
It follows that 
\begin{align}
\begin{split}
&\frac{ \lef(x_{12} t_{23} - x_{23} t_{12}\ri)^2}{t_{12}t_{23}t_{31}}
=\frac{ \lef(x_{23} t_{31} - x_{31} t_{23}\ri)^2}{t_{12}t_{23}t_{31}}
=\frac{ \lef(x_{31} t_{12} - x_{12} t_{31}\ri)^2}{t_{12}t_{23}t_{31}}
\\
=&
-\frac{x_{12}^2}{t_{12}}
-\frac{x_{23}^2}{t_{23}}
-\frac{x_{31}^2}{t_{31}}
\\
=&
\frac{x_{32}^2}{t_{32}} 
- \frac{x_{31}^2}{t_{31}}
+ \frac{x_{21}^2}{t_{21}} 
\\
=&w^2
\end{split}
\label{RFPropertyW}
\end{align}
We note that $w^2$ again is completely anti-symmetric in 
the labels $1,2,3$.
In section \ref{RSSFourPointHyperGeometric}, we used the notation $v=\frac12 w^2$.

\subsection{Comparison to the general form of three-point functions dictated 
by Schr\"odinger symmetry}

The standard form of the three-point function in a Schr\"odinger invariant theory is,
\footnote{
Our convention differs slightly from that of 
\cite{henkel_schrodinger_2003} in that 
we adopt the Euclidean statistical field theory convention 
rather than the Minkowski one. We assume $t_3>t_2>t_1$.
}
\begin{align}
\begin{split}
&\<\mO_3 (t_3, x_3)\mO_2(t_2, x_2) \mO_1(t_1, x_1)\>
\\
=&
t_{31}^{-\frac{\D_3+\D_1-\D_2}2}
t_{21}^{-\frac{\D_2+\D_1-\D_3}2}
t_{32}^{-\frac{\D_3+\D_2-\D_1}2}
e^{-\frac{|N_2|}{2}\frac{x_{21}^2}{t_{21}}
-\frac{|N_3|}{2}\frac{x_{31}^2}{t_{31}}}
F_{123}(w^2),
\end{split}
\end{align}
where $F_{123}$ is an arbitrary scaling function 
which generically is not fixed by the Schr\"odinger symmetry alone.
The quantum numbers
$N_i (i=1,2,3)$ 
are the charges of the operators $\mO_i$
associated with a U(1)-symmetry present for
theory with the Schr\"odinger symmetry.
They satisfy $N_1+N_2+N_3=0$, 
and $N_1 >0, N_2<0, N_3<0$.
For the three-point function studied here, we have $\mO_3=\Psi, \mO_2=\Psi, \mO_1 = \bPhi$
and $\D_3=\frac12, \D_2=\frac12, \D_1=\frac32+\n$,
$N_3=-1, N_2=-1, N_1=2$.

To compare with the standard form, it is convenient to rewrite 
(\ref{RFThreePointRelabelledIntegralAsIs})
using (\ref{RFPropertyW}), 
\begin{align}
\begin{split}
&\langle
\Psi(t_3, x_3)
\Psi(t_2, x_2)
\bPhi\lef(t_1, x_1\ri)
\rangle
\\
=&
\frac1{2^{\n+1}\pi^{\frac34}\sqrt{\G(\n+1)}}
\sqrt{\frac{{t_{32}}^{\n+\frac12}}{t_{31}^{\n+\frac32}t_{21}^{\n+\frac32}}}
e^{-\frac{x_{21}^2}{2t_{21}}-\frac{x_{31}^2}{2t_{31}}}
\\
&\times
e^{-\frac12 w^2}
\Bigg(
\quad \int_{0}^\infty
e^{-\frac 12 y^2+ w y}
y^{\n+\frac12}
dy
+
\int_{0}^\infty
e^{-\frac 12 y^2- w y}
y^{\n+\frac12}
dy
\Bigg).
\end{split}
\label{RFThreePointRewritten}
\end{align}
The last line combined with the numerical prefactor is the scaling function $F_{123}$.

\subsection{Three-point function in terms of  
parabolic cylinder functions and confluent hypergeometric functions
}
The integral appearing in (\ref{RFThreePointRewritten})
can be written~\cite[(12.5.1)]{olver_nist_nodate}
\begin{align}
\int_0^\infty e^{-\frac{y^2}{2}+ w y} y^{\n+\frac12} dy
=
\G\lef(\n+\frac32\ri)
e^{\frac14 w^2} 
U(\n+1, -w),
\end{align}
using the parabolic cylinder function $U(a,z)$.
Thus we obtain
\begin{align}
\begin{split}
&\langle
\Psi(t_3, x_3)
\Psi(t_2, x_2)
\bPhi\lef(t_1, x_1\ri)
\rangle
\\
=&
\frac{\G\lef(\n+\frac32\ri)}{2^{\n+1}\pi^{\frac34}\sqrt{\G(\n+1)}}
\sqrt{\frac{{t_{32}}^{\n+\frac12}}{t_{31}^{\n+\frac32}t_{21}^{\n+\frac32}}}
e^{-\frac{x_{21}^2}{2t_{21}}-\frac{x_{31}^2}{2t_{31}}}
\\
&\times
e^{-\frac14 w^2}
\lef(
U(\n+1, -w)
+
U(\n+1, w)
\ri).
\end{split}
\end{align}

We observe that the last line is even 
in $w$. 
We rewrite the above formula in terms of a parabolic hyperbolic function,
$u_1$ in the notation of \cite{olver_nist_nodate}, which is even in $w$.
We will then rewrite the formula in terms
of the confluent hypergeometric functions.
This will be useful to check against the result by Golkar and Son~\cite{golkar_operator_2014},
and also to study simplifying limits, namely the free boson limit
($\n=-\frac12$, appendix \ref{RSASFreeThreePoint}), 
and the limit $t_{32}\to 0$, which we already computed in section \ref{RSSSThreePointEqualTime}. 

From (12.4.1) and (12.2.6) of \cite{olver_nist_nodate},
\begin{align}
U\lef(a,z\ri)=&U\lef(a,0\ri)u_{1}(a,z)+U'\lef(a,0\ri)u_{2}(a,z),
\\
U\lef(a,0\ri)=&
\frac{\sqrt{\pi}}{2^{\frac{1}{2}a+\frac{1}{4}}
\Gamma\lef(\frac{3}{4}+\frac{1}{2}a\ri)},
\end{align}
where $u_1$ and $u_2$ are respectively even and odd in $z$,
we obtain 
\begin{align}
U\lef(a,z\ri)
+
U\lef(a,-z\ri)
= 
\frac{\sqrt{\pi}}{2^{\frac{1}{2}a-\frac{3}{4}}
\Gamma\lef(\frac{3}{4}+\frac{1}{2}a\ri)}
u_{1}(a,z),
\end{align}
and hence
\begin{align}
\begin{split}
&\langle
\Psi(t_3, x_3)
\Psi(t_2, x_2)
\bPhi\lef(t_1, x_1\ri)
\rangle
\\
=&
\frac{\G\lef(\n+\frac32\ri)}{2^{\frac32 \n+\frac34} \pi^{\frac14}\sqrt{\G(\n+1)}\Gamma\lef(\frac{\nu}2+ \frac{5}{4}\ri)}
\sqrt{\frac{{t_{32}}^{\n+\frac12}}{t_{31}^{\n+\frac32}t_{21}^{\n+\frac32}}}
e^{-\frac{x_{21}^2}{2t_{21}}}
e^{-\frac{x_{31}^2}{2t_{31}}}
\\
&\times
e^{-\frac14 w{}^2}
u_{1}(\nu+1,w).
\end{split}
\end{align}

One can rewrite the result 
in terms of the confluent hypergeometric function $M(a,b,x)$ using
\cite[(12.7.12)]{olver_nist_nodate}
\begin{align}
u_{1}(a,z)
=
e^{-\tfrac{1}{4}z^{2}}
M\lef(\tfrac{1}{2}a+\tfrac{1}{4},\tfrac{1}{2},\tfrac{1}{2}z^{2}\ri)
=
e^{\tfrac{1}{4}z^{2}}
M\lef(-\tfrac{1}{2}a+\tfrac{1}{4},\tfrac{1}{2},-\tfrac{1}{2}z^{2}\ri).
\end{align}
This leads, finally, to
\begin{align}
\begin{split}
&\langle
\Psi(t_3, x_3)
\Psi(t_2, x_2)
\bPhi\lef(t_1, x_1\ri)
\rangle
\\
=&
\frac{\G\lef(\n+\frac32\ri)}{2^{\frac32 \n+\frac34} \pi^{\frac14}\sqrt{\G(\n+1)}\Gamma\lef(\frac{\nu}2+ \frac{5}{4}\ri)}
\sqrt{\frac{{t_{32}}^{\n+\frac12}}{t_{31}^{\n+\frac32}t_{21}^{\n+\frac32}}}
e^{-\frac{x_{21}^2}{2t_{21}}}
e^{-\frac{x_{31}^2}{2t_{31}}}
\\
&\times
M\lef(-\frac\nu 2 -\frac14,\frac{1}{2},-\frac{1}{2}w^{2}\ri),
\end{split}
\end{align}
and,
\begin{align}
\begin{split}
&\langle
\Psi(t_3, x_3)
\Psi(t_2, x_2)
\bPhi\lef(t_1, x_1\ri)
\rangle
\\
=&
\frac{\G\lef(\n+\frac32\ri)}{2^{\frac32 \n+\frac34} \pi^{\frac14}\sqrt{\G(\n+1)}\Gamma\lef(\frac{\nu}2+ \frac{5}{4}\ri)}
\sqrt{\frac{{t_{32}}^{\n+\frac12}}{t_{31}^{\n+\frac32}t_{21}^{\n+\frac32}}}
e^{-\frac{x_{21}^2}{2t_{21}}}
e^{-\frac{x_{31}^2}{2t_{31}}}
\\&
\times
e^{-\frac{w^2}{2}}
M\lef(\frac\nu 2 +\frac34,\frac{1}{2},\frac{1}{2}w^{2}\ri).
\end{split}
\label{RFThreePointConfluentHypergeometricWithExpAppendix}
\end{align}
The latter is our final expression for the
three-point function,
quoted in the main text as \eqref{RFThreePointConfluentHypergeometricWithExp}. 

\subsection{$t_{32}\to 0$ limit}
\label{RSSACheckEqualTime}
As a consistency check, 
we consider the limit 
$t_{32}\rightarrow 0$ to compare with the result of section \ref{RSSSThreePointEqualTime}.
In this limit, we have $w^2\approx \frac{x_{32}^2}{t_{32}}\to+\infty$.
Applying the asymptotic formula of the
confluent hypergeometric function~\cite[(13.7.1) and (13.2.4)]{olver_nist_nodate}
~\footnote{The asymptotic formula 
is invalid when $a$ and $b$ are non-positive integers.
These exceptional cases are automatically avoided in 
our use of the formula. See (\ref{RFThreePointConfluentHypergeometricWithExp}) 
or equivalently \eqref{RFThreePointConfluentHypergeometricWithExpAppendix}.
}
\begin{align}
M\lef(a,b,x\ri)\sim
\frac{e^{x}x^{a-b}}{
\Gamma\lef(a\ri)
\Gamma\lef(b\ri)}
\sum_{s=0}^{\infty}\frac{{\lef(1-a\ri)_{s}}{\lef(b-a\ri)_{s}}}{s!}x^{-s},
\quad (x\to+\infty, x\in \R),
\label{RFConfluentHypergeometricAsymptotics}
\end{align}
to the three-point function (\ref{RFThreePointConfluentHypergeometricWithExp})
(equivalently \eqref{RFThreePointConfluentHypergeometricWithExpAppendix}),
we obtain
\begin{align}
\begin{split}
&\langle
\Psi(t, x_3)
\Psi(t, x_2)
\bPhi\lef(0, x_1\ri)
\rangle
\\
=&
\frac{\G\lef(\n+\frac32\ri)}
{2^{\frac32 \n+\frac14} (4\pi)^{\frac14}\sqrt{\G(\n+1)}\Gamma\lef(\frac{\nu}2+ \frac{5}{4}\ri)}
t^{-\lef(\n+\frac32\ri)}
e^{-\frac{x_{21}^2}{2t}}
e^{-\frac{x_{31}^2}{2t}}
\\ &
\times
\lef(\frac12 x_{32}^2\ri)^{\frac{\n}{2}+\frac14}\frac{\G\lef(\frac12\ri)}
{\Gamma\lef(\frac{\nu}2+\frac34 \ri)},
\end{split}
\end{align}
where we have written $t=t_{31}= t_{21}$, and put $t_1=0$.
This indeed agrees with (\ref{RFThreepointSemiEqualTime}) which we obtained
in section \ref{RSSSThreePointEqualTime} directly from 
the pairwise equal-time four-point function (\ref{RFPairwiseEqualtime4ptFunction})
since we have 
\begin{align}
\frac{\sqrt{\pi}\G\lef(\n+\frac32\ri)}
{2^{\n+\frac12} 
\G\lef(\frac\n2+\frac54\ri)
\G\lef(\frac\n2+\frac34\ri)}
=1,
\end{align}
which follows from the duplication formula~\cite[(5.5.5)]{olver_nist_nodate}
\begin{align}
\Gamma\lef(2z\ri)=\pi^{-1/2}2^{2z-1}\Gamma\lef(z\ri)\Gamma\lef(z+\tfrac{1}{2}\ri).
\end{align}

\section{Golkar and Son's analysis in Euclidean signature}
\label{RSAGolkarSonEuclid}
Golkar and Son showed~\cite{golkar_operator_2014} that
the form of the scaling function appearing in the three-point function 
in a Schr\"odinger invariant theory
is severely restricted when the scaling dimension of 
one of the operators equals the special value, $\D=\frac d 2$,
where $d$ is the number of spacelike dimensions.
The scaling function satisfies (except for a simple 
prefactor) the confluent hypergeometric equation.
Their analysis was done in Minkowski signature.
Since how the analysis takes over to Euclidean signature
is not entirely trivial, in this appendix we give 
the Euclidean version of the analysis of Golkar and Son.~\footnote{We note that the notation 
used in \cite{golkar_operator_2014} is slightly unusual. 
They call what is usually called (up to constant 
multiplication) the confluent hypergeometric function
$M(a,b,x)$ (in the notation of \cite{olver_nist_nodate})
as ``a generalised Laguerre polynomial'' $L_n{}^\a(x)$,
with $n=-a$, $b=\a+1$. 
The function is not a polynomial unless $n=-a$ is a non-negative integer.
As shown in \cite{golkar_operator_2014} 
the parameter $a$ is related to the scaling dimensions of the operators (see (\ref{RFaIntermsOfDelta})),
and is not an integer, in general.
}
In this appendix $d$ is arbitrary and we write $x=(t,\bm x)$. 

The solution to the differential equation 
contains two arbitrary parameters. In \cite{golkar_operator_2014}
it was advocated that one of the parameters vanishes 
due to the regularity conditions of the OPE, acting as the boundary conditions 
of the differential equation.
We will also give below a careful discussion of the regularity conditions,
in particular, for the case $d=1$.
We will see that for that case, the regularity conditions are weaker 
and do not imply the vanishing of the parameter.

\subsection{Preliminaries}
The operators in the Heisenberg picture are
\begin{align}
\mO(t, \bm x )
=
U^{-1}
\mO(0, \bm 0)
U,
\label{RFHeisenbergOp}
\end{align}
where 
\begin{align}
U=e^{-Ht+i\bm P \cdot \bm x}.
\end{align}
It is straightforward to verify
\begin{align}
U K_i U^{-1} =& K_i + i P_i t + N x_i,
\label{RFUKUinv}
\\
UCU^{-1} =&C+ i D t + K_i x_i - H t^2 +i P_i x^i t + \frac12 N \bm x^2,
\label{RFUCUinv}
\end{align}
using the elementary identity 
\begin{align}
e^A B e^{-A} = B+ [A,B] + \frac1{2!} [A,[A,B]]+\frac1{3!}[A,[A,[A,B]]]+\cdots,
\end{align}
and the commutation relations given in appendix \ref{RSASchrodingerSymmetry}.

A primary operator $\mO$ in a Schr\"odinger invariant theory is defined by
the conditions
\begin{align}
[K_i, \mO(0)]=&0,
\\
[C, \mO(0)]=&0.
\end{align}
It follows that 
\begin{align}
[K_i, \mO(t, \bm x)] =&-t \der_i \mO(x) + N_\mO x_i \mO(x),
\label{RFCommutatorKmO}
\\
[C, \mO(t, \bm x)]
=&
-\D_\mO t \mO - t^2 \der_t \mO -t x^i \der_i \mO + 
\frac12 N_\mO \bm x^2 \mO,
\label{RFCommutatorCmO}
\end{align}
where we used 
\begin{align}
\der_i \mO(t, \bm x)
=&
[\mO(t, \bm x), i P_i] ,
\\
\der_t \mO =& [H,\mO],
\\
[N, \mO]=& N_\mO \mO,
\\
[D, \mO(0)] =& i \D_\mO \mO(0).
\end{align}
Here $N_\mO$ and $\D_\mO$ are the U(1) charge and 
the scaling dimension of the operator $\mO$.

\subsection{OPE coefficients}

We consider general constraints on the OPE coefficients
imposed by the Schr\"odinger symmetry.
We consider the OPE  $\mO_2 \mO_1$ and 
focus on the part proportional to $\mO_3$,
where $\mO_i$ are scalar primary operators with nonzero U(1) charges.
We consider the special case, $\D_3=\frac d2$.

We write down explicitly the first few descendants of $\mO_3$,
\begin{align}
\mO_2(x) \mO_1(0) =
\lef.
\lef(
\lef(
C_0(x) + C_1^i(x)\der_i +C_2(x)\der_t
+C_3^{ij}(x)\der_i\der_j+\cdots\ri)\mO_3\ri)\ri|_{x=0}
+\cdots,
\end{align}
where $t>0$ is assumed.

By taking the commutators of $K_i, C$ with the LHS and RHS,
we obtain 
\begin{align}
-t \der_i C_0+ N_{2} x_i C_0 
=& N_3  C_1^i, \label{RFOPEKCCommutator1}
\\
-t\der_{i} C_1^j+ N_2 x_i C_1^j 
=&
- C_2 \d_{ij} +2 N_3 C_3^{ij},
\\
-\D_2 t C_0 - t^2 \der_t C_0 - t x^i \der_i C_0
+ \frac{N_2}{2} \bm x ^2 C_0
=&
-\D_3 C_2 + C_3^{jj} N_3, \label{RFOPEKCCommutator3}
\end{align}
where we write $N_i\equiv N_{\mO_i}$, $(i=1,2,3)$.
Generically, these equations express differential operators
acting on $C_0$ to give $C_1, C_2, \cdots$.
For the special case $\D_3=\frac d2$, \eqref{RFOPEKCCommutator1}-\eqref{RFOPEKCCommutator3}
imply a differential equation on the coefficient $C_0$:
\begin{align}
&
t^2\der_{i}^2 C_0
+ 2 N_3 t^2 \der_t C_0 
+ 2 N_1 t x^i \der_i C_0
+ \lef(2 N_3 \D_2 - N_{2} d\ri) t C_0 
-N_2N_1 \bm x^2 C_0
=0. \label{RFCzeroDiffEq}
\end{align}
The scale and SO($d$) invariance require $C_0$ to have the form
\begin{align}
C_0(t, \bm x) =
t^{-\frac{\D_2+\D_1-\D_3}2} f\lef(\frac{\bm x^2}t\ri).
\end{align}
Substituting this to \eqref{RFCzeroDiffEq},
we obtain, using $N_3=N_2+N_1$, 
\begin{align}
\begin{split}
0=&
4y
\frac{d^2 f}{dy^2}
+ 2d 
\frac{df}{dy}
+2(N_1 -N_2) y \frac{df}{dy}
\\&
+ 
\lef( N_3 (\D_2 -\D_1) 
+(N_1- N_2) \frac d2   
\ri)
f
-N_2N_1 y f,
\end{split}
\end{align}
where $y=\frac{\bm x^2}t$.
This differential equation becomes the
confluent hypergeometric equation
\begin{align}
z \frac{d^2 v}{dz^2}+ (b-z) \frac{dv}{dz} - a v =0,
\label{RFKummerEq}
\end{align}
with 
\begin{align}
a=& \frac12\lef(\D_1-\D_2 + \frac d2 \ri), \label{RFaIntermsOfDelta}
\\
b=&\frac{d}{2}>0, \label{RFbEqualsdhalf}
\end{align}
by a simple transformation
$f= e^{ - \frac{N_1 }2 y} v,
z= \frac{N_3}{2} y$.
Thus we have
\begin{align}
C_0(t, \bm x) =
t^{-\frac{\D_2+\D_1-\D_3}2}
e^{  -\frac{N_1 }2 \frac{\bm x^2}{t}}
v\lef(
 \frac{N_3}2 \frac{\bm x^2}{t}
\ri).
\label{RFCzeroviaV}
\end{align}
We assume, for simplicity, that $a$ is not a negative integer.
(This can always be met for example by replacing ($\mO_1, \mO_2$) with ($\bmO_2, \bmO_1$)
so that $\D_1>\D_2$.)
The standard confluent hypergeometric functions $M(a,b,z)$ and $U(a,b,z)$
(in the notation of \cite{olver_nist_nodate})
are then linearly independent. Hence any solution can be written 
\begin{align}
v=A M (a, b, z)+ B U(a,b,z), \label{RFvConfluentHypergeometric}
\end{align}
where $A, B$ are constants.

In \cite{golkar_operator_2014}, it was advocated that 
appropriate regularity conditions on the OPE coefficient imply
\begin{align}
B=0.
\end{align}
This point will be examined
in the next subsection
\ref{RSASThreePointBoundaryCondition}.

\subsection{Boundary condition}
\label{RSASThreePointBoundaryCondition}

Here we will study the regularity conditions of the OPE,
leading to boundary conditions on the function $v$ 
appearing in 
\eqref{RFCzeroviaV}.
We focus in particular on the possible restrictions
on the coefficients $A$ and $B$ in \eqref{RFvConfluentHypergeometric}. 

Firstly, we observe that the regularity of $C_0$, in the limit $t\to 0$, $\bm x \to 0$ 
with $\frac{\bm x^2}{t}$ fixed at a nonzero finite value,
does not impose any conditions on $v$.
Hence, if we wish to restrict the form of $v$ we have to consider 
the limit 
$\frac{\bm x^2}{t} \to \infty$ 
(the equal-time OPE) 
or 
$\frac{\bm x^2}{t} \to 0$ 
(the ``equal-space'' OPE, {\it i.e.} the OPE $\mO_2 (t, \bm 0) \mO_1 (0, \bm 0)$).

Let us first examine the latter limit, {\it i.e.}
the behaviour at fixed $t>0$ and $\bm x \to \bm 0$.
Then $z\to+0$ and the prefactor in \eqref{RFCzeroviaV} behaves as, $e^{  -\frac{N_1 }2 \frac{\bm x^2}{t}} \to 1$.
In this limit, and for the values of $a, b$ relevant for us \eqref{RFaIntermsOfDelta}
\eqref{RFbEqualsdhalf}, we have
\begin{align}
M(a,b,z)=& 1 + O(z),
\\
U(a,b,z)\approx& \frac{\G(b-1)}{\G(a)} z^{1-b}, \qquad \lef(b=\frac d2,\   d=1, 3, 4, 5, \cdots\ri),
\\
U(a,b,z)\approx& -\frac1{\G(a)} \lef(\log z  +\text{const.}\ri), \qquad \lef(b=\frac d2,\  d=2\ri).
\end{align}
We have to distinguish the case $d=1$ and $d=2, 3, \cdots$.

For $d=2, 3, \cdots$, $U(a,b,z)\to \infty$ as $z\to 0$ whereas $M(a,b,z)\to 1$.
Hence if we require the existence of the ``equal-space'' OPE~\footnote{
We wish to note that 
it is far from obvious whether the requirement of the existence of the regular ``equal-space'' OPE 
is mandatory or not.
}, we obtain
\begin{align} 
B=0.
\end{align}
This is the result of Golkar and Son~\cite{golkar_operator_2014}.

For $d=1$, we have $M(a,b,z) \to 1$ and $U(a, b,z) \to 0$ as $z\to 0$.
Hence even if we require the existence of the OPE in the limit $\frac{\bm x^2 }{t}\to 0$, 
the coefficients $A$ and $B$ are not constrained. (If we require further
that the OPE be non-vanishing then we get $A\neq 0$.)

Let us next examine the behaviour at $z\to +\infty$, 
which corresponds to $t\to 0+$
with fixed $\bm x$, \ie\ to the limit of the equal-time OPE.
We shall see in fact that the OPE coefficient $C_0$ 
in this limit either diverges or goes
to zero.
This is not surprising:
Consider, in the free-field theory, the part of the OPE $\Psi(t,\bm x) \bPsi\,\bPsi(0,\bm 0)$ 
proportional to $\bPsi(0, \bm 0)$.
The OPE coefficient is essentially the two-point function
$\<\Psi(t,\bm x) \bPsi(0,\bm 0)\>
\sim \frac1{\sqrt t} e^{-\frac{\bm x^2}{2t}}$
and is singular in the limit $t\to 0+$ with fixed $\bm x$.

The general argument goes as follows.
In the limit, $z\to \infty$, we have
\begin{align}
M(a,b,z)\sim& \frac{1}{\G(a)} e^z z^{a-b},
\\
U(a,b,z) \sim& z^{-a}.
\end{align}
Let us separately consider the $A$- and $B$- type solution,
\ie\
the first and the second term in \eqref{RFvConfluentHypergeometric}, respectively.
The $A$-type solution gives, in the limit $t\to 0+$
with fixed $\bm x$, 
\begin{align}
\begin{split}
C_0(t, \bm x) 
\sim
A 
e^{  -\frac{N_1 }2 \frac{\bm x^2}{t}}
e^{\frac{N_3}2 \frac{\bm x^2}{t}}
=
A 
e^{  \frac{N_2}2 \frac{\bm x^2}{t}}.
\end{split}
\end{align}
Hence if $N_2>0$, $C_0$ diverges, and if $N_2<0$, $C_0$ goes to zero.
(We only write in the above formula the leading exponential behaviour.)
For the $B$-type solution, we have
\begin{align}
\begin{split}
C_0(t, \bm x) 
\approx&
B t^{-\frac{\D_2+\D_1-\D_3}2}
e^{  -\frac{N_1 }2 \frac{\bm x^2}{t}}
\lef(\frac{\bm x^2}{t}\ri)^{-\frac12 \lef(\D_1-\D_2+\frac d2\ri)}
\\
=&
B t^{-\D_2+\frac{d}{2}}
\lef(\bm x^2\ri)^{-\frac12 \lef(\D_1-\D_2+\frac d2\ri)}
e^{  -\frac{N_1 }2 \frac{\bm x^2}{t}},
\end{split}
\end{align}
using $\D_3=\frac d2$.
Again for $N_1\neq 0$,
$C_0$ either goes to $0$ or diverges.
Hence, 
for operators with nonzero charges
and $\D_3=\frac d 2$, 
the equal-time OPE either diverges or vanishes.

To summarise this subsection, the limit $\frac{\bm x^2}{t} \to +\infty$ (the equal-time OPE) 
is singular (the OPE coefficient either diverging or vanishing) 
and does not give constraints on the coefficients $A, B$.
If we require the regularity of $\frac{\bm x^2}{t} \to 0$ (the equal-space OPE),
we obtain $B=0$ for $d=2,3,\cdots$, but no 
constraints for $d=1$.

\subsection{Three-point function}
We consider the general form of the three-point function of primary operators~\cite{henkel_schrodinger_1994},
\begin{align}
\begin{split}
&\<\bmO_3 (t_3, x_3)\mO_2(t_2, x_2) \mO_1(t_1, x_1)\>
\\
=&
t_{31}^{-\frac{\D_3+\D_1-\D_2}2}
t_{21}^{-\frac{\D_2+\D_1-\D_3}2}
t_{32}^{-\frac{\D_3+\D_2-\D_1}2}
e^{-\frac{|N_2|}{2}\frac{\bm x_{21}^2}{t_{21}}
-\frac{|-N_3|}{2}\frac{\bm x_{31}^2}{t_{31}}}
F_{123}(w^2),
\end{split}
\label{RFThreePointGeneralInAppendix}
\end{align}
where $t_3>t_2>t_1$.
Here $w$ is the Schr\"odinger 
invariant spacetime cross-ratio defined by \eqref{RFPropertyW}.
To be specific, we consider the case, 
\begin{align}
&N_1>0, N_2 <0, N_3>0, -N_3 <0,
\\
&N_1+N_2-N_3=0.
\end{align}
(This choice is consistent with the three-point function we studied in section
\ref{RSSThreePointGeneralPosition}. The comparison is done at the end of this subsection.)

To compare with the OPE coefficient, we set
\begin{align}
(t_1, \bm x_1) =& (0, \bm 0),
\\
(t_2, \bm x_2) =& (t, \bm x),
\end{align}
and consider the limit $(t, \bm x) \to (0, \bm 0)$ with $\frac{\bm x^2}{t}$ fixed.
Then we have 
\begin{align}
w^2=
\frac{(\bm x_3 -\bm x)^2}{t_3-t}
-
\frac{(\bm x_3)^2}{t_3}
+
\frac{\bm x^2}{t}
\longrightarrow
\frac{\bm x^2}{t},
\end{align}
and hence
\begin{align}
\begin{split}
&\<\bmO_3 (t_3, x_3)\mO_2(t_2, x_2) \mO_1(t_1, x_1)\>
\\
\longrightarrow&
t_{3}^{-\D_3}
e^{-\frac{N_3}{2}\frac{\bm x_{3}^2}{t_{3}}}
t^{-\frac{\D_2+\D_1-\D_3}2}
e^{-\frac{-N_2}{2}\frac{\bm x^2}{t}}
F\lef(\frac{\bm x^2}{t}\ri).
\end{split}
\label{RFThreePtLimitByF}
\end{align}

On the other hand, the OPE yields
\begin{align}
\begin{split}
&\<\bmO_3 (t_3, \bm x_3)\mO_2(t, \bm x) \mO_1(0, \bm 0)\>
\\
\to&
\<\bmO_3 (t_3, \bm x_3 )\mO_3(0,\bm 0)\>
\times
C_0(t, \bm x)
\\
=&
t_{3}^{-\D_3}
e^{-\frac{N_3}{2}\frac{\bm x_{3}^2}{t_{3}}}
\times
t^{-\frac{\D_2+\D_1-\D_3}2}
e^{  -\frac{N_1 }2 \frac{\bm x^2}{t}}
v\lef(
 \frac{N_3}2 \frac{\bm x^2}{t}
\ri),
\end{split}
\label{RFThreePtLimitByv}
\end{align}
using \eqref{RFTwoPointGeneral} and \eqref{RFCzeroviaV}.
Note that the contributions from descendants of $\mO_3$ 
vanish in this limit.

Comparing \eqref{RFThreePtLimitByF} and \eqref{RFThreePtLimitByv}, we finally obtain
\begin{align}
\begin{split}
F\lef(y\ri)
=&
e^{-\frac{N_3}2 y}
v\lef(y\ri)
\\
=&
e^{-\frac{N_3}2 y}
\lef( 
AM\lef(\frac12\lef(\D_1-\D_2+\frac d2\ri), \frac d2, \frac{N_3}{2} y \ri)
+
BU\lef(\frac12\lef(\D_1-\D_2+\frac d2\ri), \frac d2, \frac{N_3}{2} y \ri)
\ri),
\end{split}
\end{align}
using \eqref{RFvConfluentHypergeometric}, \eqref{RFaIntermsOfDelta}, \eqref{RFbEqualsdhalf}.

Substituting back to \eqref{RFThreePointGeneralInAppendix}, the three-point function is
\begin{align}
\begin{split}
&\<\bmO_3 (t_3, x_3)\mO_2(t_2, x_2) \mO_1(t_1, x_1)\>
\\
=&
t_{31}^{-\frac{\D_3+\D_1-\D_2}2}
t_{21}^{-\frac{\D_2+\D_1-\D_3}2}
t_{32}^{-\frac{\D_3+\D_2-\D_1}2}
e^{-\frac{|N_2|}{2}\frac{\bm x_{21}^2}{t_{21}}
-\frac{|-N_3|}{2}\frac{\bm x_{31}^2}{t_{31}}}
\\
\times&
e^{-\frac{N_3}2 w^2}
\lef( 
AM\lef(\frac12\lef(\D_1-\D_2+\frac d2\ri), \frac d2, \frac{N_3}{2} w^2 \ri)
+
BU\lef(\frac12\lef(\D_1-\D_2+\frac d2\ri), \frac d2, \frac{N_3}{2} w^2 \ri)
\ri).
\end{split}
\end{align}

To compare with our result presented  in section \ref{RSSThreePoint}, 
we put 
$\mO_1=\bPhi,
\mO_2=\Psi,
\bmO_3=\Psi$ with
$N_1=2, N_2=-1, N_3=1, \D_1=\frac32+\n, \D_2=\frac12, \D_3=\frac12$, 
and $d=1$. 
We see that for our model,
the result of our explicit computation \eqref{RFThreePointConfluentHypergeometricWithExp}
supports Golkar and Son's ansatz, $B=0$, even if the regularity conditions 
do not require $B=0$ in $d=1$.

\section{Free-Boson limit}
\label{RSAFree}
In this appendix,
we consider the limiting case
$\n=-\frac12$, which is the free boson theory.

\subsection{Pairwise equal-time four-point function}
Substituting
\begin{align}
I_{-\frac12}\lef(y\ri)
=&
\sqrt{\frac{2}{\pi}}\frac1{\sqrt{y}}
\cosh y
\end{align}
into the pairwise equal-time four-point function \eqref{RFPairwiseEqualtime4ptFunction}
(which is valid for $t>0, x_{21}>0, x_{43}>0$), we obtain
\begin{align}
\begin{split}
&
\langle
\Psi(t, x_4) 
\Psi(t, x_3) 
\bPsi(0, x_2) 
\bPsi(0, x_1) 
\rangle
\\
=&
e^{-\frac{
        x_{21}^2+x_{43}^2+\lef(x_{3}+x_{4}-x_{1}-x_{2}\ri)^2}
        {4t}}
\times
\sqrt{
    \frac{x_{21}x_{43}}
        {4\pi t^3}}
\sqrt{\frac{2}{\pi}}
\frac1{\sqrt{
    \frac{x_{21} x_{43}}{2t}
}}
\cosh{
    \frac{x_{21} x_{43}}{2t}
}
\\
=&
e^{-\frac{
        x_{21}^2+x_{43}^2+\lef(x_{3}+x_{4}-x_{1}-x_{2}\ri)^2}
        {4t}}
\times
\frac{1}{\pi t}
\cosh{
    \frac{x_{21} x_{43}}{2t}
}
\\
=&
\frac{1}{2\pi t}
\lef(
e^{-\frac{
        (x_{21}+x_{43})^2+\lef(x_{3}+x_{4}-x_{1}-x_{2}\ri)^2}
        {4t}}
+
e^{-\frac{
        (x_{21}-x_{43})^2+\lef(x_{3}+x_{4}-x_{1}-x_{2}\ri)^2}
        {4t}}
\ri)
\\
=&
\frac{1}{2\pi t}
\lef(
e^{-\frac{ x_{14}^2+x_{23}^2}
        {2t}}
+
e^{-\frac{ x_{13}^2+x_{24}^2}
        {2t}}
\ri)
\\
=&
K(x_4, x_1;t)
K(x_3, x_2;t)
+
K(x_4, x_2;t)
K(x_3, x_1;t),
\end{split}
\end{align}
where $K(x',x;t)$ in the last line is the free propagator \eqref{RFFreePropagator}.
This is the expected result for free bosons.

\subsection{Three-point function}
\label{RSASFreeThreePoint}
Substituting $\n=-\frac12$ into 
\eqref{RFThreePointConfluentHypergeometricWithExp},
and then using $M(1, b, z)=1$ for general $b$, we obtain 
\begin{align}
\langle
\Psi(t_3, x_3)
\Psi(t_2, x_2)
\bPhi\lef(t_1, x_1\ri)
\rangle
=&
\frac1{\sqrt{\pi}}
\sqrt{\frac{1}{t_{31}t_{21}}}
e^{-\frac{x_{21}^2}{2t_{21}}}
e^{-\frac{x_{31}^2}{2t_{31}}}.
\end{align}
Here $\bPhi=\sqrt\pi \bPsi^2$ for the free-field theory;
the normalisation condition is fixed by the two-point function \eqref{RFTwoPointPhibPhi}.
This reproduces the free theory result.

%DRAFT
\ifdraft
\nocite{*}
\fi
%/DRAFT
\bibliography{z}
\bibliographystyle{JHEP-2}

%%%%%%%%%%%%%%%%%%%%%%%%%%%%%%%%%%%%%%%%%%%%%%%%%%%%%%%%%%%%%%%%%%%%%%%%%%%%
%%%%%%%%%%%%%%%%%%%%%%%%%%%%%%%%%%%%%%%%%%%%%%%%%%%%%%%%%%%%%%%%%%%%%%%%%%%%

\end{document}